\documentclass[12pt]{article}
\newlength{\dinwidth}
\newlength{\dinmargin}
\usepackage{graphicx}          
\setkeys{Gin}{keepaspectratio} 

\newcommand {\pom}  {I\hspace{-0.2em}P}
\newcommand {\reg}  {I\hspace{-0.2em}R}
\begin{document}

\def\ap#1#2#3   {{\em Ann. Phys. (NY)} {\bf#1} (#2) #3}   
\def\apj#1#2#3  {{\em Astrophys. J.} {\bf#1} (#2) #3.} 
\def\apjl#1#2#3 {{\em Astrophys. J. Lett.} {\bf#1} (#2) #3.}
\def\app#1#2#3  {{\em Acta. Phys. Pol.} {\bf#1} (#2) #3.}
\def\ar#1#2#3   {{\em Ann. Rev. Nucl. Part. Sci.} {\bf#1} (#2) #3.}
\def\cpc#1#2#3  {{\em Computer Phys. Comm.} {\bf#1} (#2) #3.}
\def\epj#1#2#3  {{\em Europ. Phys. J.} {\bf#1} (#2) #3}
\def\err#1#2#3  {{\it Erratum} {\bf#1} (#2) #3.}
\def\ib#1#2#3   {{\it ibid.} {\bf#1} (#2) #3.}
\def\jmp#1#2#3  {{\em J. Math. Phys.} {\bf#1} (#2) #3.}
\def\ijmp#1#2#3 {{\em Int. J. Mod. Phys.} {\bf#1} (#2) #3}
\def\jetp#1#2#3 {{\em JETP Lett.} {\bf#1} (#2) #3}
\def\jpg#1#2#3  {{\em J. Phys. G.} {\bf#1} (#2) #3.}
\def\mpl#1#2#3  {{\em Mod. Phys. Lett.} {\bf#1} (#2) #3.}
\def\nat#1#2#3  {{\em Nature (London)} {\bf#1} (#2) #3.}
\def\nc#1#2#3   {{\em Nuovo Cim.} {\bf#1} (#2) #3.}
\def\nim#1#2#3  {{\em Nucl. Instr. Meth.} {\bf#1} (#2) #3.}
\def\np#1#2#3   {{\em Nucl. Phys.} {\bf#1} (#2) #3}
\def\pcps#1#2#3 {{\em Proc. Cam. Phil. Soc.} {\bf#1} (#2) #3.}
\def\pl#1#2#3   {{\em Phys. Lett.} {\bf#1} (#2) #3}
\def\prep#1#2#3 {{\em Phys. Rep.} {\bf#1} (#2) #3}
\def\prev#1#2#3 {{\em Phys. Rev.} {\bf#1} (#2) #3}
\def\prl#1#2#3  {{\em Phys. Rev. Lett.} {\bf#1} (#2) #3}
\def\prs#1#2#3  {{\em Proc. Roy. Soc.} {\bf#1} (#2) #3.}
\def\ptp#1#2#3  {{\em Prog. Th. Phys.} {\bf#1} (#2) #3.}
\def\ps#1#2#3   {{\em Physica Scripta} {\bf#1} (#2) #3.}
\def\rmp#1#2#3  {{\em Rev. Mod. Phys.} {\bf#1} (#2) #3}
\def\rpp#1#2#3  {{\em Rep. Prog. Phys.} {\bf#1} (#2) #3.}
\def\sjnp#1#2#3 {{\em Sov. J. Nucl. Phys.} {\bf#1} (#2) #3}
\def\shep#1#2#3 {{\em Surveys in High Energy Phys.} {\bf#1} (#2) #3}
\def\spj#1#2#3  {{\em Sov. Phys. JEPT} {\bf#1} (#2) #3}
\def\spu#1#2#3  {{\em Sov. Phys.-Usp.} {\bf#1} (#2) #3.}
\def\zp#1#2#3   {{\em Zeit. Phys.} {\bf#1} (#2) #3}

\title{\vspace{5cm}
\bf{ The structure of the Troika: Proton, Photon and Pomeron, as seen at 
HERA~\footnote{Talk presented at the XXI International Workshop on the
fundamental problems of High Energy Physics and Field Theory, 23--25
June 1998, Protvino, Russia.}
}
\vspace{2cm}}

\author{
 {\bf Aharon Levy} \\ 
{\small \sl School of Physics and Astronomy}\\ {\small \sl Raymond and 
Beverly Sackler Faculty of Exact Sciences}\\
  {\small \sl Tel Aviv University, Tel Aviv, Israel}
}
\date{ }
\maketitle

\begin{abstract}
HERA, the electron-proton collider, enables to probe the proton with a
high resolving power due to the deep inelastic scattering reactions at
high $Q^2$ values.  In the low $Q^2$ region, one can study the
properties of the photon. The large fraction of diffractive events
found both in the low and high $Q^2$ region allows the study of the
Pomeron. A review of what we have learned from HERA so far about the
structure of these three objects is presented.
\end{abstract}

\vspace{-18cm}
\begin{flushleft}
TAUP 2531--98 \\
November 1998 \\
\end{flushleft}

\setcounter{page}{0}
\thispagestyle{empty}
\newpage  

%

%
%
%
%
%
%
%



\section{Introduction}

The ultimate goal of high energy physics is to search for the fundamental 
constituents of matter and to understand their interactions. This view 
was already expressed by Newton in the introduction to his book on Optics:

{\it
Now the smallest particles of matter cohere by the strongest attraction, 
and compose bigger particles of weaker virtue; and many of these may 
cohere and compose bigger particles whose virtue is still weaker, and so 
on for diverse successions, until the progression ends in the biggest 
particles on which the operations in chemistry, and the colors of 
natural bodies depend, and which by cohering compose bodies of a sensible 
magnitude. There are therefore agents in nature able to make the 
particles of bodies stick together by very strong attractions. And it is 
the business of experimental philosophy to find them out.}

There are two ways of studying structure of matter: the static way and
the dynamic one. In the first approach, symmetry arguments like the
ones used by Gel-Mann and Neeman led to the construction of a
`Mendeleev table' of the known particles, which eventually brought
Gel-Mann and Zweig to postulate the existence of quarks. In the
dynamic way one tries to `look' at the particles. This is the
`Rutherford way' in which one bombards the target with particles of
known identity and searches for structure through the study of the
outcome of the bombardment. This was used in the electron--proton deep
inelastic scattering (DIS) experiments at SLAC. The underlying
assumption was that one uses a projectile whose properties are well
known, who behaves like a pointlike structureless particle. Any
structure that is being observed following the collision is assigned
to the proton and its constituents. This way, the study of the SLAC
DIS experiment showed that the DIS cross section behaves like that
expected from the interaction of electrons with pointlike particles,
called partons, which were later on shown to have the expected
properties of quarks, namely spin $\frac{1}{2}$ and fractional
charge. This is the quark--parton model (QPM).

\subsection{DIS Kinematics}

\begin{figure}[hbt]
\begin{center}
  \includegraphics [width=\hsize,totalheight=8cm]    
  {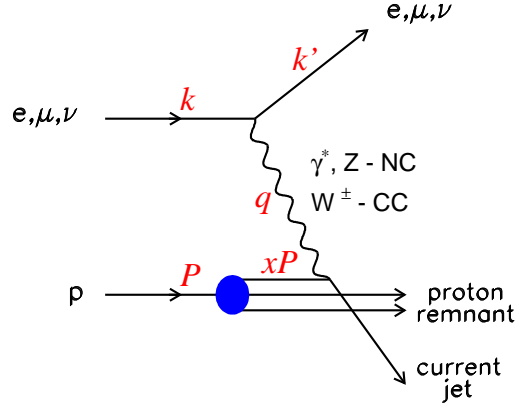}
\end{center}
\vspace{-1.5cm}
\caption {\it
{Diagram describing a DIS process on a proton.
}}
\label{fig:kinematics}
\end{figure}

The SLAC DIS experiment introduced the use of some important kinematic
variables relevant to the notion of `looking' at the structure of a
particle. In figure~\ref{fig:kinematics} a lepton with mass $m_l$ and
four-vector $k(E_l,\vec{k})$ interacts with a proton with mass $m_p$
and four-vector $P(E_p,\vec{p})$ through the exchange of a gauge vector
boson, which can be $\gamma$, $Z^0$ or $W^\pm$, depending on the
circumstances. The four-vector of the exchanged boson is $q(q_0,\vec{q})$.

With these notations one can define the following variables,
\begin{eqnarray}
         q    &=&     k - k^\prime     \\
       \nu  &\equiv&  \frac{P\cdot q}{m_p}      \\
         y  &\equiv&  \frac{P\cdot q}{P\cdot k} \\
       W^2    &=&     (P + q)^2                 \\
         s    &=&     (k + P)^2.
\end{eqnarray}
The meaning of the variables $\nu$ and $y$ is most easily realized in
the rest frame of the proton. In that frame $\nu$ is the energy of the
exchanged boson, and $y$ is the fraction of the incoming lepton energy
carried by the exchanged boson.  The variable $W^2$ is the squared
center of mass energy of the gauge--boson proton system, and thus also
the squared invariant mass of the hadronic final state. The variable
$s$ is the squared center of mass energy of the lepton proton system.

The four momentum transfer squared at the lepton vertex can be approximated
as follows (for $m_l, m_l^\prime \ll E, E^\prime$),
\begin{equation}
        q^2 = (k-k^\prime)^2
            = m_l^2+{m_l^\prime}^2-2kk^\prime
            \approx -2EE^\prime (1-\cos\theta)
            < 0 \ .
\label{eq:q2}
\end{equation}
The scattering angle $\theta$ of the outgoing lepton is defined
with respect to the incoming lepton direction.
The variable which is mostly used in DIS is the negative value of the four
momentum transfer squared at the lepton vertex,
\begin{equation}
                        Q^2\equiv-q^2 \ .
\end{equation}
One is now ready to define the other variable most frequently used in
DIS, namely the dimensionless scaling variable $x$,
\begin{equation}
                   x \equiv \frac{Q^2}{2 p\cdot q} \ .
\end{equation}
To understand the physical meaning of this variable, one goes to a
frame in which masses and transverse momenta can be neglected - the
so-called infinite momentum frame. In this frame the variable $x$ is
the fraction of the proton momentum carried by the massless parton
which absorbs the exchanged boson in the DIS interaction. This
variable, defined by Bjorken, is duly referred to as Bjorken-$x$.

The diagram in figure~\ref{fig:kinematics} describes both the
processes in which the outgoing lepton is the same as the incoming
one, which are called neutral current reactions (NC), as well as those
in which the nature of the lepton changes (conserving however lepton
number) and which are called charged current processes (CC). In the NC
DIS reaction, the exchanged boson can be either a virtual photon
$\gamma^*$, if $Q^2$ is not very large and then the reaction is
dominantly electromagnetic, or can be a $Z^0$ which dominates the
reaction at high enough $Q^2$ values and the process is dominated by
weak forces. In case of the CC DIS reactions, only the weak forces are
present and the exchange bosons are the $W^\pm$.

\subsection{The proton structure function $F_2$}

The inclusive Born cross section of a NC DIS reaction can be expressed
(for $Q^2 \ll m_Z^2$) as,
\begin{equation}
\frac{d^2\sigma^{Born}}{dx dQ^2} = \frac{4\pi\alpha^2}{xQ^4}
\left[\frac{y^2}{2}2xF_1 + (1-y)F_2\right],
\end{equation}
where $\alpha$ is the electromagnetic coupling constant. The two
structure functions $F_1$ and $F_2$ are related to the transverse and
longitudinal $\gamma^*p$ cross sections~\cite{leader-predazzi}.

\begin{figure}[hbt]
\begin{center}
\includegraphics [width=\hsize,totalheight=12cm]    
  {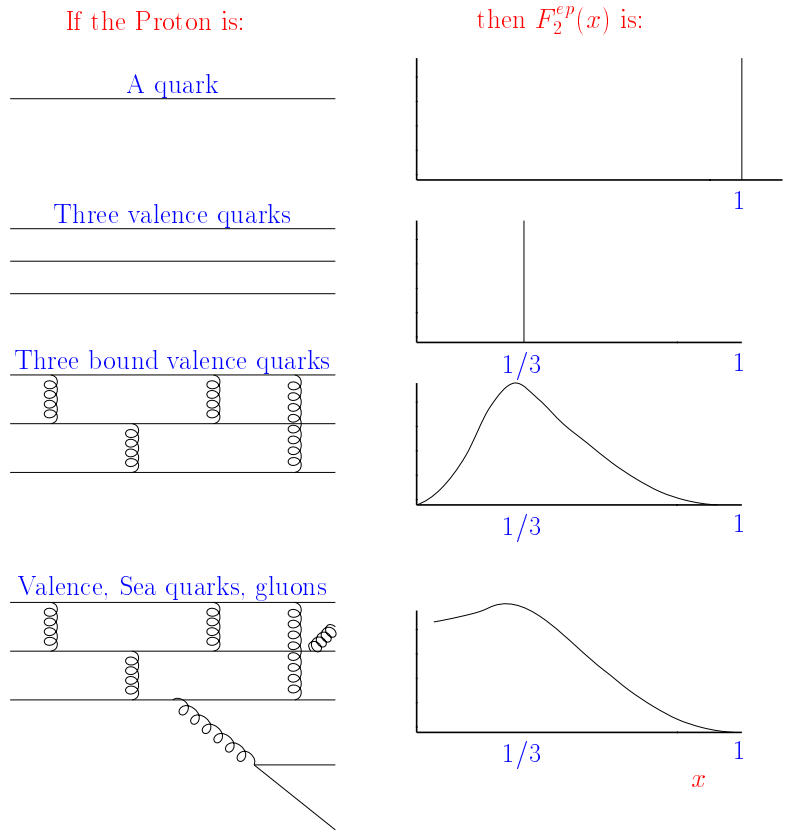}
\end{center}
\vspace{-.5cm}
\caption {\it
{The expected dependence of $F_2$ on $x$ given a certain scenario of
the structure of the proton.}}
\label{fig:f2-meaning}
\end{figure}

The relation between the values of $F_2$ and their meaning as far as
the structure of the proton is concerned can be best seen in a figure
adopted from the book of Halzen and Martin~\cite{halzen-martin}. In
figure~\ref{fig:f2-meaning} one sees what are the expectations for the
distribution of $F_2$ as function of $x$ given a certain picture of
the proton. 
The static approach mentioned above could explain most properties of
the known particles with the proton being composed of three valence
quarks. The first measurements of $F_2$~\cite{first-f2} indeed
confirmed this picture and the QPM was constructed. Later
measurements~\cite{scaling-violation} showed that sea quarks and
gluons are also present in the proton, as the bottom part of
figure~\ref{fig:f2-meaning} shows.

Clearly in order to have a good picture of the structure of the proton
one needs to `see' the partons and thus needs the means to have a good
resolving power. If we denote by $\Delta$ the sizes one can resolve
inside the proton, the higher the virtuality of the exchanged gauge
boson in figure~\ref{fig:kinematics}, the smaller $\Delta$ gets,
\begin{equation}
\Delta \sim \frac{\hbar c}{\sqrt{Q^2}} = \frac{0.197\ {\rm GeV \ fm}}
{\sqrt{Q^2}}.
\end{equation}
Thus for $Q^2$ = 4 GeV$^2$, $\Delta$ = $10^{-14}$cm; for $Q^2$ = 400
GeV$^2$, $\Delta$ = $10^{-15}$cm; and for $Q^2$ = 40000 GeV$^2$,
$\Delta$ = $10^{-16}$cm. 

\subsection{The HERA collider}

\begin{figure}[hbt]
\begin{center}
  \includegraphics [bb= 23 147 572 666,width=\hsize,totalheight=12cm]      
  {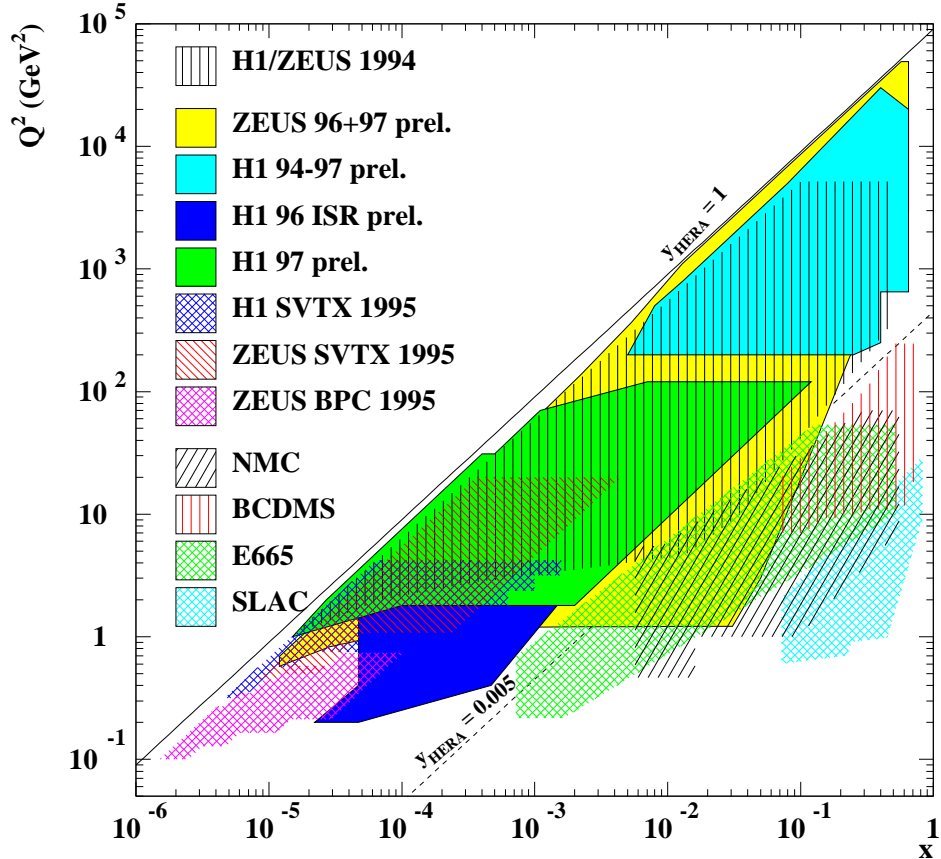}
\end{center}
\vspace{-.5cm}
\caption {\it
{The $x$--$Q^2$ kinematic plane of some of the fixed target and of the  
HERA collider DIS experiments.}}
\label{fig:x-q2}
\end{figure}
How does one achieve high $Q^2$ values? One can show that the
following relation holds between $Q^2$, $x$, $y$ and $s$,
\begin{equation}
Q^2 \approx x y s,
\end{equation}
which means that $Q^2_{max} \approx s$. Therefore in order to reach
large $Q^2$ values one needs to build a large $s$ $ep$ collider,
which is what was done at DESY with the HERA collider.

HERA~\cite{hera} is the first $ep$ collider, where a beam of 27.5 GeV
electrons (or positrons) collides with a beam of 820 GeV protons
yielding a center of mass energy of 300 GeV, or $s \approx$ 90000
GeV$^2$.~\footnote{Presently the proton beam energy was increased to
  920 GeV, increasing the center of mass energy to 318 GeV and $s$ to
  101200 GeV$^2$.} It has increased the available kinematic $x$--$Q^2$
plane by two orders of magnitude going up in $Q^2$ and down in $x$.
This can be seen in figure~\ref{fig:x-q2} which shows the range of
existing measurements of some fixed target DIS experiments
(SLAC~\cite{slac}, BCDMS~\cite{bcdms}, E665~\cite{e665},
NMC~\cite{nmc}) together with the HERA measurements by the
H1~\cite{h1f2} and ZEUS~\cite{zeusf2} collaborations.

During the period 1994--1997 the HERA collider has delivered an
integrated luminosity of more than 70 pb$^{-1}$ out of which about 47
bp$^{-1}$ could be used for physics analyses. At present much effort
is concentrated on a luminosity upgrade program, to come into effect
in the year 2000, which will deliver an integrated luminosity of about
1 fb$^{-1}$ till the year 2005.

\subsection{Low-$x$ at HERA}

\begin{figure}[hbt]
\begin{center}
\includegraphics [width=\hsize,totalheight=7cm]      
  {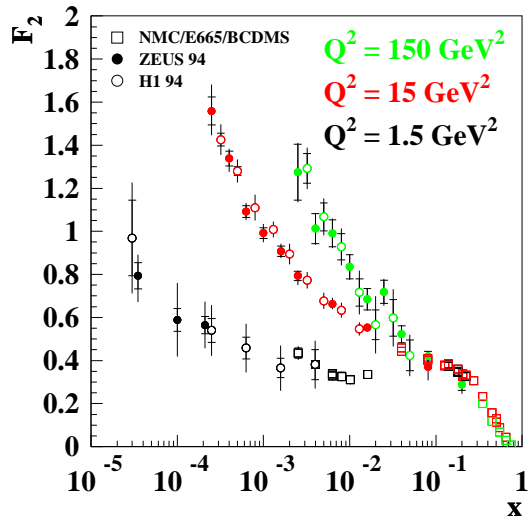}
\end{center}
\vspace{-.5cm}
\caption {\it
{The proton structure function $F_2$ as function of $x$ for three $Q^2$ values.
}}
\label{fig:f2-x-rise}
\end{figure}

A closer look at figure~\ref{fig:x-q2} reveals two facts, one obvious
and the other quite surprising. The two HERA collaborations strive to
$Q^2$ values as high as possible. With the high statistics 1996--1997
data, the experiments have measured some DIS events with $Q^2 \sim$
40000 GeV$^2$. However, surprisingly, there is an effort also to go to
as low $Q^2$ as possible, which also allows measuring at very low $x$
values. The reason for trying to reach very low $Q^2$ and low $x$ values
can be seen from figure~\ref{fig:f2-x-rise}. In this figure, the
dependence of of the proton structure function $F_2$ on Bjorken $x$ is
shown for three values of $Q^2$.
One sees a clear rise of the structure function with decreasing
$x$. However, as $Q^2$ gets smaller this rise is less steep. What does
this plot tell us? In order to understand it, let us first look at the
variable $x$. It is related to $Q^2$ and to $W$ (the $\gamma^* p$
center of mass energy) as,
\begin{equation}
W^2 = Q^2 (\frac{1}{x} - 1) + m_p^2 \approx \frac{Q^2}{x},
\end{equation}
where the approximate relation is good for low $x$ values. Thus for
fixed $Q^2$, going in the low $x$ directions means increasing $W$.

The proton structure function $F_2$ can be related to the total
$\gamma^* p$ cross section $\sigma_{tot}^{\gamma^* p}$ through the
relation,
\begin{equation}
F_2 = \frac{Q^2 (1 - x)}{4\pi^2\alpha}\frac{Q^2}{Q^2 + 4m_p^2x^2}
\sigma_{tot}^{\gamma^* p} \approx \frac{Q^2}{4\pi^2\alpha}
\sigma_{tot}^{\gamma^* p},
\label{eq:f2-sig}
\end{equation}
where we have used the Hand~\cite{hand} definition of the flux of
virtual photons, and again the approximate expression holds for low
$x$ values. Thus, the behaviour seen in figure~\ref{fig:f2-x-rise} can
be interpreted as a rising $\gamma^* p$ cross section with increasing
$W$, where the increase gets steeper as $Q^2$ increases. How does this
steepness decrease as one goes to lower and lower values of $Q^2$? Is
there a sharp or a smooth transition? What happens at $Q^2$ = 0 when
the photon is real?

\subsection{Low $Q^2$ at HERA}

These questions motivated the HERA experimentalists to try to measure
the behaviour of the structure function at low $Q^2$ and also to
measure the real photoproduction cross section in the high $W$ region
of HERA. How does one do low $Q^2$ physics in a machine which was
built to reach highest possible $Q^2$ values? A look at
equation~(\ref{eq:q2}) shows that the value of $Q^2$ is determined by
the energies of the incoming ($E$) and the outgoing ($E^\prime$)
electrons and by the scattering angle $\theta$ of the outgoing
electron with respect to the incoming one,
\begin{equation}
Q^2 = 2 E E^\prime (1 - \cos\theta).
\end{equation}
To get to low values of $Q^2$, the angle $\theta$ has to be small and
therefore the scattered electron remains in the beam pipe. However, if
one can arrange to measure the outgoing electron at very low
scattering angles in a special detector, one has a handle of measuring
low $Q^2$ photons, with the possibility to go down to the quasi--real
photon case for extremely small angles.

\begin{figure}[hbt]
\begin{center}
 \includegraphics [bb = 1 1 581 301,totalheight=6cm,width=\hsize]
 {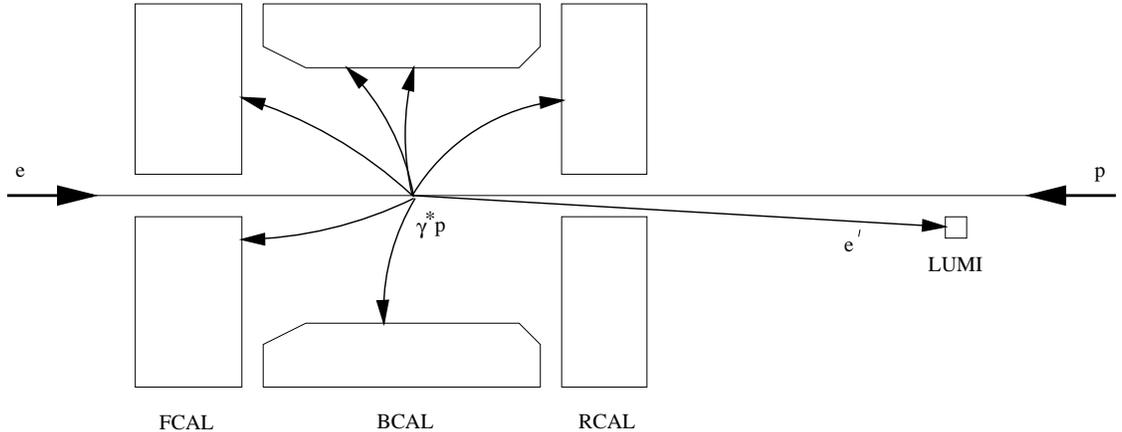}
\end{center}
\caption {\it
{
A diagrammatic example of a photoproduction event in the ZEUS detector,
where the scattered electron is detected in the small angle
electron calorimeter LUMI.
}}
\label{fig:gpevent}
\end{figure}

The two experiments, H1 and ZEUS, have each built a small calorimeter
at a distance of about 30 m from the interaction point which allows to
detect electrons which were scattered by less than 5 mrad with respect
to the incoming electron direction. This ensures that the virtuality
of the exchanged photons is in the range $10^{-8} < Q^2 < 0.02$
GeV$^2$, with the median $Q^2 \approx 10^{-5}$ GeV$^2$.  A diagrammatic
example of an event produced by a quasi--real photon, denoted as a
photoproduction event, is shown in figure~\ref{fig:gpevent}.
In this event the scattered electron is detected in the electron
calorimeter. This calorimeter is part of the luminosity detector,
which includes also a photon detector at a distance of about 100 m
from the interaction point.

\begin{figure}[hbt]
\begin{center}
 \includegraphics [totalheight=6cm,width=\hsize]
 {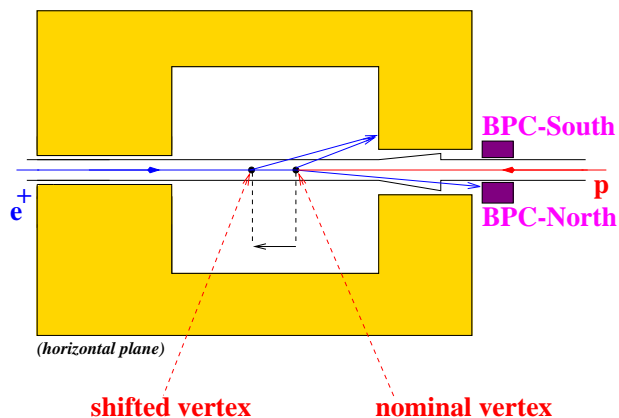}
\end{center}
\caption {\it
{ A diagrammatic description of the ZEUS detector showing the
beam--pipe calorimeter (BPC), and the nominal and the shifted vertex
positions. 
}}
\label{fig:bpc}
\end{figure}

The way to tag events with $Q^2$ in the range of 0.1-1 GeV$^2$ is
through two methods. One methods is based on moving the position of
the interaction vertex towards the incoming electron beam. By shifting
the vertex in this direction one increases the possibility to measure
low-angle scattered electrons in the rear part of the main
calorimeter. The other method is similar to that in the
photoproduction case described above. It consists of building a
special calorimeter to detect the small-angle scattered electron. This
was done by building two parts of a small calorimeter around the beam
pipe which accordingly was named the beam-pipe calorimeter (BPC). Both
methods are diagrammatically described in figure~\ref{fig:bpc}.
 
It is thus clear from the above discussion that HERA has also become a
source of high $W$ quasi-real photons. In fact, the highest $W$ photon
beams before HERA were in the range of 20 GeV and HERA has increased
this by one order of magnitude. This allows among other things to
study the structure of the photon at low $x$ values. 

\subsection{The concept of the structure of the photon}

\begin{figure}[hbt]
\begin{center}
 \includegraphics [bb= 47 40 514 484,totalheight=5cm,width=\hsize]
 {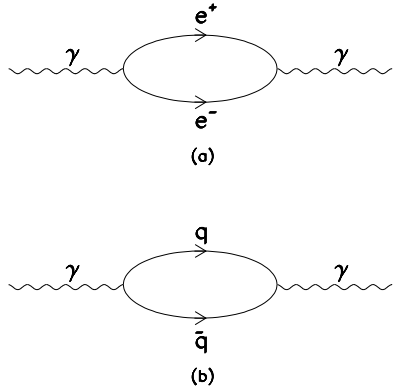}
\end{center}
\vspace{-.5cm}
\caption {\it
{ Fluctuation of a photon into (a) an $e^+e^-$ pair, (b) a $q\bar{q}$ pair. 
}}
\label{fig:fluc}
\end{figure}

What do we mean by `the structure of the photon'? The photon is the
gauge particle mediating the electromagnetic interactions and thus one
would expect it to be an elementary point-like particle. How can one
talk then about the structure of the photon? We know from low $W$ data
that when the photon interacts with hadrons it behaves like a
hadron. This property is well described by the vector dominance model
(VDM)~\cite{vdm} in which the photon turns first into a hadronic
system with the quantum numbers of a vector meson before it interacts
with the target hadron. The justification of this picture was given by
Ioffe~\cite{ioffe} who used time arguments. Just like a photon can
fluctuate in QED into a virtual $e^+e^-$ pair (figure~\ref{fig:fluc}a)
it can also fluctuate into a $q\bar{q}$ pair
(figure~\ref{fig:fluc}b). 
As long as the fluctuation time $t_f$ is
small compared to the interaction time $t_{int}$ the photon will
interact directly with the hadron. However if $t_f \gg t_{int}$ the
interaction will be between the $q\bar{q}$ pair and the hadron and
will look like a hadronic interaction. The fluctuation time of a
photon with energy $E_{\gamma}$ which is large compared to the
hadronic mass $m_{q\bar{q}}$ into which the photon fluctuates
($E_{\gamma} \gg m_{q\bar{q}}$) is given by,
\begin{equation}
t_f \simeq \frac{2E_{\gamma}}{m^2_{q\bar{q}}}.
\end{equation}
This is the case for a real photon. For a virtual photon $\gamma^*$
the fluctuation time is given by,
\begin{equation}
t_f \simeq \frac{2E_{\gamma}}{m^2_{q\bar{q}} + Q^2}.
\label{eq:tfluc-virtg}
\end{equation}
The interaction time with a proton is of the order of its radius,
$t_{int} \approx r_p \sim$ 1 fm. Thus while a high energy real photon
develops a structure due to its long fluctuation time compared to the
interaction time, a highly virtual photon has no time to acquire a
structure before probing the proton.

\begin{figure}[hbt]
\begin{center}
  \includegraphics [width=\hsize,totalheight=6cm] 
{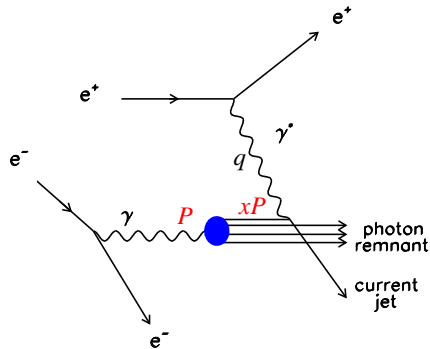}
\end{center}
\vspace{-.5cm}
\caption {\it
{Diagram describing a DIS process on a quasi-real photon using the
reaction $e^+ e^- \to e^+ e^- X$.}}
\label{fig:dis-photon}
\end{figure}

The structure of real photons has been indeed studied in $e^+e^-$
interactions where the photon structure function $F_2^\gamma$ has been
measured in a similar DIS type of experiment as on the proton. A
diagram describing this is shown in figure~\ref{fig:dis-photon} where
the proton target is replaced by a quasi-real photon target at the
vertex where the electron has a very small scattering angle.
The $x$ values reached in these experiments were not small due to the
fact that the available $W$ of the $\gamma^* \gamma$ system was
relatively small. 

As stated above, also at HERA one can study the photon structure. The
exchanged photon, which at high $Q^2$ is a probe, can change its role
at very low $Q^2$ and become a quasi-real photon target. It can be
probed by a high transverse momentum parton from the proton. We shall
discuss this in more details in section~\ref{sec:photon}.

The high $W$ values attained at HERA give a large lever arm to study
the energy behaviour of the total photoproduction cross section
$\sigma_{tot}({\gamma p})$. Does it show the same behaviour as the
total hadron-hadron cross sections? The latter were shown by Donnachie
and Landshoff~\cite{dl} to have a simple behaviour, independent on the
incoming hadron, and well described by the Regge model.

\begin{figure}[hbt]
\begin{center}
  \includegraphics [width=\hsize,totalheight=6cm] 
{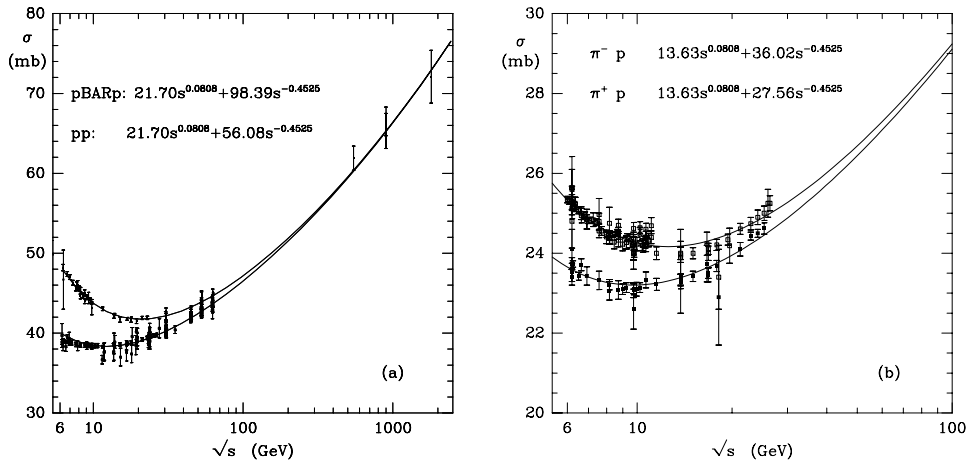}
\end{center}
\vspace{-.5cm}
\caption {\it
{The total cross section data of $\bar{p} p, p p$ and $\pi^\pm p$ as
function of the center of mass energy $\sqrt{s}$. The DL
parameterization is shown as the solid lines.  }}
\label{fig:hhtot}
\end{figure}

Donnachie and Landshoff (DL) succeeded to describe all available
$\bar{p} p$, $p p$, $K^{\pm} p$, and $\pi^{\pm} p$ total cross section
values by a simple parameterization of the form $\sigma_{tot} = X
s^{0.0808} + Y s^{-0.4525}$, where $s$ in the square of the total
center of mass energy and $X$ and $Y$ are parameters depending on the
interacting particles. The value of $X$ is constrained to be the same
for particle and anti-particle beams to comply with the Pomeranchuk
theorem~\cite{Pom}. The power of the first term is connected in the
Regge picture to the intercept of the exchanged Pomeron at $t$ = 0,
($\alpha_{\pom}(0)$ = 1.08), while the second term comes from the
intercept of the Reggeon ($\alpha_{\reg}(0)$ = 0.5475). The total
cross section data of $\bar{p} p, p p$ and $\pi^\pm p$ are shown in
figure~\ref{fig:hhtot} together with the DL parameterization.

One of the first measurements at HERA was that of the total $\gamma p$
cross section $\sigma_{tot}({\gamma p})$. 
\begin{figure}[hbt]
\begin{center}
  \includegraphics [width=\hsize,totalheight=6cm] 
{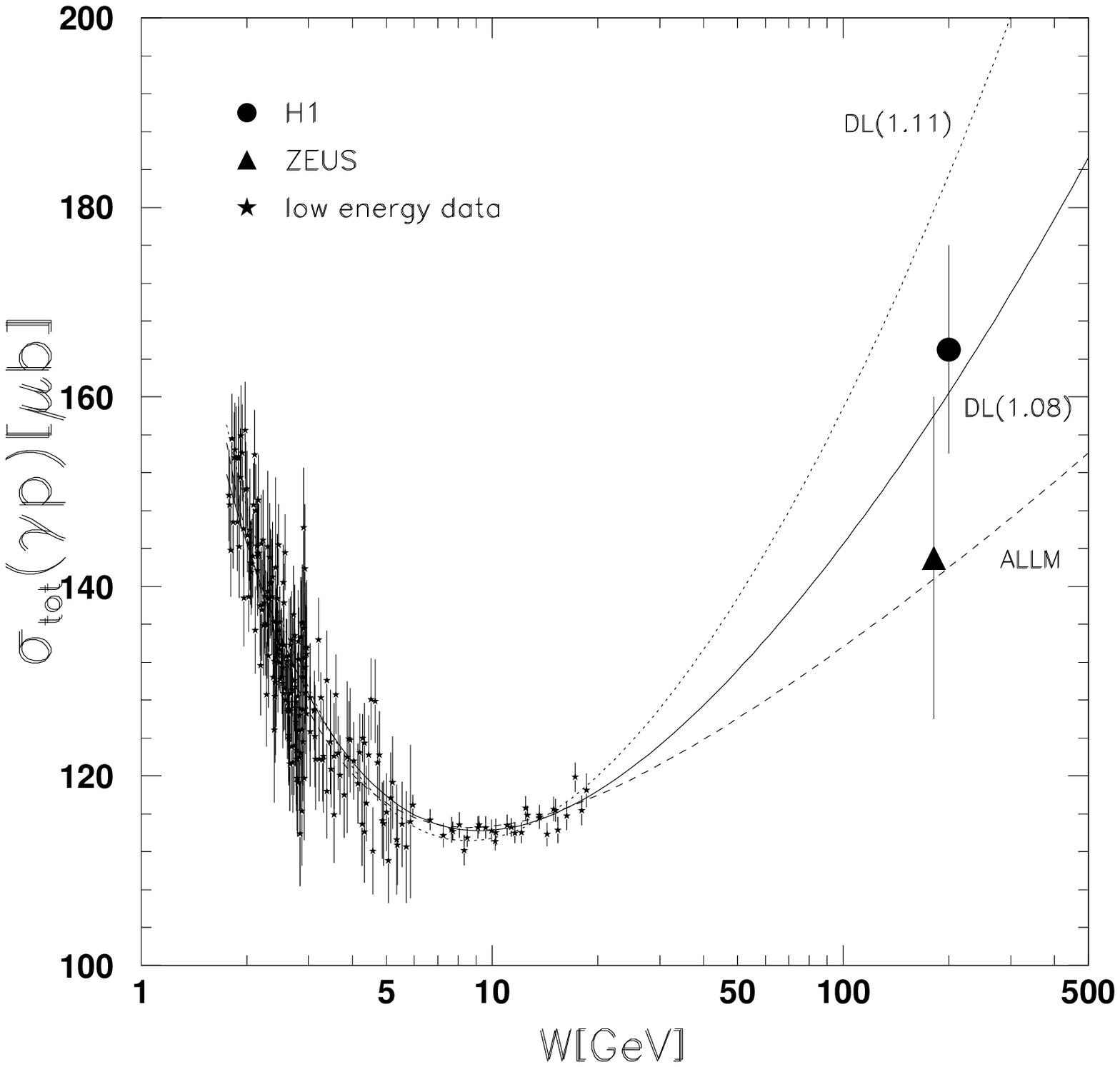}
\end{center}
\vspace{-.5cm}
\caption \protect {\it
{The total photoproduction cross section data measured by H1 and ZEUS
together with the fixed target data as function of $W$. The full line
and the dotted line are the DL parameterization prediction for
$\alpha_{\pom}$=1.08 and 1.1, respectively. The dashed line is that
of the ALLM~\cite{allm} parameterization. 
}}
\label{fig:gptot}
\end{figure}
\noindent
The measurement showed that the hadronic behaviour of the photon,
observed at lower energies, holds also in the HERA $W$ range. The
measurements of H1~\cite{h1-gptot} and ZEUS~\cite{zeus-gptot}, shown
in figure~\ref{fig:gptot}, agree well with the expectations of the DL
parameterization for photoproduction.

\subsection{Diffraction in photoproduction and DIS - the Pomeron}

If the photon behaves like a hadron, one expects to see diffractive
processes at HERA energies. Indeed it turns out that about 40 \% of
the photoproduction events are due to diffractive processes. The
diffractive reactions are described by diagrams in which the exchange
carries the quantum number of the vacuum, is a colorless object and is
referred to in the Regge language as the Pomeron trajectory. The
existence of such a trajectory was first suggested by
Gribov~\cite{gribov-pom} in order to avoid contradictions with
unitarity in the crossed channel.  The trajectory was named after
Pomeranchuk by Gel-Mann. In a reaction in which a Pomeron~\cite{levin}
is exchanged the proton remains intact or is being diffractively
dissociated into a state with similar quantum numbers (Gribov-Morrison
rule~\cite{gribov-morrison}). Thus there is a large rapidity gap
between the proton or its dissociated system and the hadrons belonging
to the system into which the photon diffracted. These large rapidity
gap events were observed in the photoproduction sample at HERA.

\begin{figure}[hbt]
\begin{center}
  \includegraphics [bb= 100 265 487 588,width=\hsize,totalheight=4cm] 
{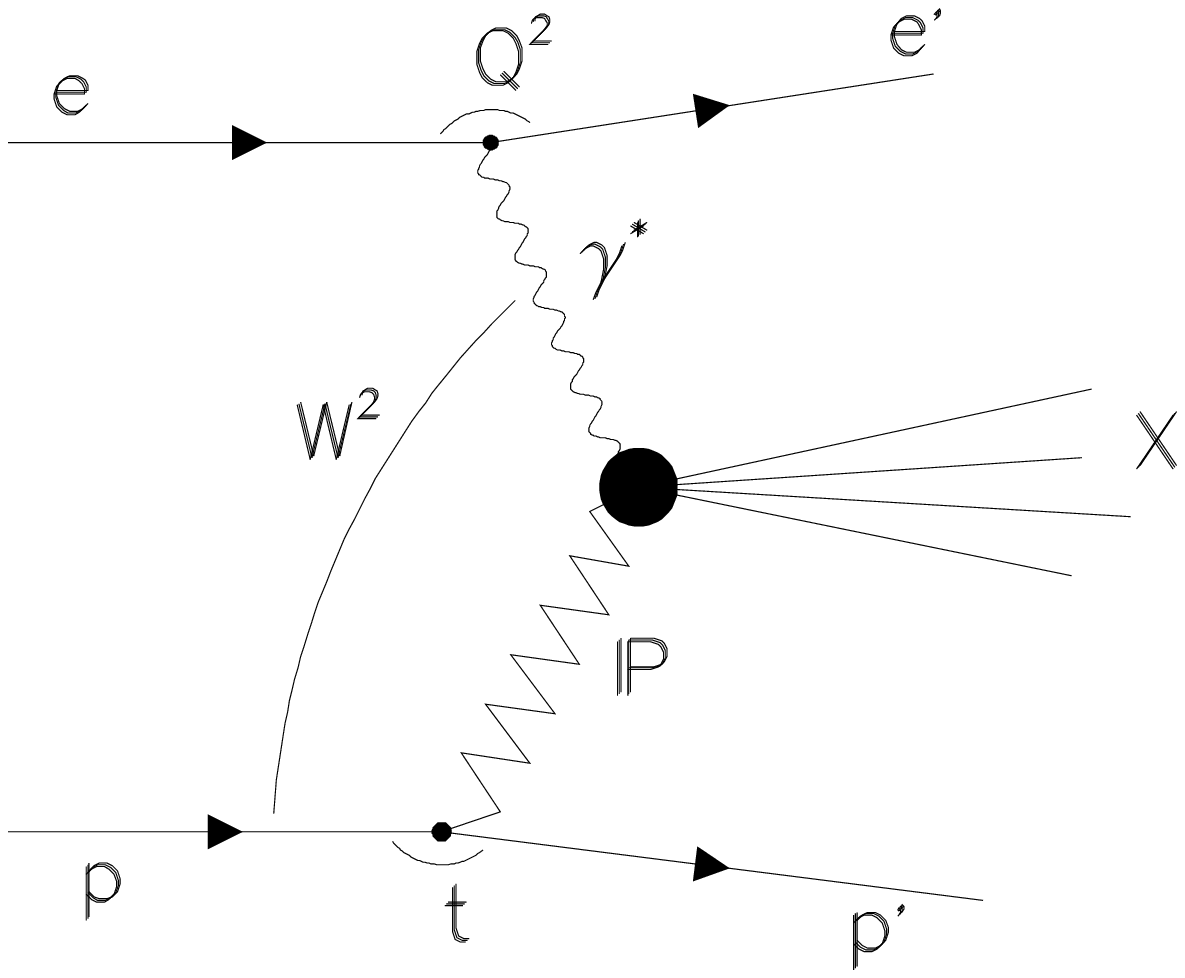}
\end{center}
\vspace{-.5cm}
\caption \protect {\it
{Diagram of a diffractive DIS event in which a photon of virtuality
$Q^2$ diffracts into a system $X$ at a $\gamma^* p$ center of mass
energy $W$ and where the four momentum transfer squared at the proton
vertex is $t$.  }}
\label{fig:dis-pomeron}
\end{figure}
One of the big surprises at HERA were the observation of large
rapidity gap events also in the DIS events~\cite{zeus-lrg,h1-lrg}. The
existence of such events meant that also a virtual photon can
diffract. This indicated that a process of DIS which is believed to be
a hard process because of the presence of a large scale, $Q^2$, can
also possess properties like diffraction which are expected in a soft,
low scale reactions. This interplay~\cite{afs} of soft and hard
processes will be discussed later. The observation of diffractive
processes in the DIS sample opened up the possibility of studying the
structure of the Pomeron in a DIS type experiment as depicted in
figure~\ref{fig:dis-pomeron}.

\begin{figure}[hbt]
\begin{center}
\includegraphics [bb= 5 231 582 823,clip,width=\hsize,totalheight=16.cm] 
{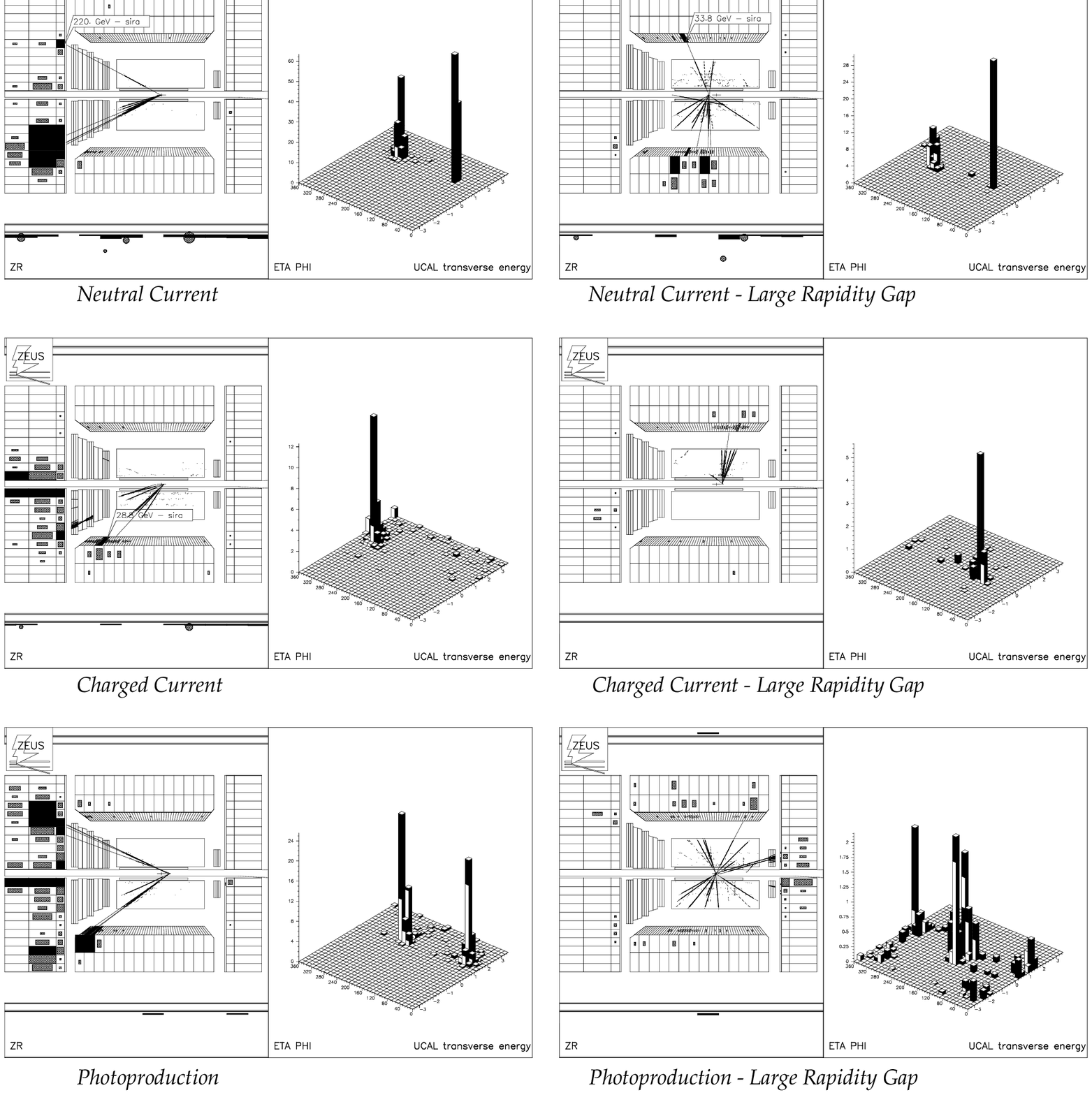}
\end{center}
\vspace{-.5cm}
\caption {\it
{Some events as seen in the ZEUS detector at HERA (see text).}}
\label{fig:hera-events}
\end{figure}

Let us finish this section by figure~\ref{fig:hera-events} which 
shows events resulting from electron--proton interactions,
as seen in the ZEUS detector (left part of each picture). The initial
electron and proton are in the beam pipe and not seen in the
detector. The electron enters the detector from the left and the
proton from the right. The right part of each picture shows a lego
plot of the transverse energy flow as function of the spatial
angle. 

The three events depicted on the left side of the page are three
different processes: NC DIS (top)
in which the scattered electron performs an almost U--turn and one of
the partons of the proton emerges as a jet; CC DIS
(center), where the initial electron turns into a
neutrino which is undetected and one of the hit partons from the
proton emerges as a jet, thus producing an unbalance in the transverse
energy; photoproduction reaction (bottom), a process where the
scattered electron emerges at a very small angle and thus remains
undetected in the beam pipe and the quasi--real photon interacts with
one of the partons of the proton producing two high transverse
momentum jets. The three events on the right hand side of the page are
similar processes, respectively, with the distinction that the proton
remains intact also after the interaction, producing a large rapidity
gap in the forward part of the detector, indicating that the reaction
is diffractive in nature and pointing to the presence of the Pomeron.

\subsection{The `Fathers'}

We have so far introduced the concept of the structures of the proton,
the photon and the Pomeron, all of which can be studied at HERA, and
details of which will be described in the next sections. We will
conclude this lengthy introductory section with the pictures of the
`fathers' of these three objects: Rutherford (proton), Einstein
(photon) and Gribov (Pomeron). Also shown is a picture of Pomeranchuk
who gave his name to the Pomeron and made remarkable contributions to
the theory of hadron-hadron interactions.
\begin{figure}[hbt]
\begin{minipage}{4.1cm}
\begin{center}
  \includegraphics [width=\hsize,totalheight=5cm] 
{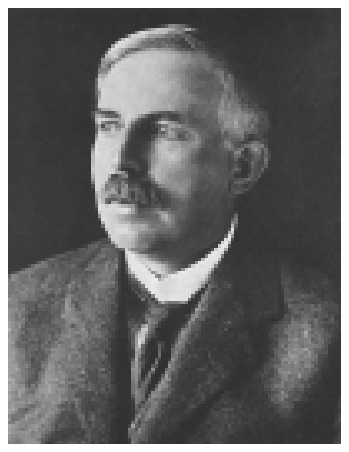}
\end{center}
\vspace{-.5cm}
\caption {\it
{Rutherford}}
\label{fig:rutherford}
\end{minipage}
\begin{minipage}{4.1cm}
\begin{center}
  \includegraphics [width=\hsize,totalheight=5cm]    
  {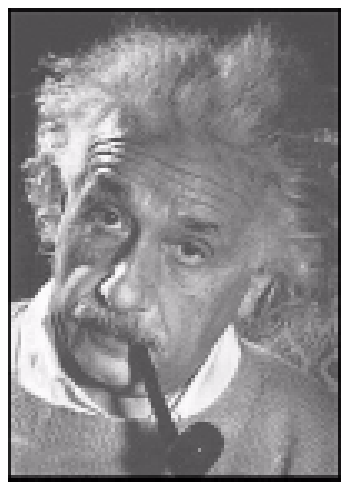}
\end{center}
\vspace{-.5cm}
\caption {\it
{Einstein}}
\label{fig:einstein}
\end{minipage}
\begin{minipage}{4.1cm}
\begin{center}
  \includegraphics [width=\hsize,totalheight=5cm]    
  {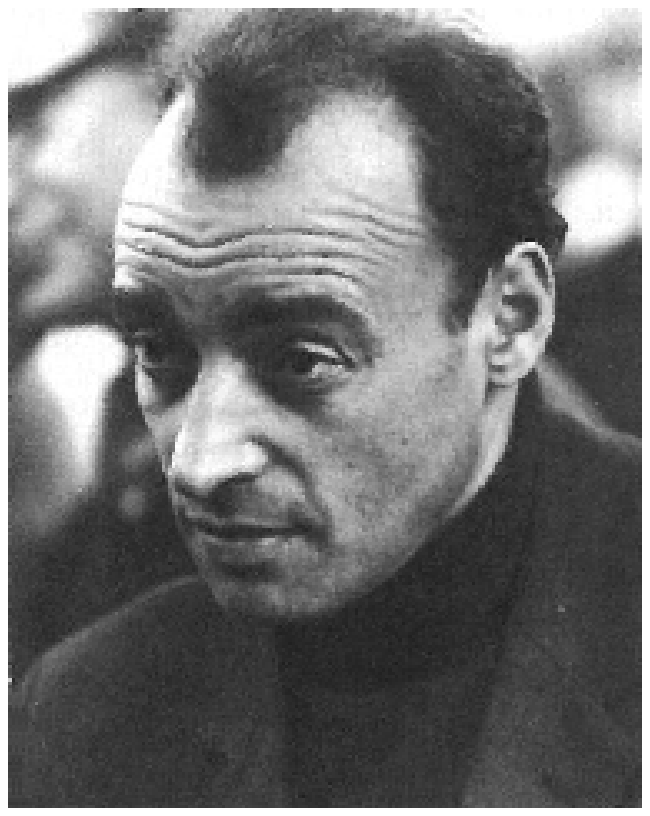}
\end{center}
\vspace{-.5cm}
\caption {\it
{Gribov}}
\label{fig:gribov}
\end{minipage}
\begin{minipage}{4.1cm}
\begin{center}
\vspace{3.5mm}
  \includegraphics [width=\hsize,totalheight=5cm]    
  {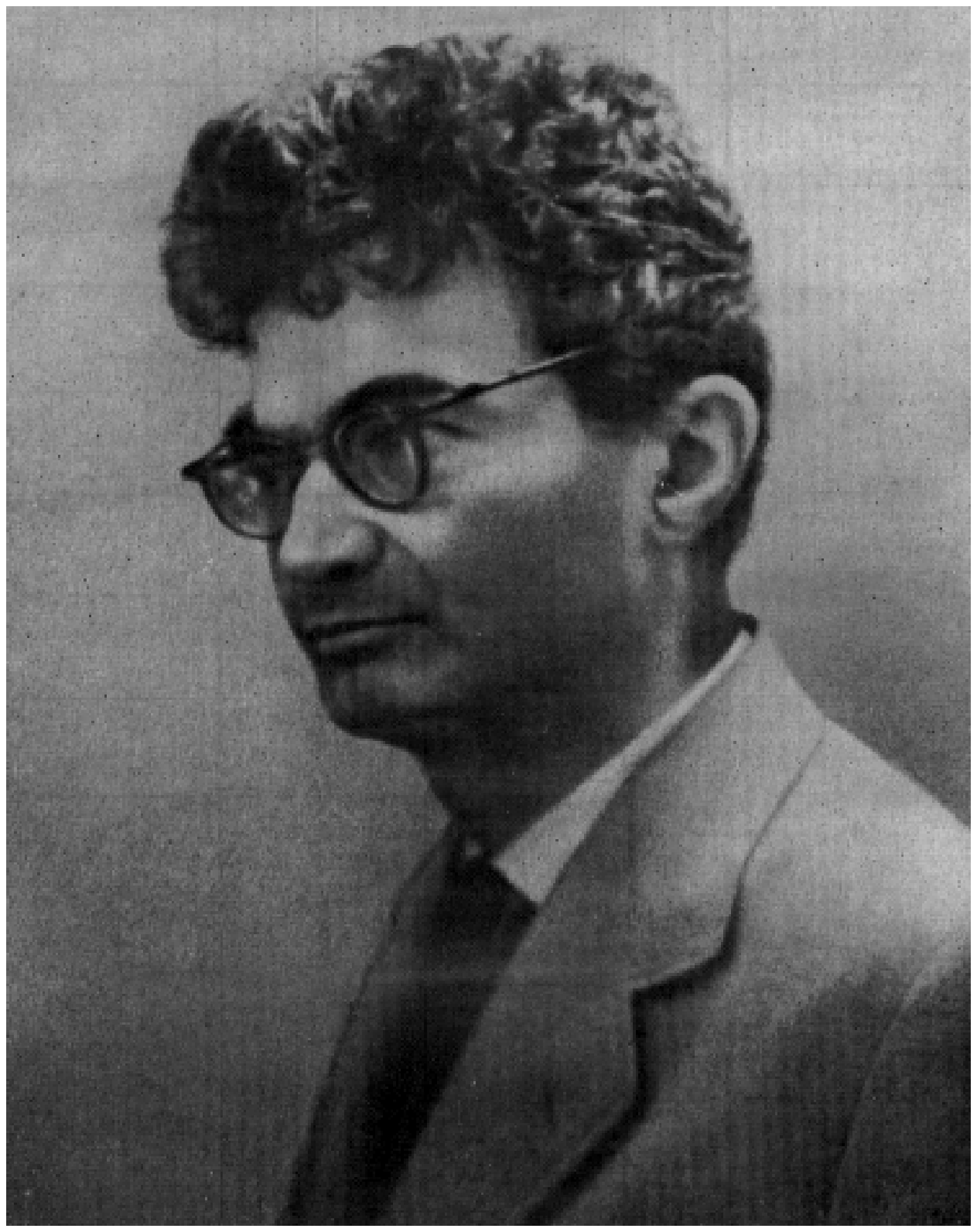}
\end{center}
\vspace{-.5cm}
\caption {\it
{Pomeranchuk}}
\label{fig:pomeranchuk}
\end{minipage}
\end{figure}

\section{The structure of the proton}
\label{sec:proton}

As mentioned in the introduction, HERA was built foremost to look
deeper inside the proton by providing very high $Q^2$ DIS
interactions~\cite{rmp}. How will we know that we see something new?
One way would be for instance to discover a lepto-quark, which is a
particle which - if it exists - may be seen at HERA as an s-channel
resonance in the $e$-$q$ system. Experimentally, such a particle would
show up as a peak in the $x$ distribution, where the position of the
peak is related to the lepto-quark mass $m_{LQ}$ as,
\begin{equation}
m^2_{LQ} = x s.
\end{equation}
The present lower limits of a lepto-quark from the Tevatron are larger
than 200 GeV, which means that for the available $s$ at HERA, a peak
would appear at high $x >$ 0.5. A look at figure~\ref{fig:x-q2} shows
that in order to reach such high values of $x$ one needs very large
$Q^2$ interactions. With the present luminosities the data statistic
are insufficient for a clear observation. 

\subsection{NC and CC cross sections}

\begin{figure}[hbt]
\begin{center}
  \includegraphics [width=\hsize,totalheight=8cm] 
{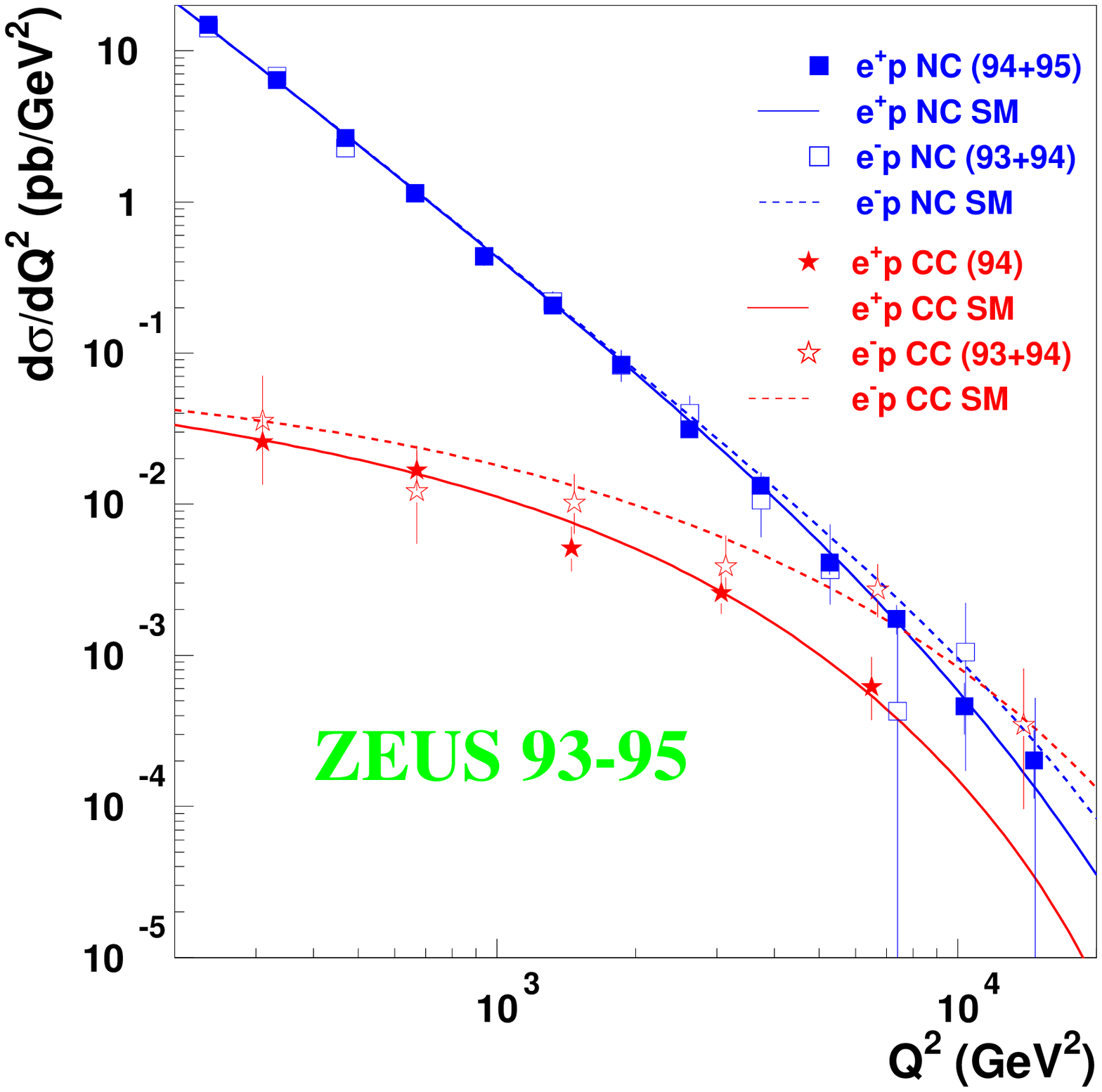}
\end{center}
\vspace{-.5cm}
\caption {\it
{The cross section of NC and CC events as function of $Q^2$ for $e^+
p$ and $e^- p$ interactions. The solid and dashed lines are
predictions of the Standard model for $e^+ p$ and $e^- p$
interactions, respectively. }}
\label{fig:nc-cc}
\end{figure}

Another way to search for new physics is to look for deviations from
the Standard Model predictions of values like the NC and CC cross
sections at high $Q^2$. Such a comparison was done with the data from
1993--1995, which can be seen in figure~\ref{fig:nc-cc}.
This figure is very educative, a `textbook' type of figure, and thus
deserves a detailed discussion. In the NC case one has for the low
$Q^2$ region the dominance of the electromagnetic interaction mediated
by the photon which predicts a behaviour like 
\begin{equation}
\frac{d\sigma}{dQ^2} \sim \frac{1}{Q^4}. 
\end{equation}
At high $Q^2$, the weak force mediated in the NC case by the $Z$ boson
is important and contributes,
\begin{equation}
\frac{d\sigma}{dQ^2} \sim \frac{1}{Q^4} \left(\frac{Q^2}{Q^2 + m_Z^2}\right)^2.
\end{equation}
In case of the CC interactions only the weak force mediated by the $W$
boson contributes,
\begin{equation}
\frac{d\sigma}{dQ^2} \sim \frac{1}{Q^4} \left(\frac{Q^2}{Q^2 + m_W^2}\right)^2.
\label{eq:sigcc}
\end{equation}
Thus at low $Q^2$, ($Q^2 \ll m_Z^2$) one expects
$\frac{d\sigma}{dQ^2}$(NC) $\gg \frac{d\sigma}{dQ^2}$(CC). In the
region $Q^2 \sim m_Z^2$ one expects the two cross sections to be of
the same order, $\frac{d\sigma}{dQ^2}$(NC) $\sim
\frac{d\sigma}{dQ^2}$(CC). These predictions of the Standard Model are
nicely borne out by the data shown in figure~\ref{fig:nc-cc}.  A
closer look into the exact formulae~\cite{leader-predazzi} shows that
the cross sections are higher for the $e^- p$ interactions than for
the $e^+ p$ ones, again in agreement with the data.

\subsection{$W$ mass determination}

\begin{figure}
\begin{center}
  \includegraphics [width=\hsize,totalheight=6cm] {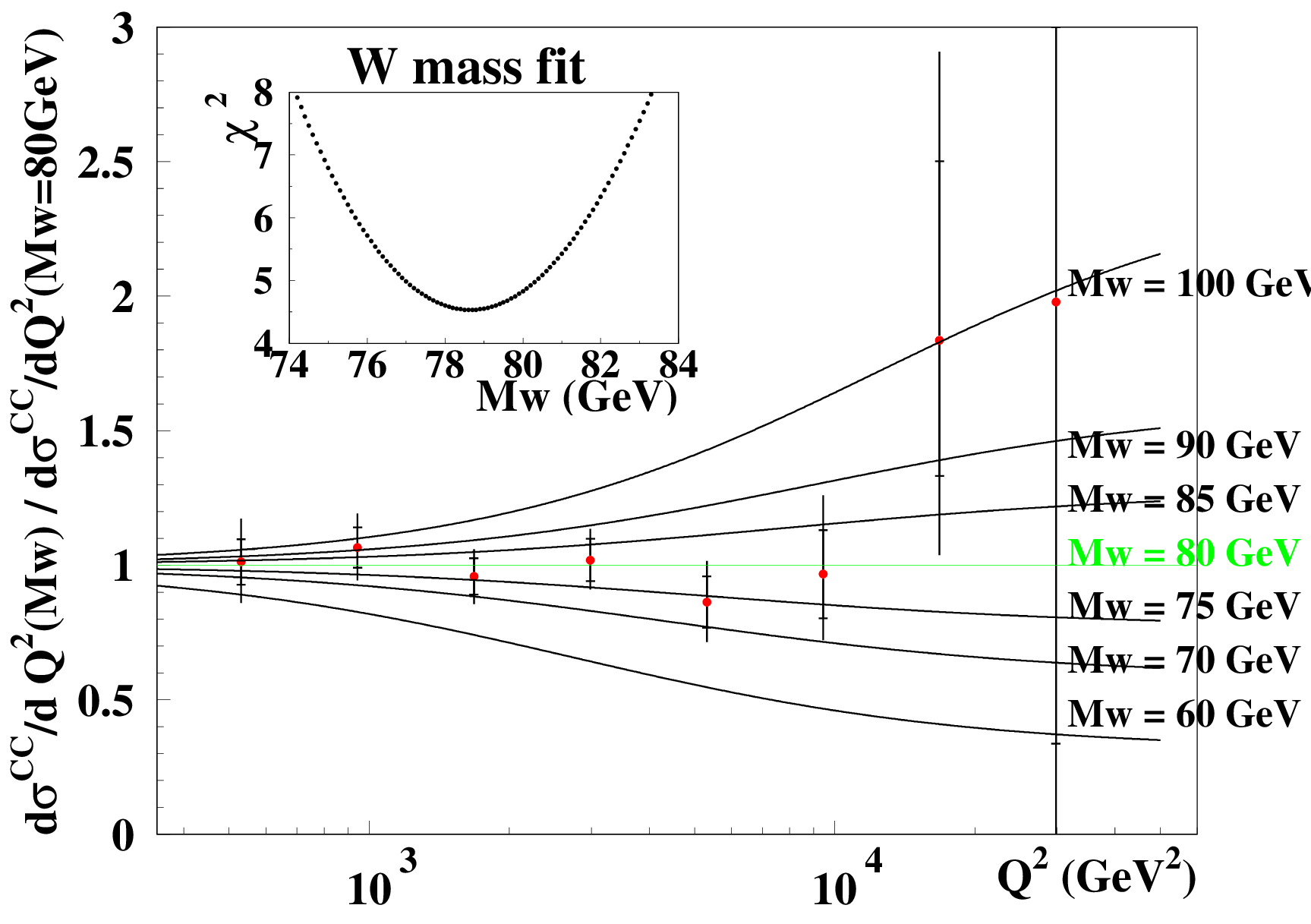}
\end{center}
\vspace{-.5cm}
\caption {\it
{The ratio of the CC differential cross section at a value of $m_W$
given at the left of the figure, to that at $m_W$ = 80 GeV, as
function of $Q^2$. The insert gives the $\chi^2$ as function of
$m_W$.}}
\label{fig:wmass}
\end{figure}
Another nice result comes from the CC events. It is clear from
equation (\ref{eq:sigcc}) that the cross section depends on the $W$
mass and thus by fitting the CC differential cross section one can in
principle determine $m_W$. A preliminary attempt of such a
determination is shown in figure~\ref{fig:wmass} which results
in~\cite{zeus-Wmass}
\begin{equation}
m_W = 78.6^{+2.5}_{-2.4}({\rm stat})^{+3.3}_{-3.0}({\rm syst}) \ {\rm GeV}.
\end{equation}
This value is in good agreement with the world average
one~\cite{pdg}. Clearly the large error on the mass will be reduced
once the high luminosity run will increase the statistics of the data
in this high $Q^2$ region.

\subsection{Signs for new physics?}

\begin{figure}[h]
\begin{minipage}{8cm}
\begin{center}
  \includegraphics [width=7cm,totalheight=8cm] 
{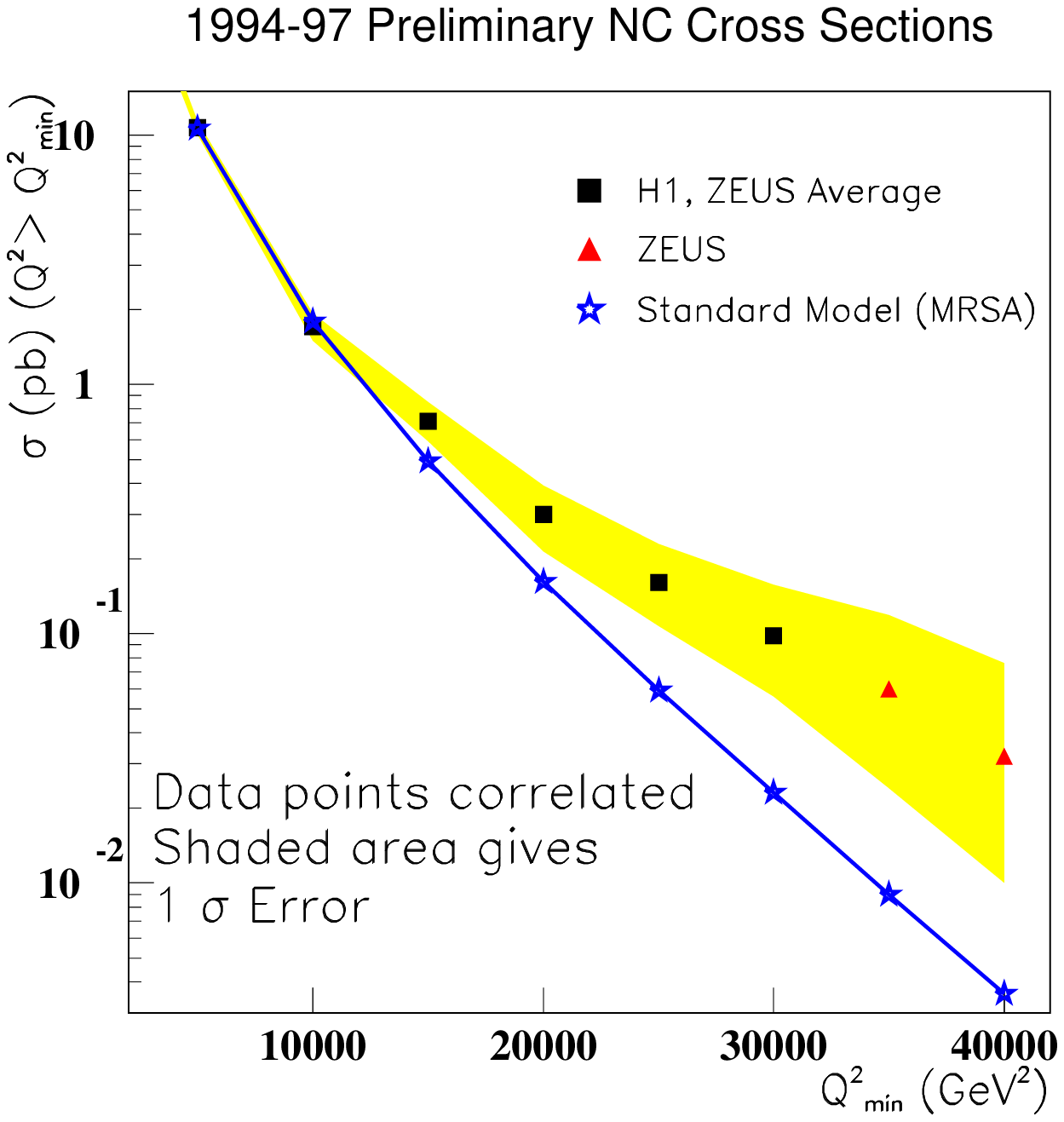}
\end{center}
\vspace{-.5cm}
\caption {\it 
{The cross section of NC and CC events as function of $Q^2$ for $e^+
p$ and $e^- p$ interactions. The solid and dashed lines are
predictions of the Standard model for $e^+ p$ and $e^- p$
interactions, respectively. }}
\label{fig:highx-q2}
\end{minipage}
\hspace{3mm}
\begin{minipage}{8cm}
\begin{center}
\vspace{5mm}
  \includegraphics [width=7cm,totalheight=8cm] 
{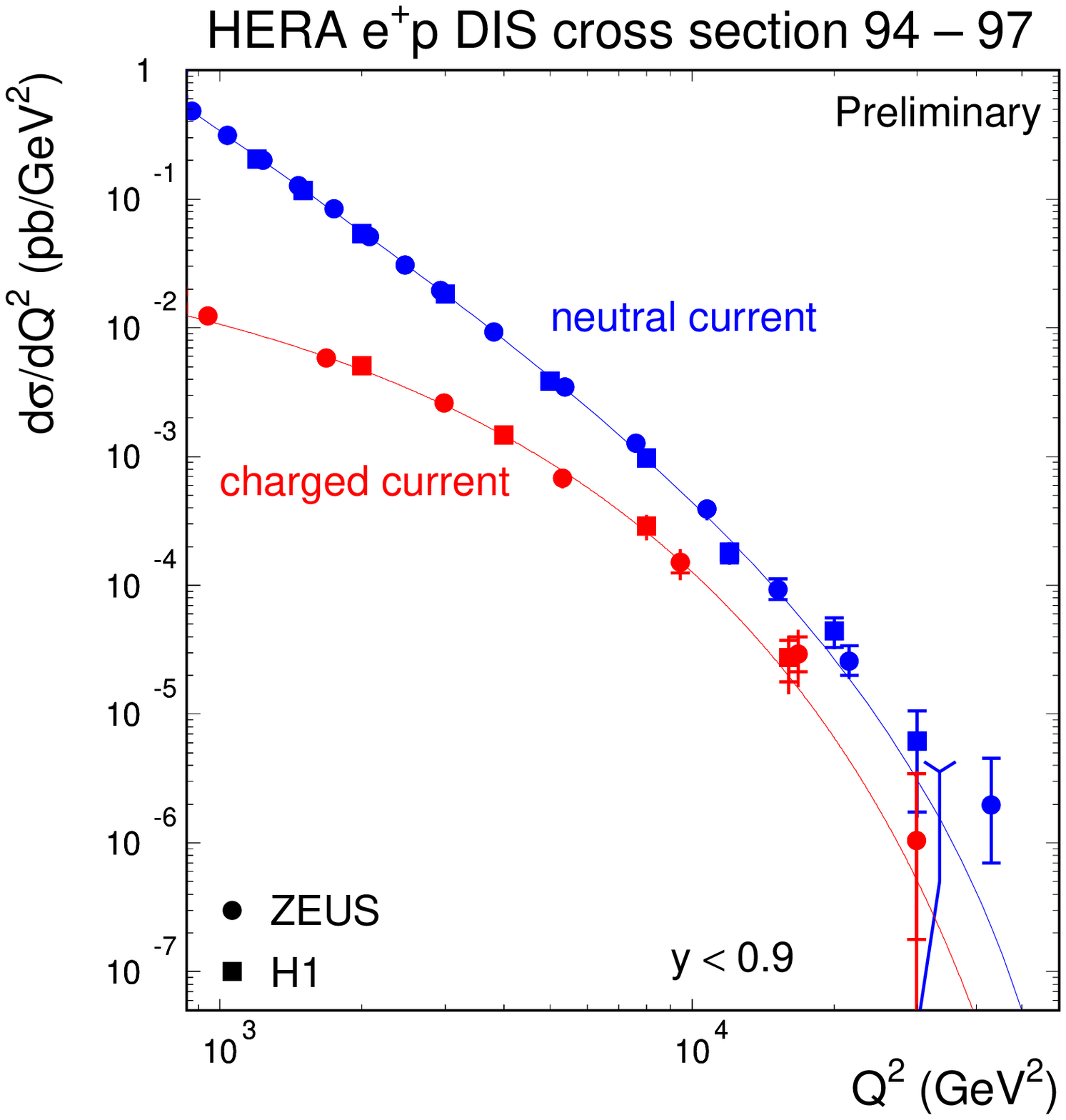}
\end{center}
\vspace{-.5cm}
\caption {\it
{The differential cross section $\frac{d\sigma}{dQ^2}$ of
NC and CC events from the 1994--1997 data as function of $Q^2$ for
$e^+ p$ interactions as measured by the H1 and ZEUS
collaborations.. The solid lines are predictions of the Standard
model.}}
\label{fig:nc-cc-new}
\end{minipage}
\end{figure}
All the above was true for the 1993--1995 data. About a year ago,
after adding the 1996 data, there was a big excitement due to a
possible sign of deviation from the predictions of the Standard
Model. Both the H1~\cite{h1-highx-q2} and ZEUS~\cite{zeus-highx-q2}
collaborations observed an excess of events at very high
$Q^2$. Preliminary results shown at the Lepton--Photon Symposium in
1997~\cite{bruce-lp97} indicated a deviation from the Standard Model
predictions which seemed to increase with $Q^2$, as can be seen in
figure~\ref{fig:highx-q2}~\cite{wolf}. The figure shows the NC cross
section of the events with $Q^2$ above a minimum value of $Q^2_{min}$
which is in excess of the expectations of the Standard Model. The
shaded area gives the 1 standard deviation error of the data.

In the meantime both collaborations finished to analyze their 1997
data and published preliminary results of $\frac{d\sigma}{dQ^2}$ of
the NC and CC data accumulated during
1994--1997~\cite{h1-vc,zeus-vc,zeus-Wmass}, displayed in
figure~\ref{fig:nc-cc-new}. The following observations can be made:
\begin{itemize}
\item 
There is good agreement between the data of the H1 and ZEUS 
collaborations.
\item 
The cross section for NC and CC events are of the same order for 
$Q^2 \geq m^2_{W,Z}$.
\item 
In the region of $Q^2 \geq 10^4$ GeV$^2$ the cross section
measurements are still statistics limited.
\item 
Both NC and CC cross sections seem to show some excess over the 
Standard Model predictions at the highest measured $Q^2$ point.
\end{itemize}
We will have to wait for much higher statistics in order to evaluate
the significance of the excess at high $Q^2$.

\begin{figure}[hbt]
\begin{center}
  \includegraphics [width=7cm,totalheight=8cm] 
{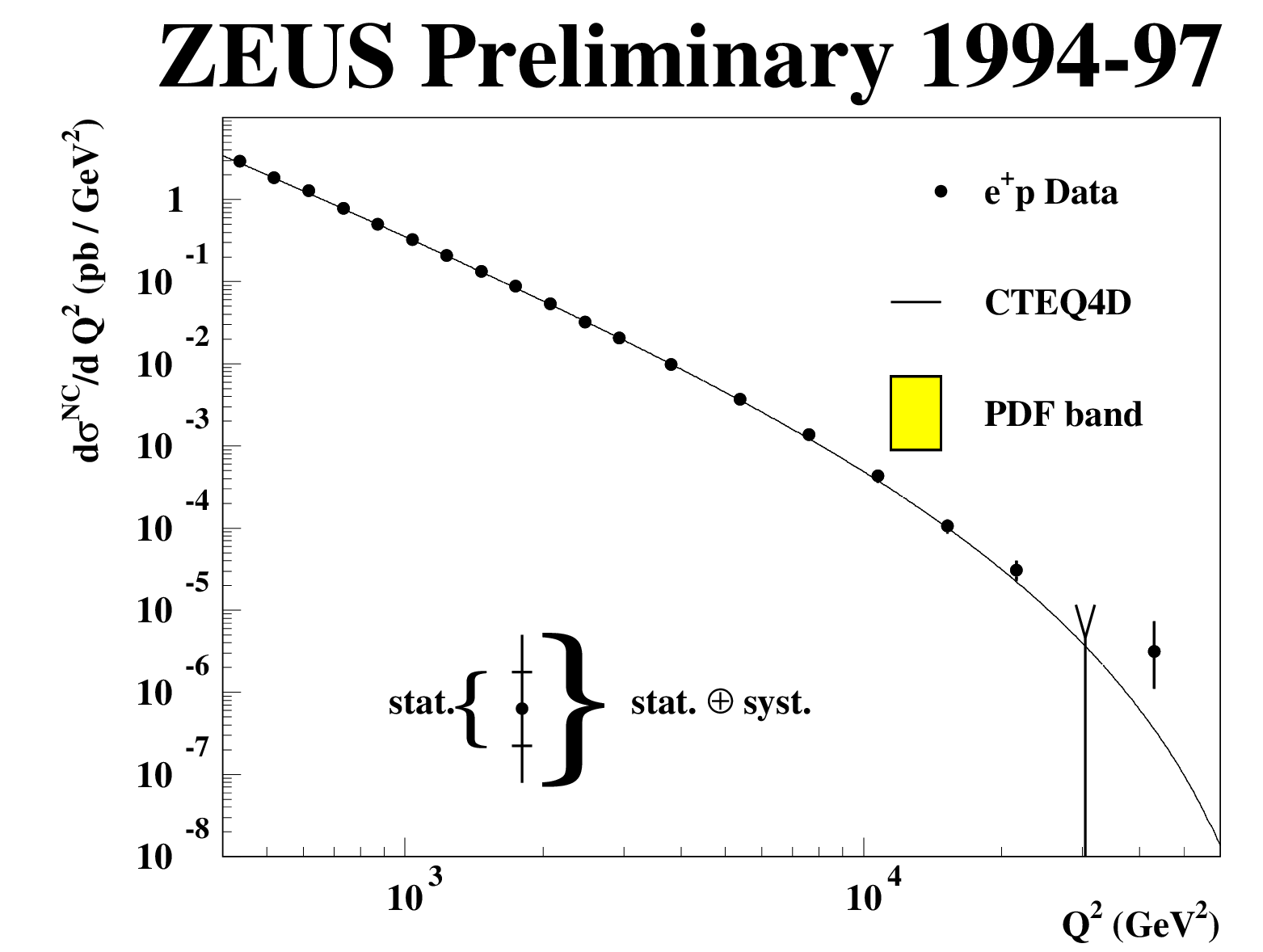}
  \includegraphics [width=7cm,totalheight=8cm] 
{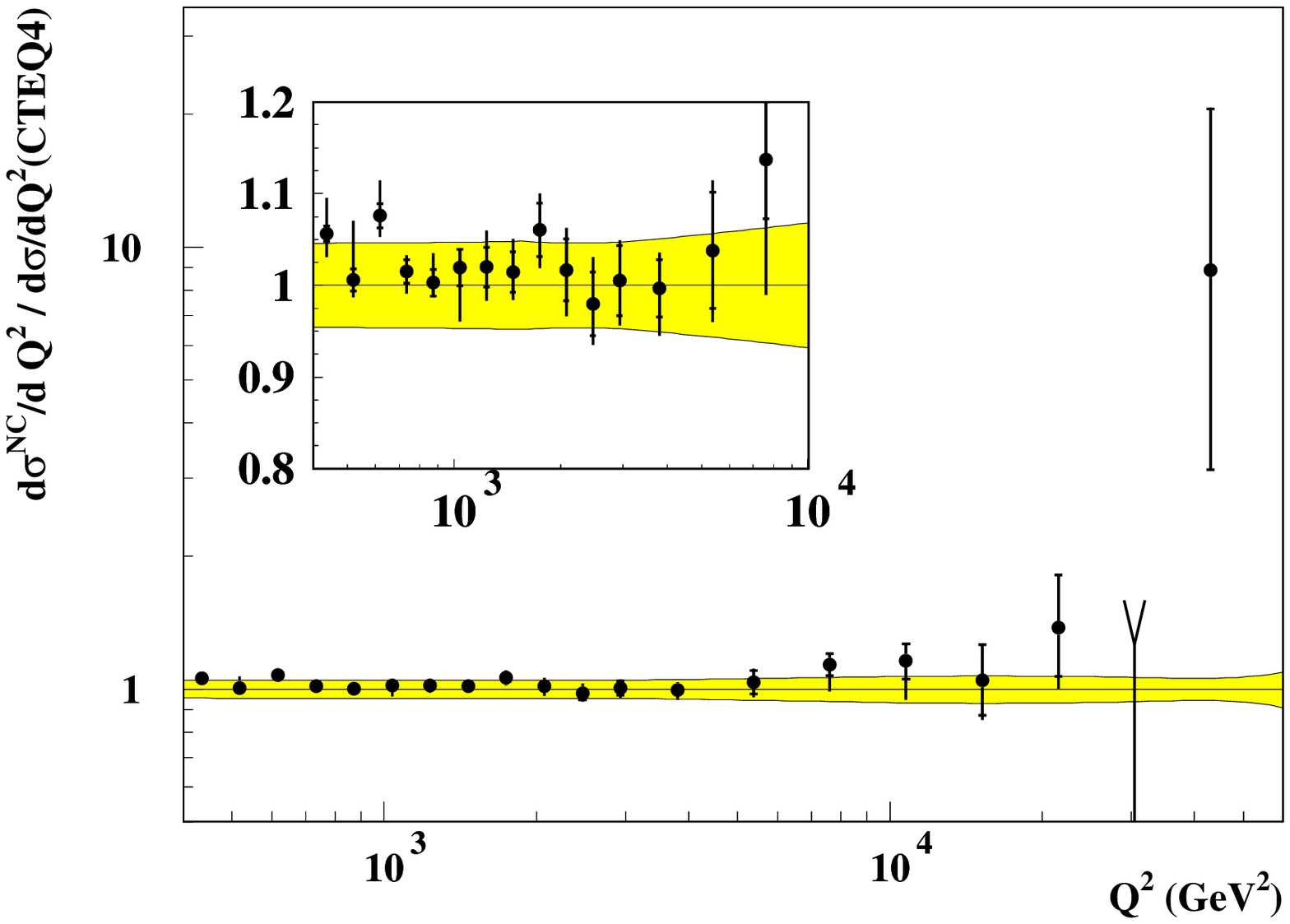}
\end{center}
\vspace{-.5cm}
\caption {\it
{Left part: $d\sigma/dQ^2$ of the 94--97 NC data. The line is the
predictions of the Standard Model.  Right part: the ratio of the data
to the prediction using for the latter the CTEQ4D parameterization of
the parton distribution function (PDF). The band, shows the
uncertainty of the ratio due to the uncertainty in the parton
distributions.
}}
\label{fig:ncdsdq2}
\end{figure}

It is clear that in order to make any claim of a disagreement with the
Standard Model, one first has to be able to state how well the
predictions are known. This was
estimated~\cite{h1-highx-q2,zeus-highx-q2} to be 6.5 \%, where the
main uncertainty comes from the knowledge of the parton distributions
in the proton. This can also be seen in the left-hand side of
figure~\ref{fig:ncdsdq2} where the NC differential cross section of
the 94--97 data are shown together with the predictions of the
Standard Model.  The right-hand side shows the ratio of the data to
the prediction, where in the prediction a particular representation
(CTEQ4D~\cite{cteq4d}) of parton distribution parameterizations has
been used. The band, labeled as PDF band, shows the uncertainty of the
prediction due to the uncertainty in the parton distributions.

\subsection{Parton distributions}

Why should there be an uncertainty in the parton distributions? Is not
QCD a theory which describes the interactions of quarks and gluons? As
we all know, the answer to this rhetoric question is connected to the
behaviour of the strong coupling $\alpha_S(Q^2)$ which becomes too
large at low $Q^2$ to enable a reliable calculation. Only when the
scale $Q^2$ is large enough one can do a perturbative QCD (pQCD)
calculation. This fact does not allow to calculate the parton
distribution functions from first principles. We can nevertheless
predict the parton distribution at a larger scale once we know them at
a lower scale, due to the QCD factorization
theorem~\cite{qcd-factorization}. This theorem allows to factorize the
cross section into short distance phenomena, fully calculable in pQCD
due to the presence of a large scale, and long distance phenomena
which include the flux of universal incoming partons. The latter
cannot be calculated perturbatively and has to be taken from
experimental data. The Dokshitzer-Gribov-Lipatov-Altarelli-Parisi
(DGLAP)~\cite{dglap} equations enable the evolution of the parton
distribution to any $Q^2$ once they are given at a starting value
$Q^2_0$.  How low can one go with the value of the starting scale
$Q^2_0$? Will will try to answer this somewhat later. Before doing
that, it is educative to point out what are the assumptions used in
the DGLAP evolution equations.

\begin{figure}[hbt]
\begin{center}
  \includegraphics [width=7cm,totalheight=8cm]
  {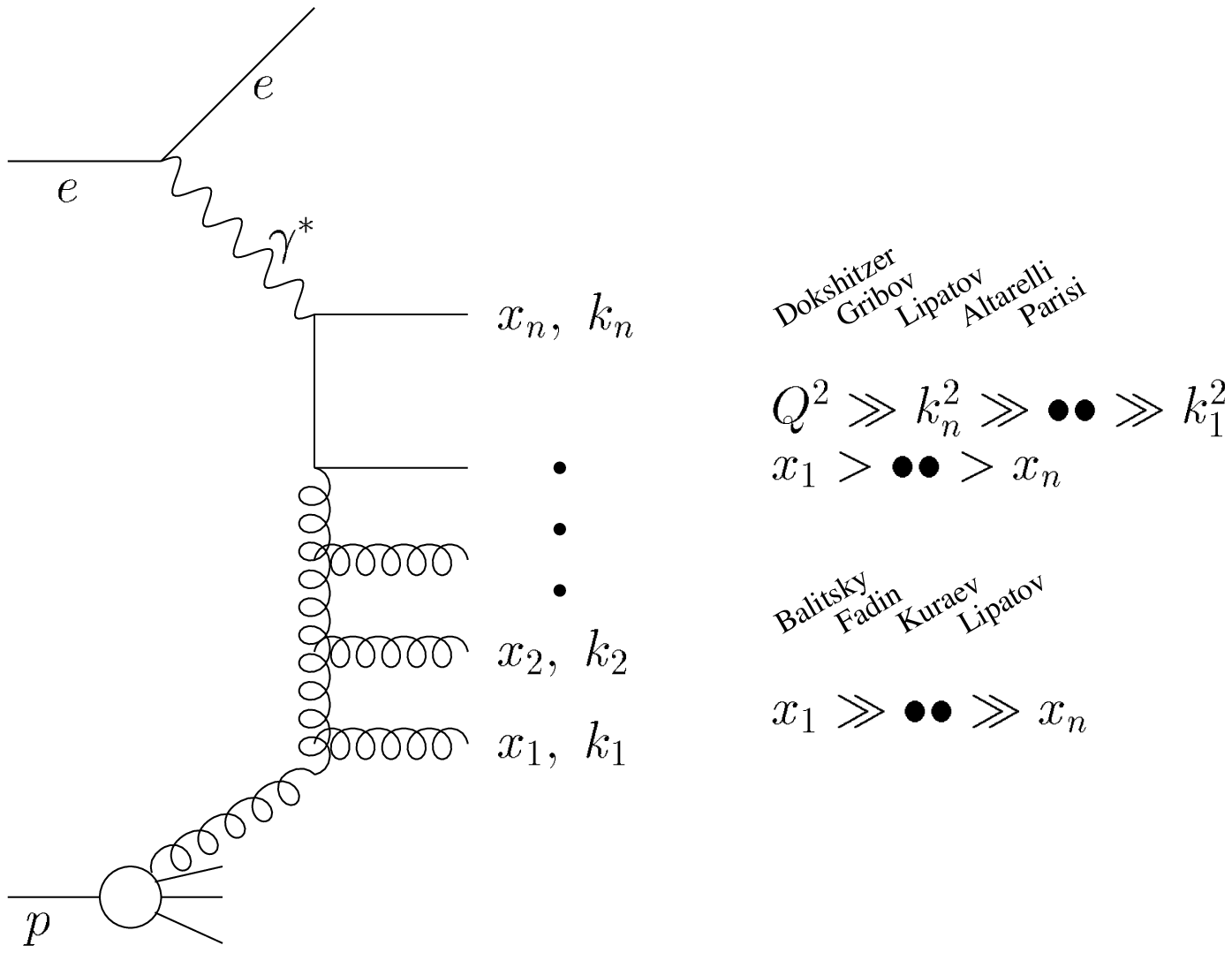} \includegraphics
  [width=7cm,totalheight=8cm] {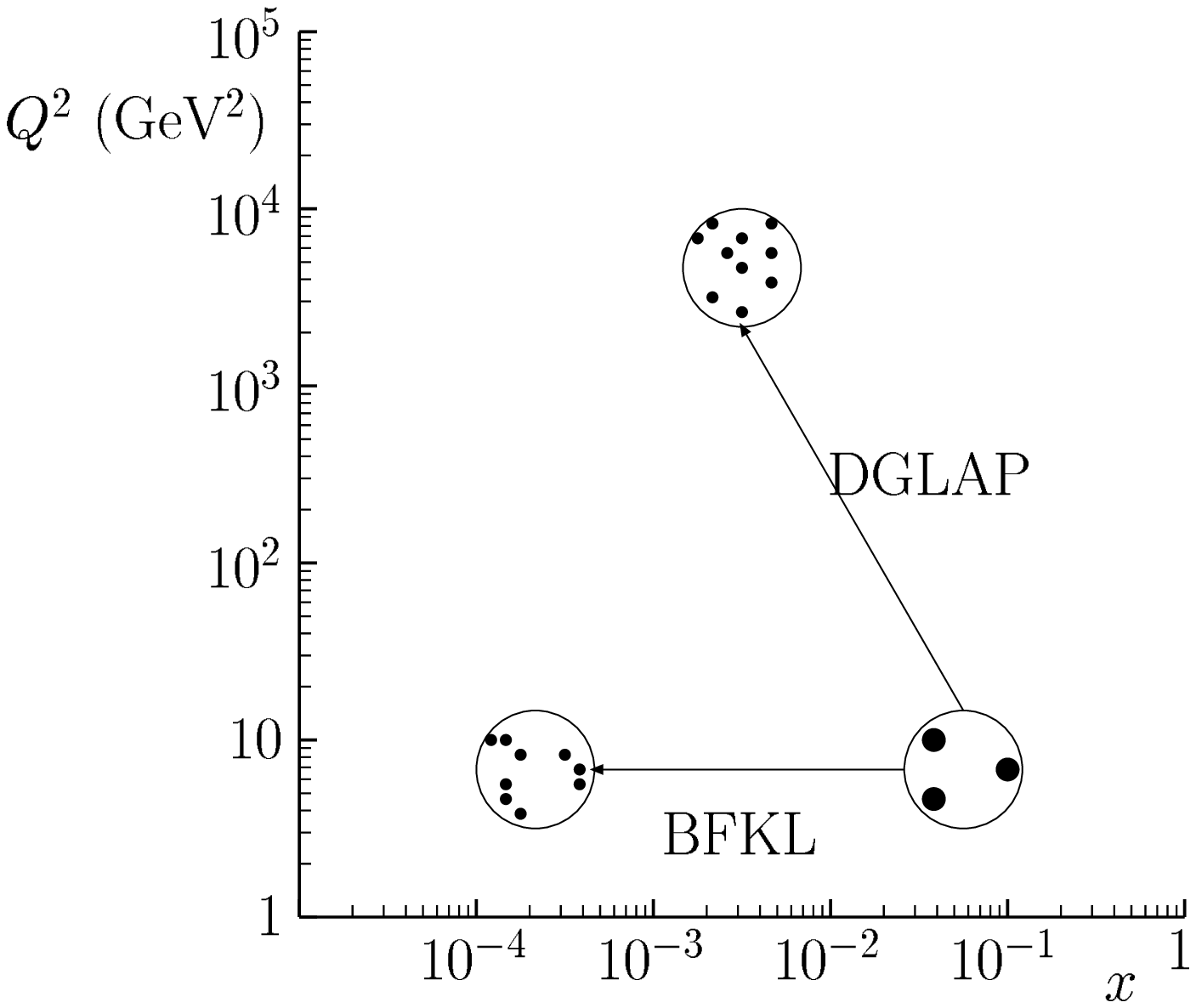}
\end{center}
\vspace{-.5cm}
\caption {\it
{Left part: A diagram describing the development of a parton chain in
an $e p$ interaction. Each parton has a value of $x_i$ and of
transverse momentum $k_i$. Also given are the assumptions about these
values for the DGLAP and the BFKL evolution equation. Right part: the
path taken in the $x, Q^2$ plane when evolving through the DGLAP and
the BFKL equations. 
}}
\label{fig:dglap-bfkl}
\end{figure}
The DGLAP evolution equations allow us to go from a given point
($x_0,Q_0^2$) to another point ($x_1,Q_1^2$) so that $x_1 < x_0$ and
$Q^2_1 > Q^2_0$. During this evolution there is strong ordering in the
transverse momentum of the parton chain and also ordering in
their $x$ values (see figure~\ref{fig:dglap-bfkl}). Another evolution
equation has been suggested by Balitzki-Fadin-Kuraev-Lipatov
(BFKL)~\cite{bfkl} which deals with cases in which there is only
evolution in $x$. There is no ordering in the transverse momentum of
the partons but only strong ordering in their $x$ values
(figure~\ref{fig:dglap-bfkl}). We will not discuss it further in this
talk.

\subsection{Scaling violation}

\begin{figure}[hbt]
\begin{minipage}{8cm}
\begin{center}
  \includegraphics [width=8cm,totalheight=16cm] 
{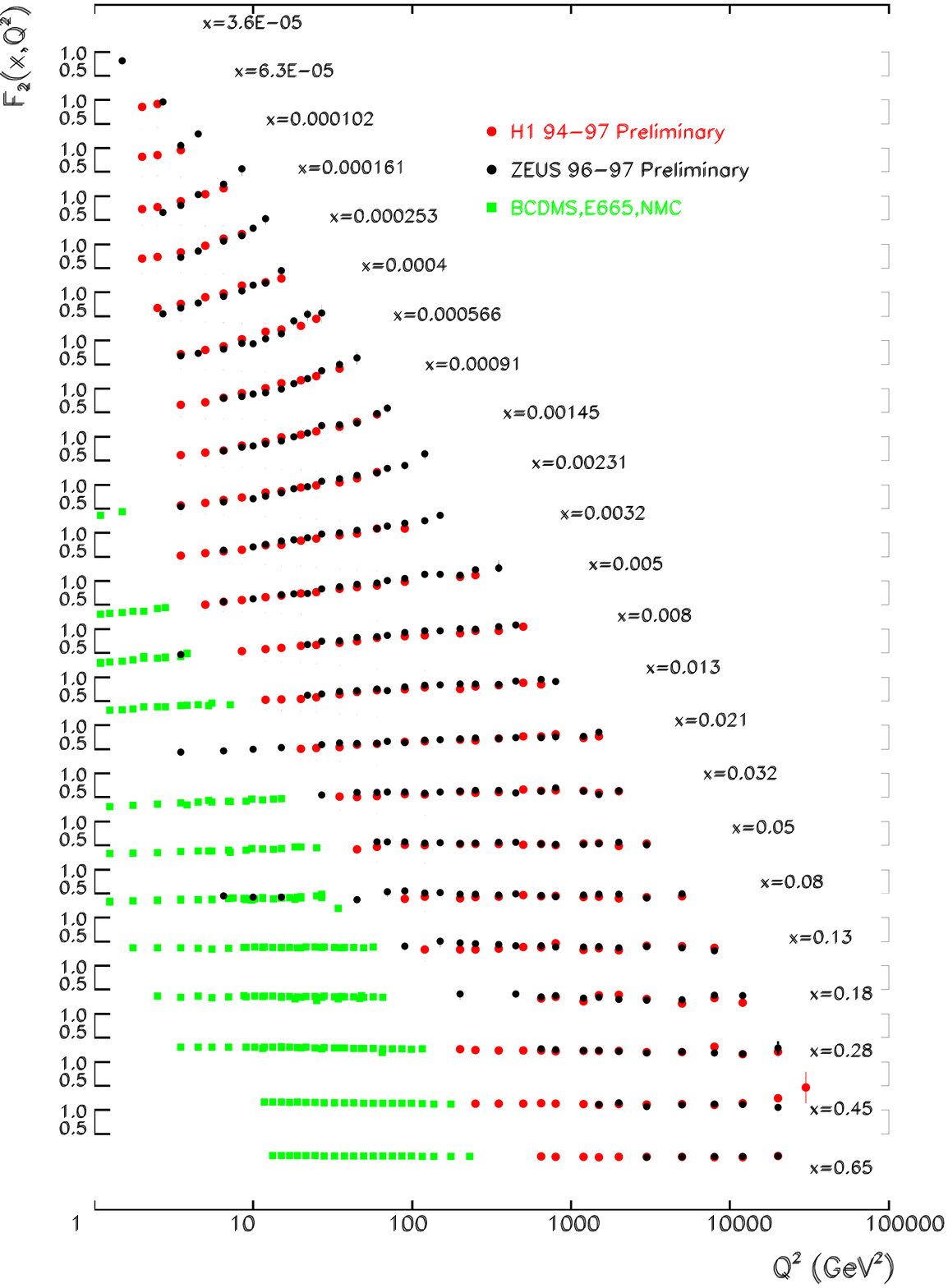}
\end{center}
\vspace{-.5cm}
\caption {\it
{ The dependence of the proton structure function $F_2$ on $Q^2$ for
fixed values of $x$.  }}
\label{fig:f2-q2}
\end{minipage}
\begin{minipage}{8cm}
\begin{center}
  \includegraphics [width=7cm,totalheight=7cm] 
{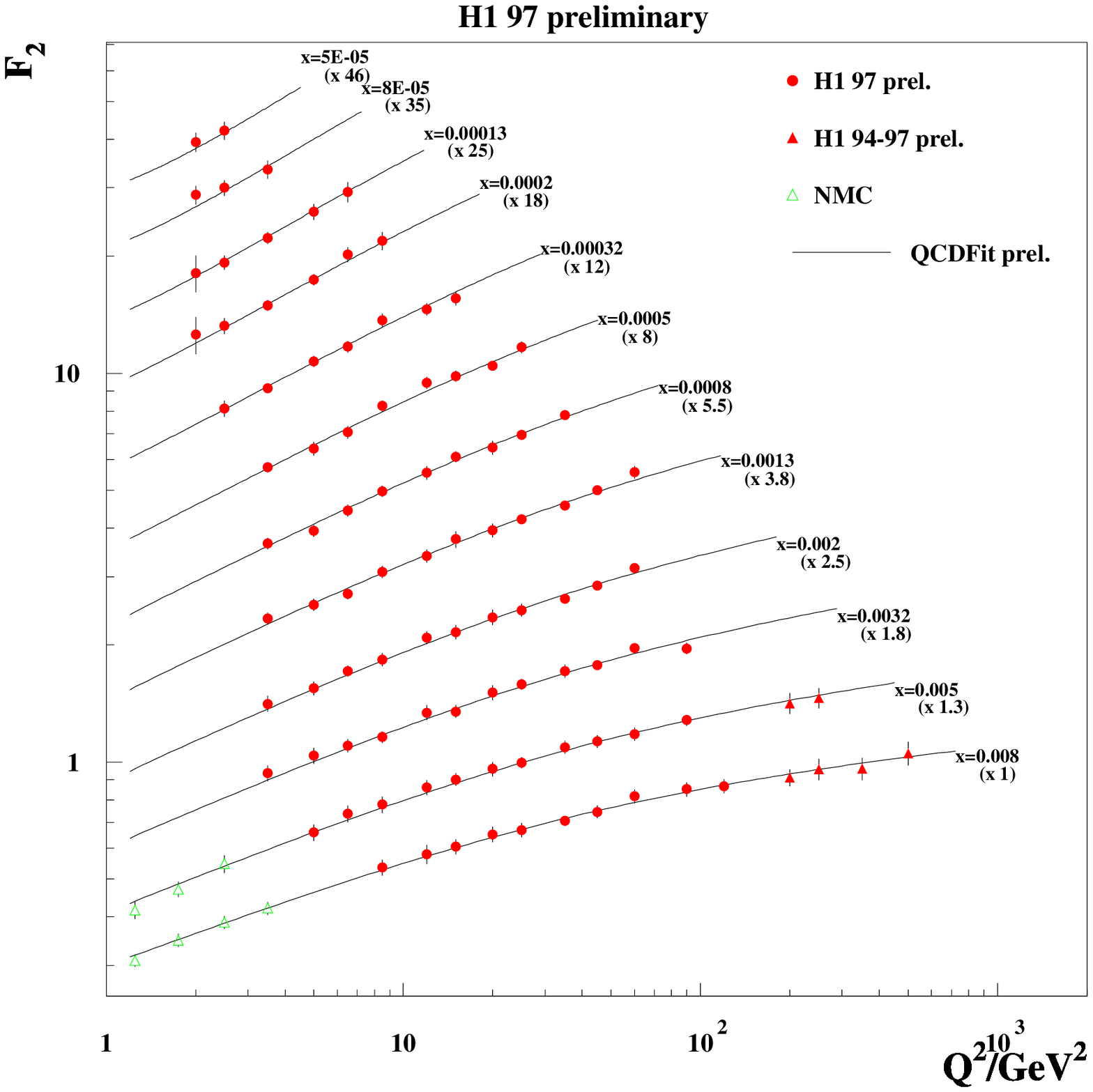}
  \includegraphics [width=7cm,totalheight=7cm] 
{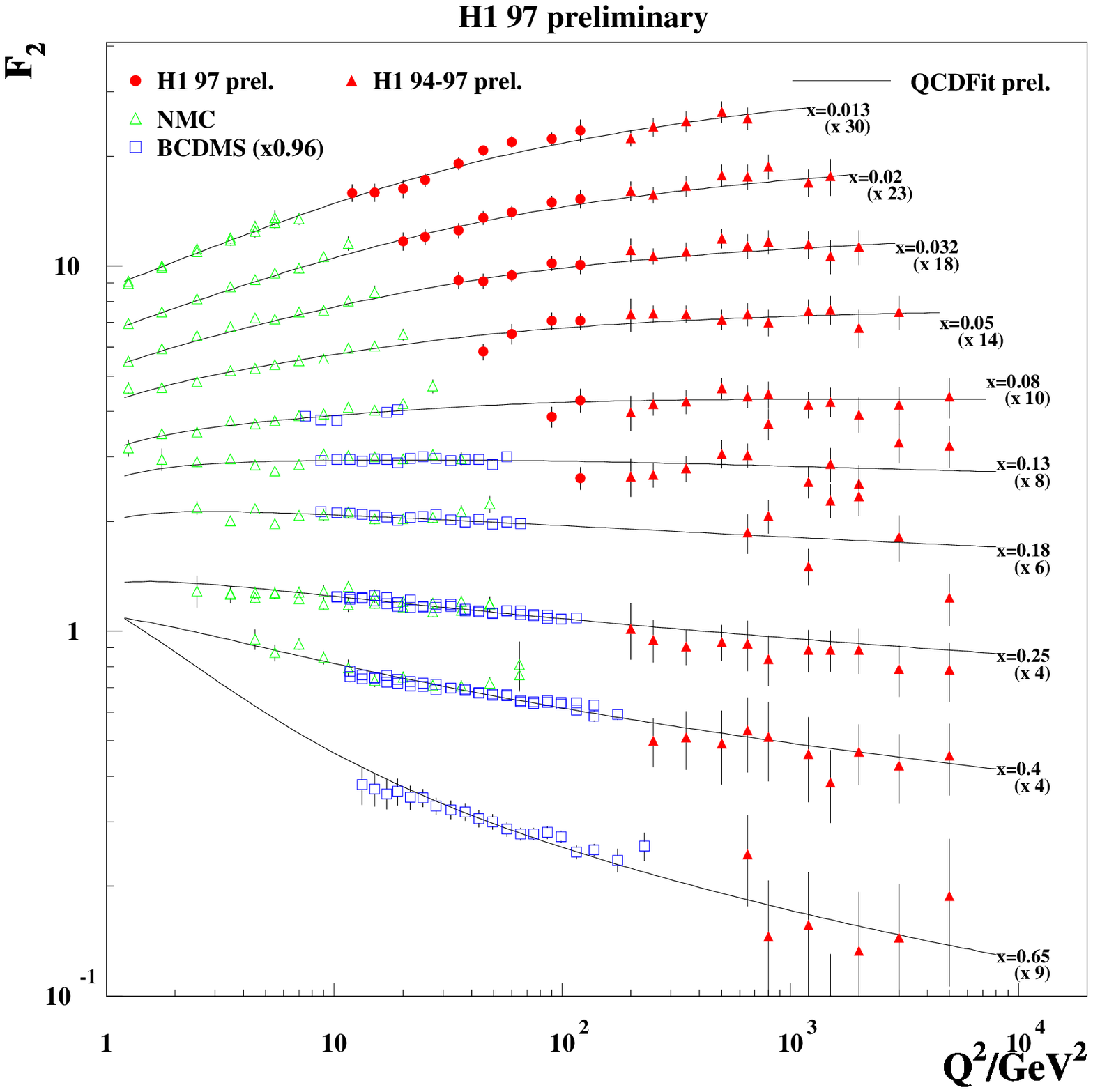}
\end{center}
\vspace{-.5cm}
\caption {\it
{ Comparison of the scaling violation behaviour of $F_2$ with the results of a
next-to-leading order (NLO) DGLAP evolution equation.
 }}
\label{fig:scalingviolation}
\end{minipage}
\end{figure}
Back to the data. In figure~\ref{fig:f2-q2} one sees the dependence of
the proton structure function $F_2$ on $Q^2$ for fixed values of
$x$~\cite{f2review}. The data come from the two HERA experiments and
from some of the lower energy fixed target experiments. One sees a
remarkable agreement between the data of different experiments. One
can also see the positive scaling violation at low $x$ and the
approximate scaling at intermediate $x$. This can be seen in more
details, together with the negative scaling violation at high $x$, in
figure~\ref{fig:scalingviolation}.  In this figure one also sees that
a next-to-leading-order (NLO) QCD fit using the DGLAP equations can
describe the data up to quite high $Q^2$ values.

\subsection{Global QCD fits}

\begin{figure}[hbt]
\begin{center}
  \includegraphics [width=\hsize,totalheight=8cm] 
{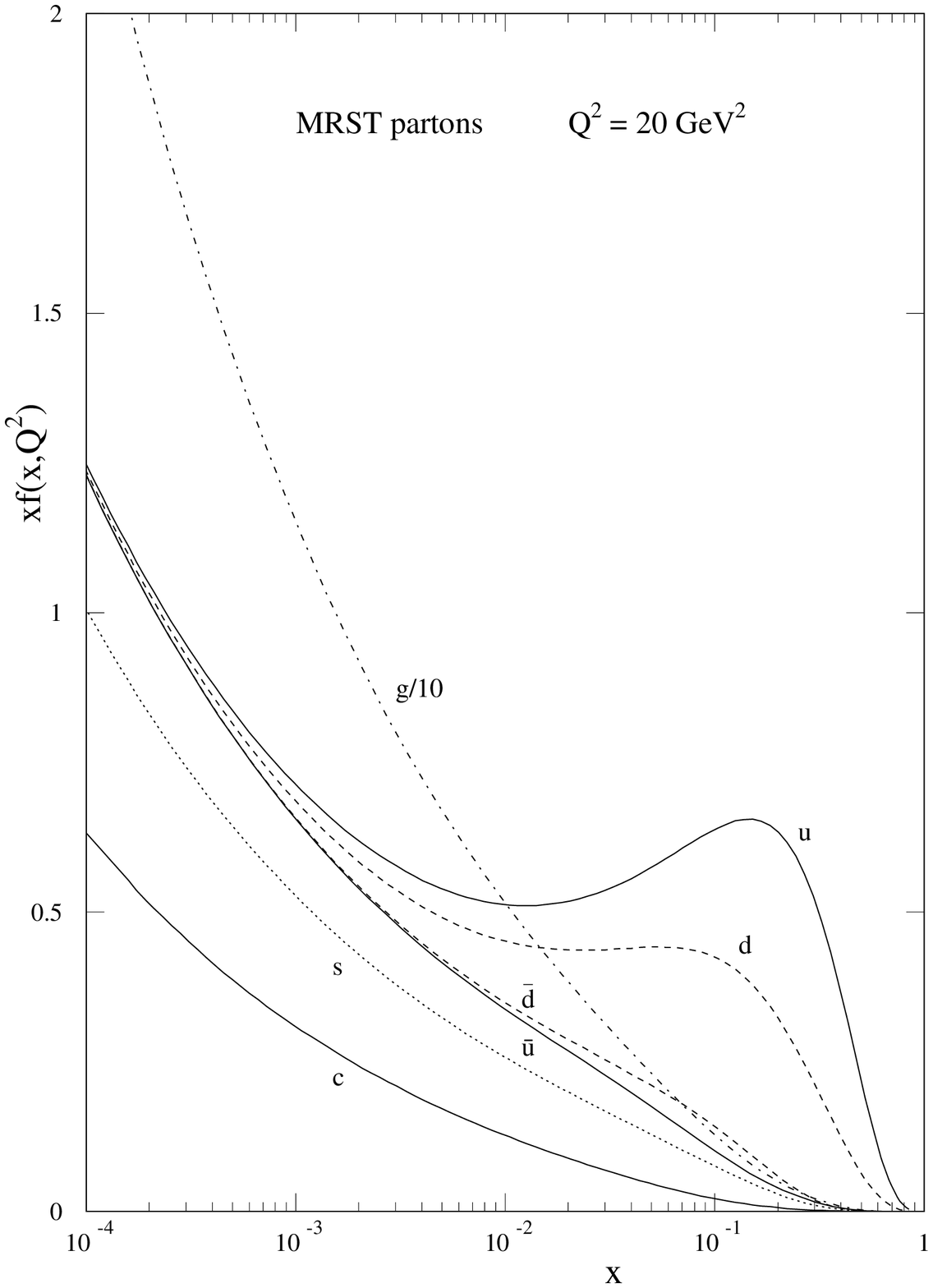}
\end{center}
\vspace{-.5cm}
\caption {\it
{ Parton density distributions as function of $x$ of the MRST global
QCD fit at a scale of $Q^2$ = 20 GeV$^2$.}}
\label{fig:mrst-partons}
\end{figure}

How does one do a QCD fit to the data and gets out of it parton
distributions? One assumes a mathematical form in the $x$ variable for
the different parton distributions at a given $Q^2_0$. One uses the
DGLAP equations to evolve the parton distributions to any other $Q^2$
value where measurable quantities exist. By fitting the calculations
to the data, the parameters of the initial parton distributions which
give the best fit are thus determined. Measurable quantities do not
necessarily have to be only structure functions. Drell-Yan cross
sections, W assymetries, direct photon production yields, inclusive
jet cross sections - all such quantities are included in a global
fit. As an example we show in table 1 a list of such variables and the
number of data points used in a recent global QCD fit by
Martin-Roberts-Stirling-Thorne (MRST)~\cite{mrst}. They use in total
close to 1500 data points and obtain fits with $\chi^2$ of the order of
$\sim$ 1-1.2 per degree of freedom.
\begin{table}[hbt]
\begin{center}
\begin{tabular}{|c|c|c|c|}
\hline
Process & Experiment & Measurable & Data Points
\\ \hline\hline
DIS & BCDMS & $F_{2\ H}^\mu, F_{2\ D}^\mu $ & 324 \\ \hline
    & NMC & $F_{2\ H}^\mu, F_{2\ D}^\mu, F_{2\ n/p}^\mu $ & 297  \\ \hline
    & SLAC & $F_{2\ H}^e $ & 70  \\ \hline
    & E665 & $F_{2\ H}^\mu, F_{2\ D}^\mu $ & 70  \\ \hline
    & H1 & $F_{2\ H}^e $ & 172  \\ \hline
    & ZEUS & $F_{2\ H}^e $ & 179   \\ \hline\hline
    & CCFR & $F_{2\ Fe}^\nu, x\ F_{3\ Fe}^\nu $ & 126  \\ \hline
Drell-Yan & E605 & $sd\sigma /d\sqrt{\tau }dy$ & 119  \\ \hline
          & E772 & $sd\sigma /d\sqrt{\tau }dy$ & 219  \\ \hline
          & NA-51 & $A_{DY}$ & 1  \\ \hline
          & E886  & $pd/pp$ & 11  \\ \hline\hline
W-prod. & CDF & Lepton asym. & 9  \\ \hline\hline
Direct $\gamma $ & WA70 & $Ed^3\sigma /d^3p$ & 8  \\ \hline
       & UA6 & $Ed^3\sigma /d^3p$ & 16 \\ \hline
       & E706 & $Ed^3\sigma /d^3p$ & 19 \\ \hline\hline
Incl. Jet & CDF & $d\sigma /dE_t$ & 36  \\ \hline
          & D0 & $d\sigma /dE_t$ & 26  \\ \hline
\end{tabular}
\caption {\it
{A list of measurables used in a global QCD fit to determine parton
distribution functions.
}}
\end{center}
\label{table:mrst}
\end{table}
The resulting parton distributions of the MRST global QCD fit at a
scale of $Q^2$ = 20 GeV$^2$ are shown in
figure~\ref{fig:mrst-partons}. 
At large $x$ only the valence, $u$ and $d$, quarks contribute
appreciably. The sea quark densities rise sharply as $x$
decreases. However the rise of the gluon density is much sharper and
in order to fit into the same figure with all the quarks it had to be
suppressed by an order of magnitude. The physical meaning of the
parton density functions is the number of partons per unit of
rapidity. Thus according to the MRST parameterization there are more
than 20 gluons per unit of rapidity at $x$ = $10^{-4}$ at a scale of
$Q^2$ = 20 GeV$^2$. It is also interesting to see how the proton
momentum is distributed among the partons according to the MRST
parameterization at different $Q^2$ values. This can be seen in table
2 from which one learns that the gluon carries close to 40 \% of the
proton momentum and this fraction increases with $Q^2$, while the
momentum fraction carried by the valence quarks decreases with $Q^2$.
\begin{table}[hbt]
\begin{center}
\begin{tabular} {|c|c|c|c|c|c|c|c|c|} \hline
$Q^2$(GeV$^2$) & $u_v$ & $d_v$ & $2\bar{u}$ & $2\bar{d}$ & $2\bar{s}$ &
$2\bar{c}$ & $2\bar{b}$ & g \\ \hline \hline
2 & 0.310 & 0.129 & 0.058 & 0.075 & 0.037& 0.001 & 0.000& 0.388 \\ \hline
20 & 0.249 & 0.103 & 0.063 & 0.077 & 0.046 & 0.017 & 0.000 & 0.439 \\ \hline
20000& 0.178 & 0.074 & 0.070 & 0.080 & 0.058 & 0.036 & 0.026 & 0.472 \\ \hline
\end{tabular}
\caption {\it
{The momentum distribution among partons according to the MRST
parameterization.  }}
\end{center}
\label{table:mrst-mom}
\end{table}

\subsection{Behaviour of $F_2$ at low $x$}
 
\begin{figure}[hbt]
\begin{center}
 \includegraphics [width=\hsize,totalheight=12cm] 
{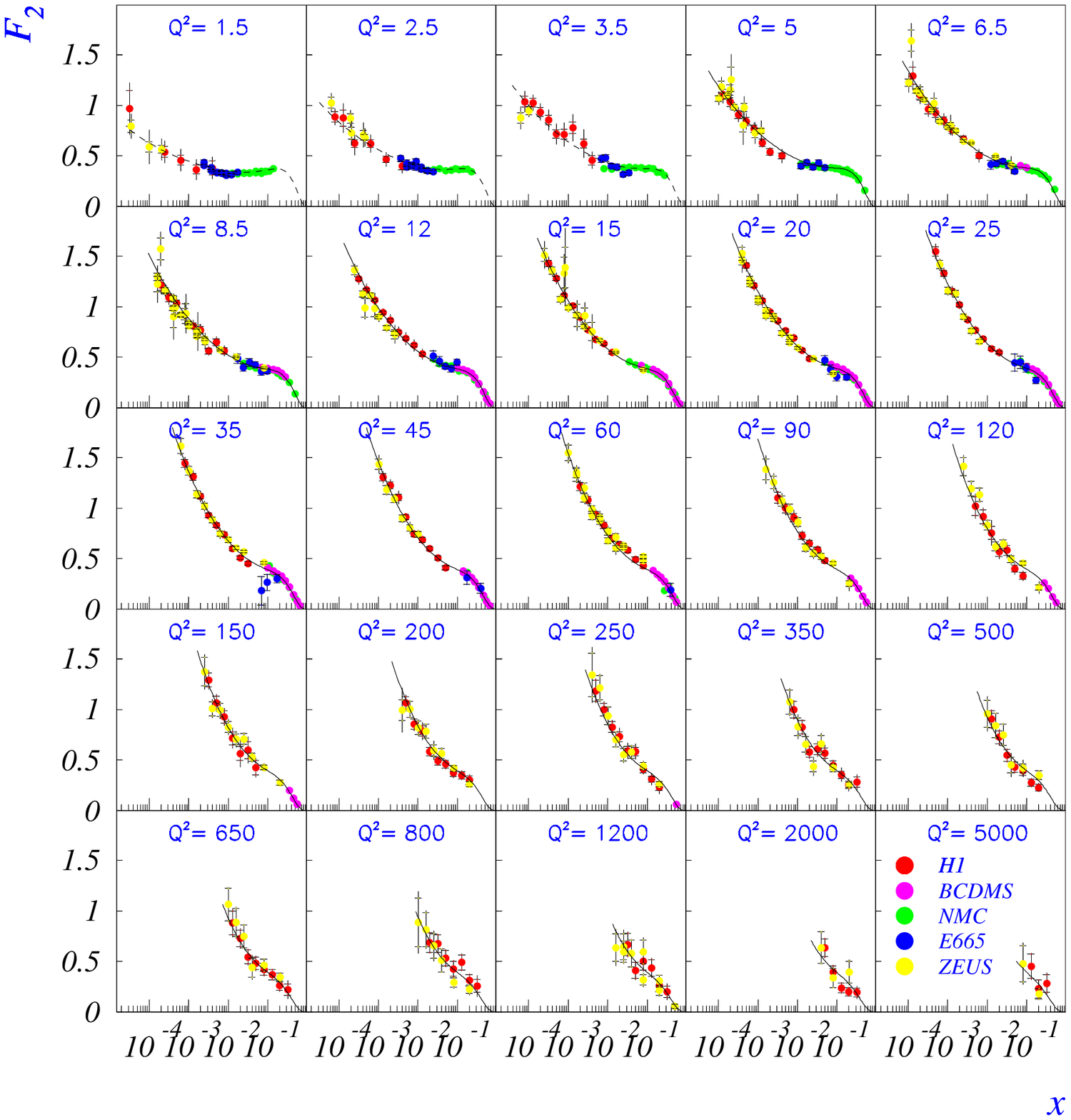}
\end{center}
\vspace{-.5cm}
\caption {\it
{ $F_2$ as function of $x$ for fixed $Q^2$ values (in GeV$^2$) as
indicated in the figure, for the HERA 94 data together with some fixed
target data. The curves are the result of a NLO QCD fit.
}}
\label{fig:f2-94}
\end{figure}

Back to the $F_2$ behaviour.  The good agreement with the QCD
evolution calculations can also be seen in figure~\ref{fig:f2-94}
where this time the $F_2$ data are shown as function of $x$ for fixed
$Q^2$ values.
We see again the good agreement between the HERA and the fixed target
data.  Another interesting observation from this figure is the fact
that the NLO DGLAP equations seem to give a good description of the
data down to $Q^2$ values as low as 1.5 GeV$^2$.  However most
striking is the strong rise of $F_2$ with decreasing $x$ in all $Q^2$
bins. The steepness of this rise seems to be $Q^2$ depended; it gets
shallower as $Q^2$ decreases. This behaviour seems to be embedded in
the QCD evolution equations.

\begin{figure}[hbt]
\begin{center}
  \includegraphics [width=\hsize,totalheight=6cm] 
{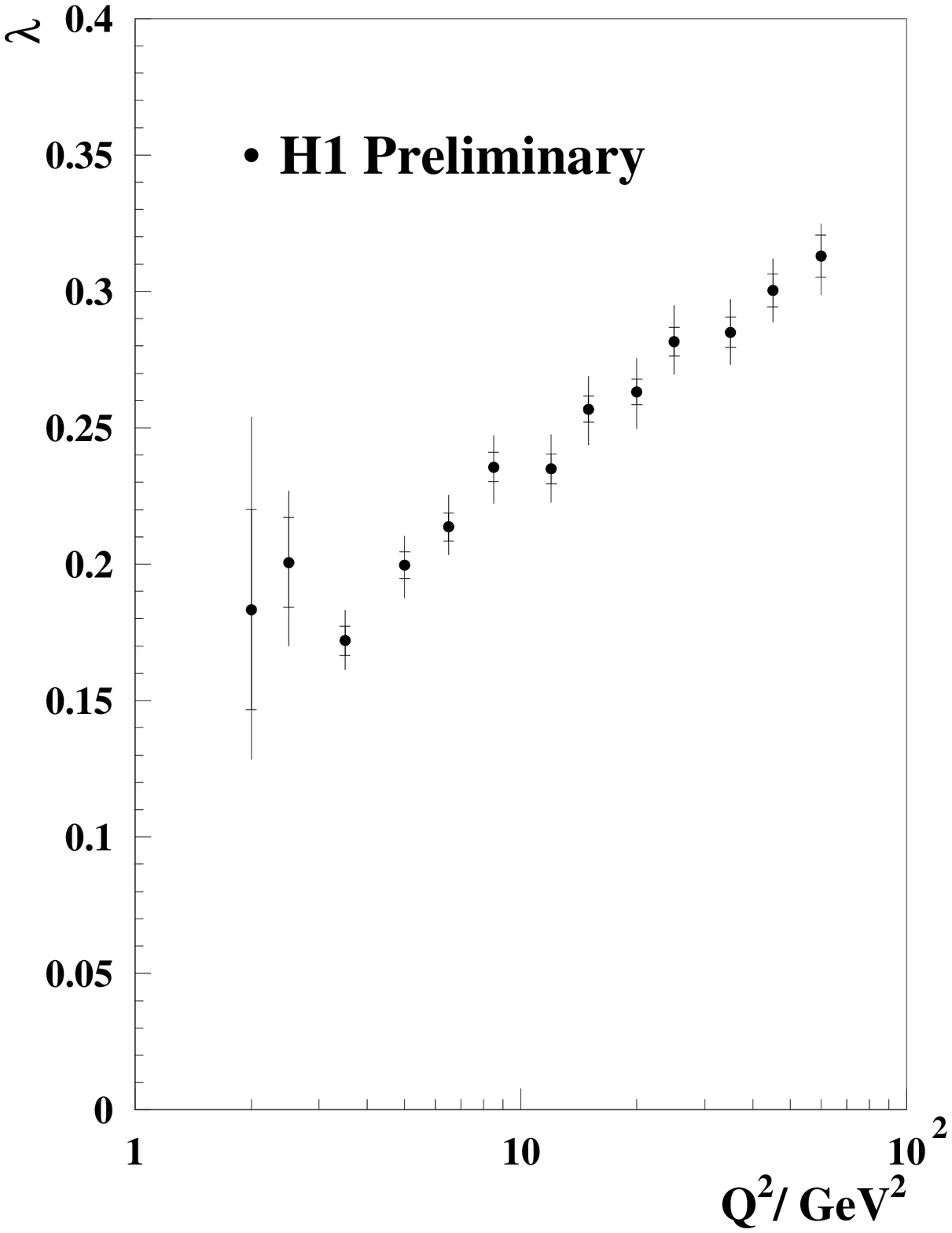}
\end{center}
\vspace{-.5cm}
\caption {\it
{ The behaviour with $Q^2$ of the exponent $\lambda$ from fits of the form
$F_2 \sim x^{-\lambda}$ at fixed $Q^2$ values and $x < $0.1.
}}
\label{fig:h1-lambda}
\end{figure}

Let us try to quantify the rise using a phenomenological argument. In
order to do so it is easier to return to equation
(\ref{eq:f2-sig}) which relates $F_2$ to the $\gamma^* p$ total cross
section and which, for low $x$, reads
\begin{equation}
F_2 \approx
\frac{Q^2}{4\pi^2\alpha}\sigma_{tot}^{\gamma^* p}. 
\end{equation}
Regge theory, which gives a good description of the behaviour of the
total cross section with energy, expects 
\begin{equation}
\sigma_{tot} \sim
s^{\alpha(0)-1}, 
\end{equation}
where $s$ is the squared center of mass energy and $\alpha(0)$ is the
intercept of the leading trajectory. Since in the $\gamma^* p$ case
the squared center of mass energy is $W^2$ and since for low $x$ we
have $W^2 \approx Q^2/x$, one would expect that for fixed $Q^2$
\begin{equation}
F_2 \sim x^{-\lambda},
\label{eq:f2-x}
\end{equation}
where $\lambda = \alpha(0) - 1$. One thus restricts the data to the low
$x$ region, say $x <$ 0.1, and for each $Q^2$ region fits equation
(\ref{eq:f2-x}) to the data and obtains $\lambda(Q^2)$. From the
introductory section where we mentioned the total photoproduction
cross section behaviour we know that $\lambda(Q^2=0)$ = 0.08. What
happens at higher $Q^2$? This we can see in figure~\ref{fig:h1-lambda}
which shows the result of such a fit done by the H1
collaboration~\cite{h1-lambda}.
In spite of the large error bars, one sees the trend of increasing
$\lambda$ from a value of about 0.15 at low $Q^2$ to about 0.3 at
intermediate $Q^2$. The precision of the data does not allow any
conclusion about the continuation of the rise at higher $Q^2$. Since
in the limit of $Q^2$ = 0 the value should decrease to 0.08, it is of
interest to see what is happening between 0.15 and 0.08. Is the
transition sharp or smooth?

\subsection{The low $Q^2$ region}

What is so interesting about the transition? The total photoproduction
cross section shows the same behaviour as the total hadron-hadron
cross sections. They are both dominated by low transverse momentum
interactions, `soft' interactions, and can be described in the Regge
language by the exchange of a DL type Pomeron trajectory. The DIS
domain is well described by QCD, the physics of which is believed to be
dominated by `hard' interactions. Clearly not all of the `soft' region
is only soft just as not all of the `hard' regime is completely
hard. There is an interplay between soft and hard interactions
(see~\cite{afs}). We would like to find where this transition occurs
and maybe learn more about the properties of these two domains. This
is what drove the two HERA experiments to measure also in the very low
$Q^2$ region by methods described in the introductory section.

\begin{figure}[hbt]
\begin{center}
  \includegraphics [width=\hsize,totalheight=10cm] 
{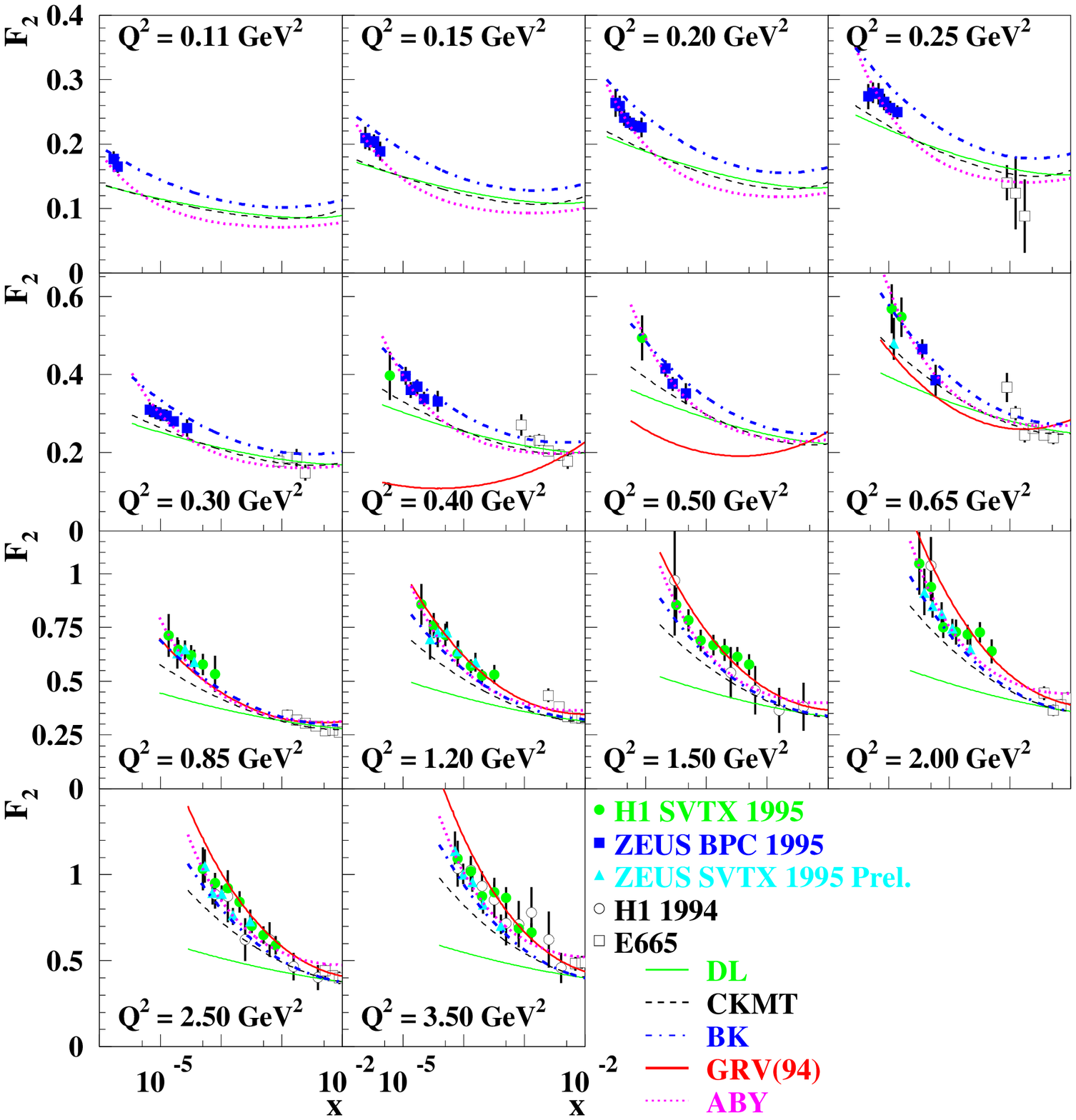}
\end{center}
\vspace{-.5cm}
\caption {\it
{ 
$F_2$ as function of $x$ for fixed values of $Q^2$ as indicated in the figure.
The different lines are the expectations from different parameterizations.
}}
\label{fig:f2-lowq2}
\end{figure}
Figure~\ref{fig:f2-lowq2} shows the behaviour of $F_2$ as function of
$x$ at fixed $Q^2$ values, in the low $Q^2$ region. As one sees, $F_2$
keeps rising with decreasing $x$ even at $Q^2$ as low as 0.11
GeV$^2$. The rise in all $Q^2$ regions seems to be steeper than
expected in the Regge based DL model. The QCD based GRV(94)
model~\cite{grv94} seems to have the correct behaviour around $Q^2
\sim$ 1 GeV$^2$, but overshoots the data at higher $Q^2$ values. For a 
detailed discussion of the other models shown in the figure
(CKMT~\cite{ckmt}, BK~\cite{bk}, ABY~\cite{aby}) see~\cite{al-lowq2}.

\begin{figure}[hbt]
\begin{center}
  \includegraphics [width=\hsize,totalheight=6cm]
  {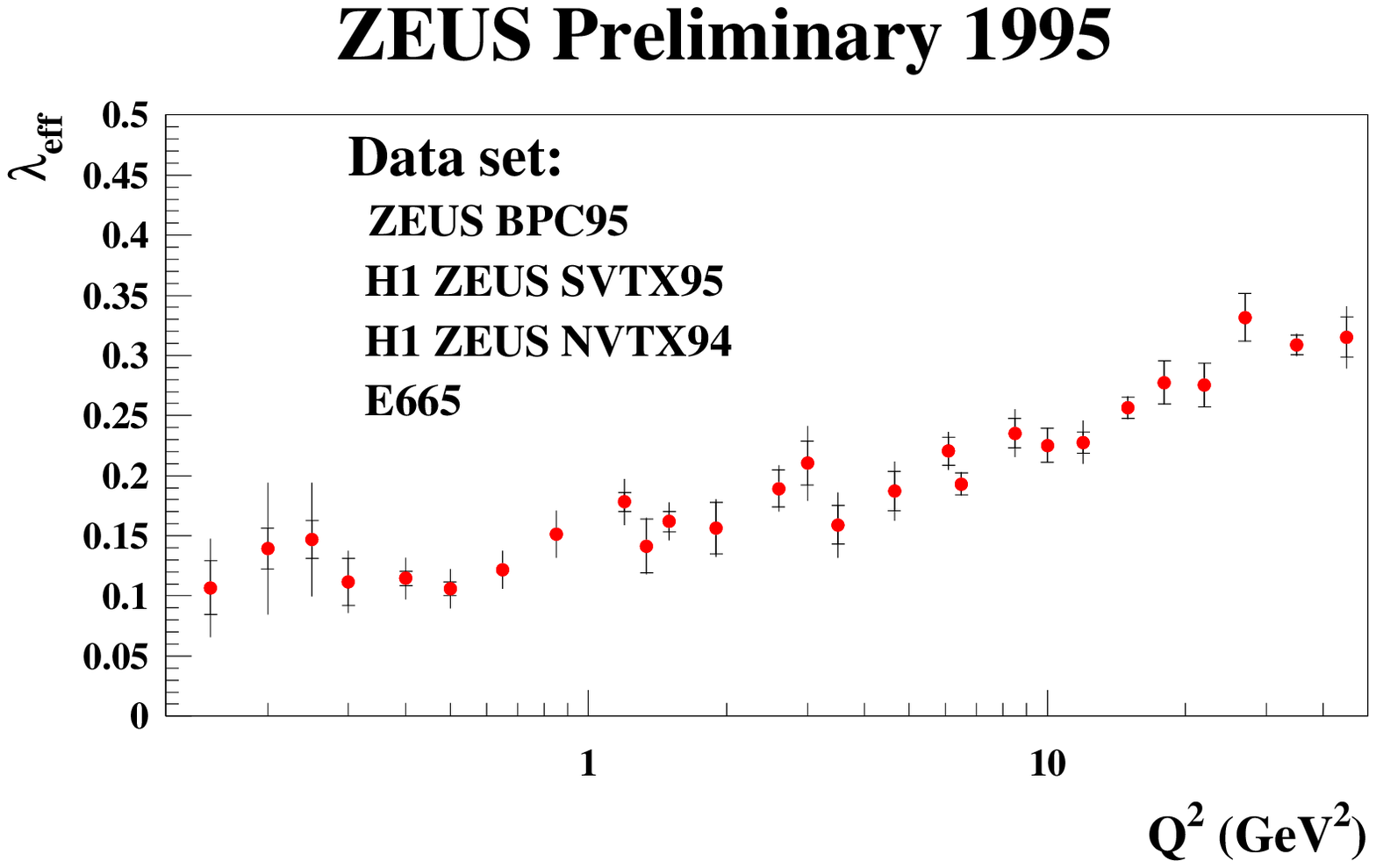}
\end{center}
\vspace{-.5cm}
\caption {\it
{ The behaviour with $Q^2$ of the exponent $\lambda_{eff}$ from fits
of the form $F_2 \sim x^{-\lambda_{eff}}$ at fixed $Q^2$ values and $x <
$0.1.  }}
\label{fig:zeus-lambda}
\end{figure}

One can quantify the rise of $F_2$ with $x$ also in the low $Q^2$
region as was done in figure~\ref{fig:h1-lambda}. This is shown in
figure~\ref{fig:zeus-lambda} for the $Q^2$ range 0.11 - 50 GeV$^2$.
The rise of the exponent (denoted in this figure as $\lambda_{eff}$)
with $Q^2$ seems to be a smooth one. A warning is however in place at
this point. The exponent is sensitive to the range in $x$ over which
the fit is being carried out. There is not always a large enough lever
arm in $x$ from lower energy data to get a good reliable fit for all
values of $Q^2$. Thus further analysis is needed to give an accurate
behaviour of the slope at the low $Q^2$ values.

\begin{figure}[hbt]
\begin{center}
  \includegraphics [width=\hsize,totalheight=10cm] {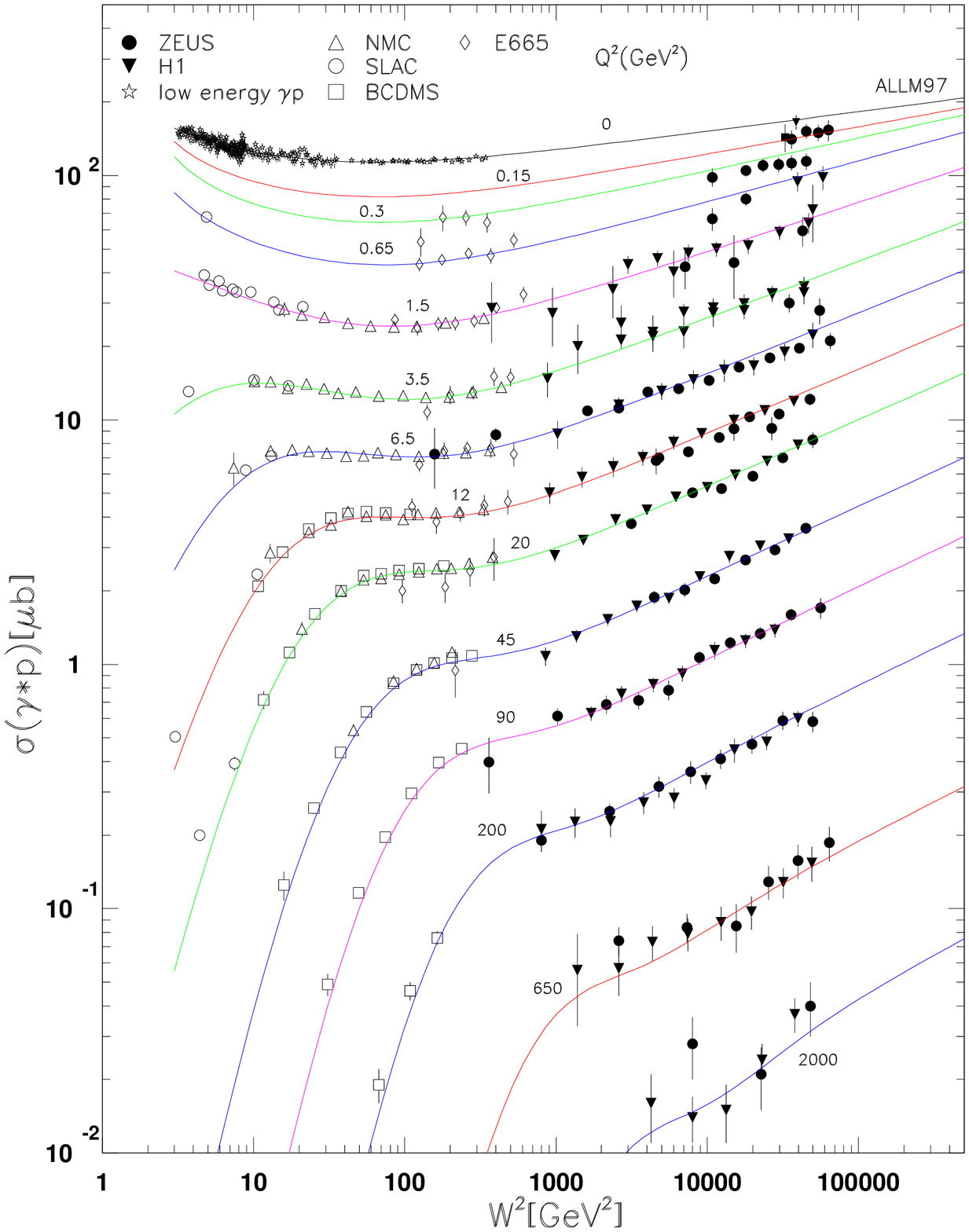}
\end{center}
\vspace{-.5cm}
\caption {\it
{ 
The $\gamma^* p$ total cross section $\sigma(\gamma^* p)$ as
function of $W^2$ for fixed values of $Q^2$, including the total
photoproduction cross section ($Q^2$ = 0). The curves are the results
of the ALLM97 parameterization.
  }}
\label{fig:allm97}
\end{figure}

\begin{figure}[hbt]
\begin{center}
  \includegraphics [width=\hsize,totalheight=10cm] 
{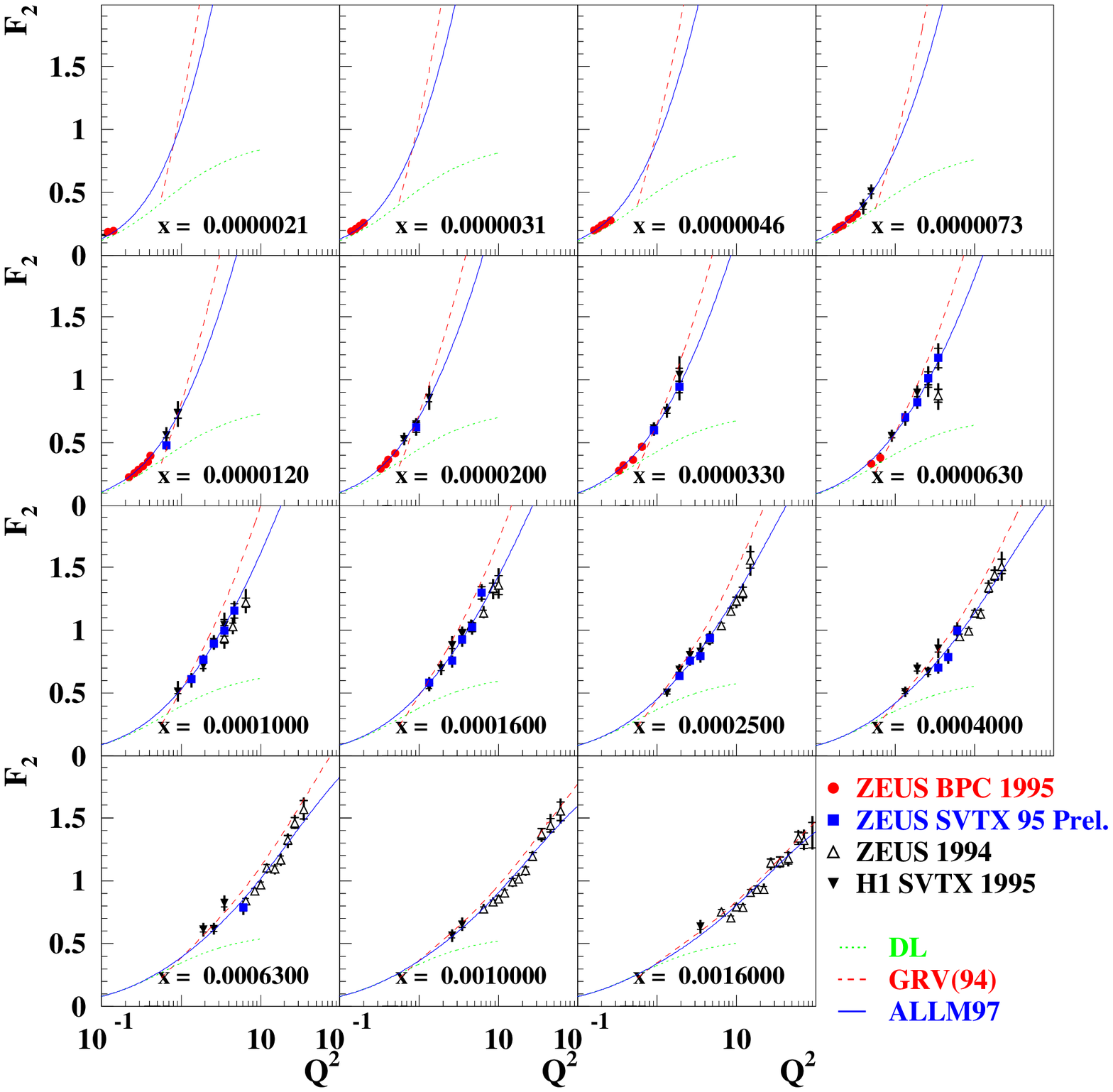}
\end{center}
\vspace{-.5cm}
\caption {\it
{ The behaviour with $Q^2$ of the exponent $\lambda$ from fits of the form
$F_2 \sim x^{-\lambda}$ at fixed $Q^2$ values and $x < $0.1.
}}
\label{fig:f2-q2-lowx}
\end{figure}

There are quite a few parameterizations which attempt to describe the
data both at low and at high $Q^2$ including the transition region. In
figure~\ref{fig:allm97} the $\gamma^* p$ total cross section
$\sigma(\gamma^* p)$ is plotted as function of $W^2$ for fixed values
of $Q^2$, including the total photoproduction cross section ($Q^2$ =
0). The curves are the results of the ALLM97~\cite{allm97}
parameterization which gives a good fit to all the available data over
the whole kinematic region.  The parameterization is based on a Regge
motivated approach, similar to that used earlier by Donnachie and
Landshoff~\cite{dlq2}, extended into the large $Q^2$ regime in a way
compatible with QCD expectations. The transition from the Regge to the
QCD regime can be seen in figure~\ref{fig:f2-q2-lowx} which
shows the $F_2$ data as function of $Q^2$ for fixed $x$ values, a
comparison with the Regge based DL model, the QCD GRV(94) model and
the ALLM97 parameterization which moves from the DL at low $Q^2$ to
GRV at higher $Q^2$, following the data.

\subsection{What have we learned about the proton?}

To summarize this section let us recap what we have learned about the
structure of the proton from the HERA data. There are two clear points
that one can make:
\begin{itemize}
\item
The density of partons increases with decreasing $x$. At $Q^2$ values
of 10-20 GeV$^2$ and low $x$, the proton is dominated by gluons with a
density of more than 20 gluons per unit of rapidity.
\item
The rate of increase of the parton density is $Q^2$ dependent. This
dependence at high $Q^2$ is as expected from `hard' interactions
described by pQCD and at low $Q^2$, as coming from `soft' interactions
described by Regge phenomenology.
\end{itemize}

What does it imply as far as the actual picture of the structure of
the proton is concerned? Can we say for instance that these dense
partons are concentrated somewhere in the interior of the proton? We
will come back to this and other questions at the end of the talk.

\section{The structure of the photon}
\label{sec:photon}

In what follows, whenever we talk about a photon, we mean either a
real photon or a quasi-real one. We will also discuss separately the
structure of virtual photons.  Let us recap what we said in the
introduction about the photon structure.  A high energy photon can
develop a structure when interacting with another object. This happens
because the photon can fluctuate into $q\bar{q}$ pairs and as long as
the fluctuation time is large compared to the interaction time we can
talk about the structure of the photon. Due to this phenomena a photon
has a probability to interact as a photon directly with the other
object, or first resolve itself into partons which subsequently take
part in the interaction. We call the first case a {\bf direct} photon
interaction and the second one, a {\bf resolved} photon interaction.

\subsection{The behaviour of the photon structure function}

\begin{figure}[hbt]
\begin{center}
  \includegraphics [bb = 11 68 524 731,width=\hsize,totalheight=11cm]
  {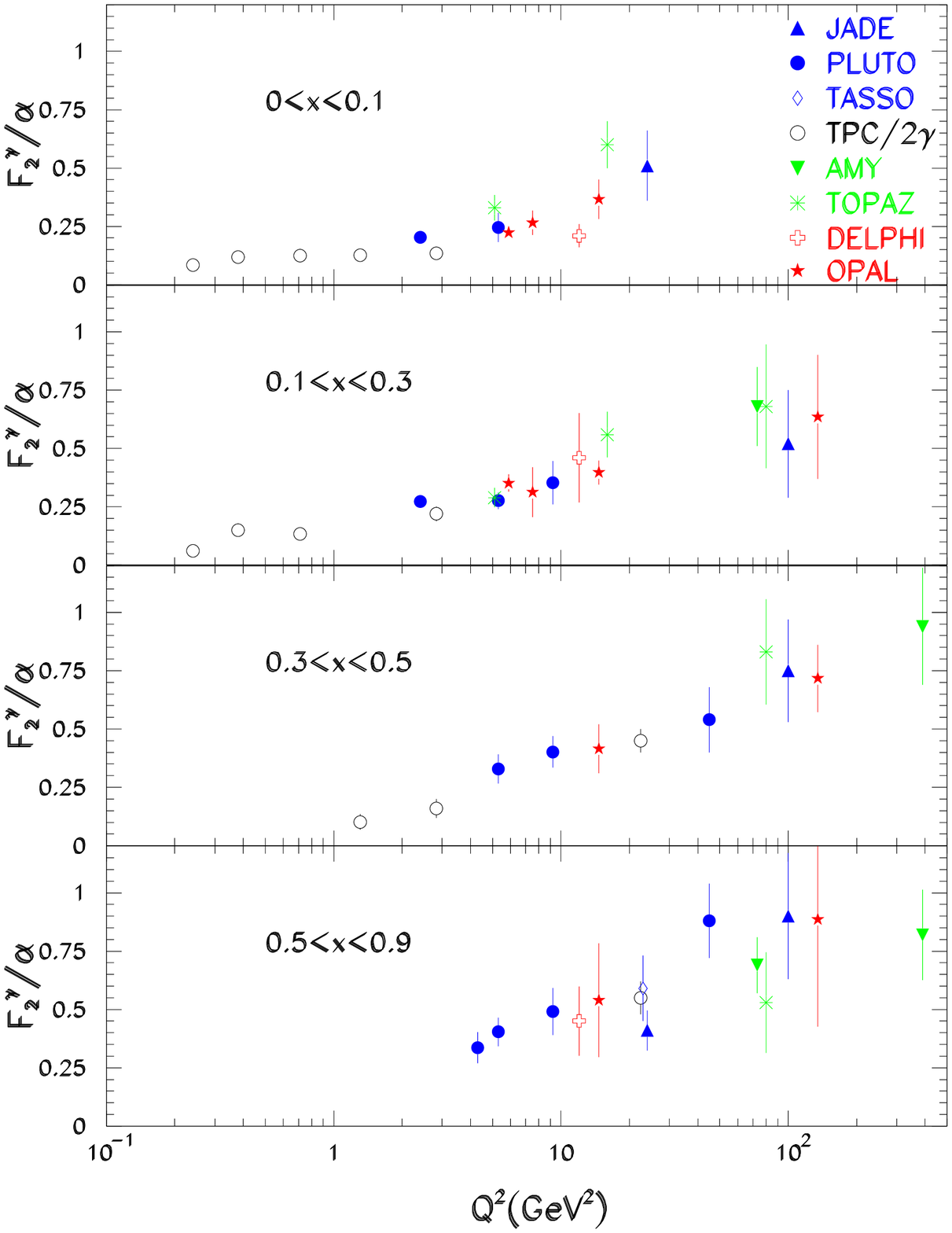}
\end{center}
\vspace{-.5cm}
\caption {\it
{ $F_2^\gamma$ as function of $Q^2$ for $x$ intervals as given in the
figure.  }}
\label{fig:f2g-q2}
\end{figure}
The photon structure function $F_2^\gamma$ was measured in DIS type of
$e^+ e^-$ collisions as described in figure~\ref{fig:dis-photon}. From
the structure function measurements one could obtain information on
the parton distributions in the photon in a similar way to that of the
proton, using the DGLAP equations. There is however one difference
when viewing the DGLAP equations for the photon more closely. In the
proton case, when considering the evolution equations, one takes into
account the splitting of the partons in the following way: a quark of
a given $x$ can split into a gluon and a quark, $q \to q + g$, both
with lower $x$ but the sum of their $x$'s equal to the `parent' $x$; a
gluon can split into a pair of quark-antiquark, $g \to q + \bar{q}$; a
gluon can split into two gluons, $ g \to g + g$. However in the photon
case one has in addition to all above splittings the possibility of a
photon to split into a quark-antiquark pair, $\gamma \to q + \bar{q}$.
This changes the DGLAP evolution equations from homogeneous to
inhomogeneous ones. One of the consequences is that the scaling
violations for the photon case are positive for all $x$ values,
contrary to the proton case which had positive scaling violation for
low $x$ and negative for high $x$. This can be seen in
figure~\ref{fig:f2g-q2}, where the photon structure function
$F_2^\gamma$ is plotted as function of $Q^2$ for different $x$
intervals, and shows positive scaling violation in all $x$ bins. This
behaviour should be contrasted to figure~\ref{fig:scalingviolation}
for the proton case.

\begin{figure}[hbt]
\begin{center}
  \includegraphics [bb = 23 160 537 670,width=\hsize,totalheight=13cm]
  {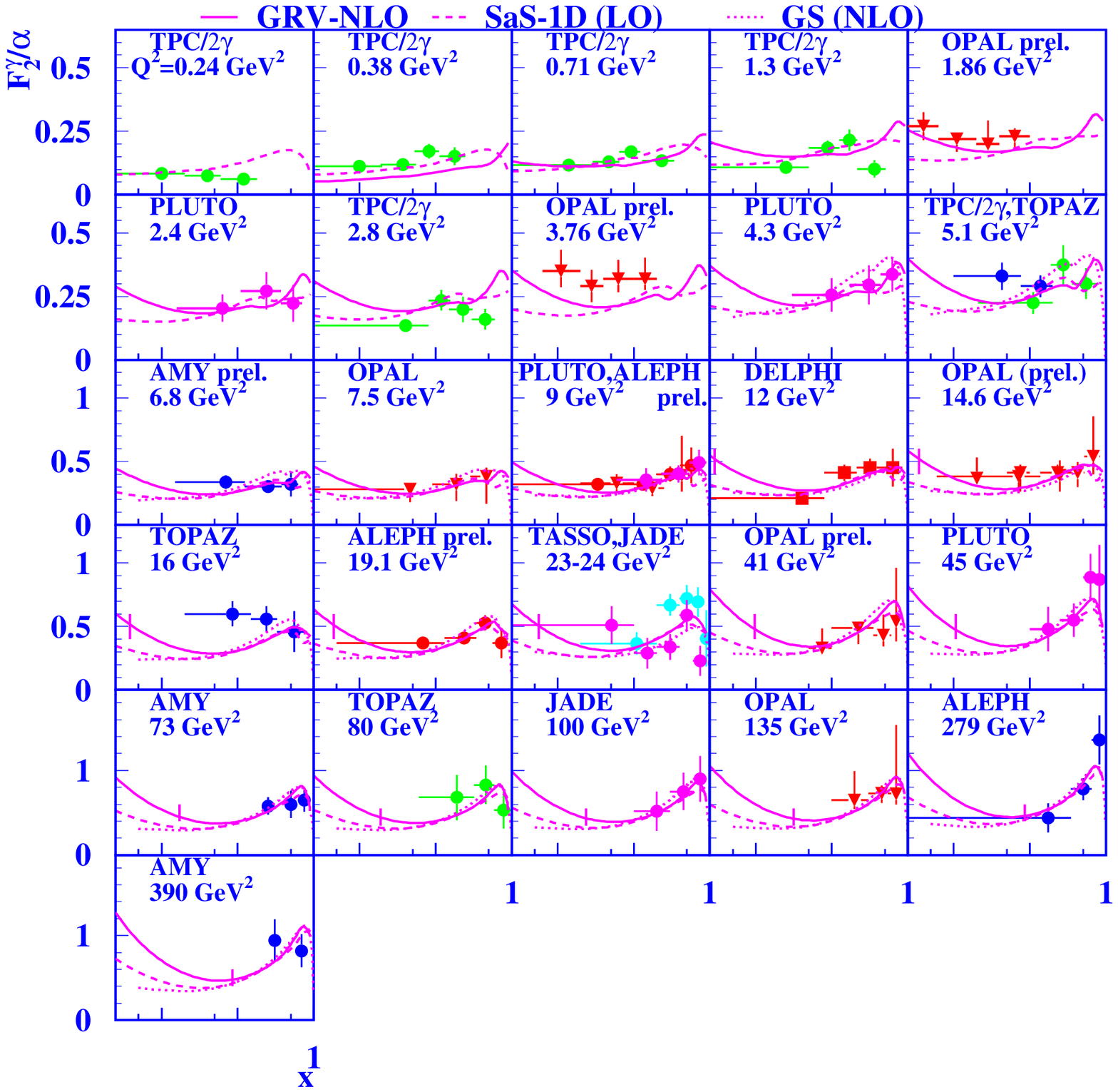}
\end{center}
\vspace{-.5cm}
\caption {\it 
{ $F_2^\gamma$ as function of $x$ for fixed $Q^2$ as given in the
figure. The curves are the expectations of different parameterizations
of parton distributions in the photon.  }}
\label{fig:f2g-lp97}
\end{figure}

A compilation of the photon structure function, as presented at the
Lepton Photon Symposium LP97~\cite{stefan-lp97}, is shown in
figure~\ref{fig:f2g-lp97} as function of $x$ for fixed $Q^2$.
The measurements have much larger errors than in the proton case. One
of the main reasons is connected to the fact that it is difficult to
determine the $x$ value of a given event in $\gamma \gamma$
interactions. The value of $x$ can in principle be calculated through
the relation
\begin{equation}
x = \frac{Q^2}{Q^2 + W^2}.
\end{equation}
While $Q^2$ can be obtained by measuring the scattered electron (see
figure~\ref{fig:dis-photon}), usually with an accuracy of better than
10 \%, it is difficult to reconstruct the true $W$ value just from the
measured final state kinematics. Another observation about
figure~\ref{fig:f2g-lp97} is that there is little data at low
$x$. This is a consequence of not reaching, so far, high values of $W$
in $\gamma \gamma$ reactions. The curves in the figure are the
calculations of some parameterizations~\cite{f2g-para} of parton
distributions in the photon. While they give approximately the same
results in the region where data exist, their predictions are quite
different for the low $x$ region which is not constrained by
data. Lately there has been much activity in the LEP
community~\cite{lep-gg} to get more accurate data of $F_2^\gamma$ and
extend the measurements to lower $x$ values (see
also~\cite{maria-photon}).

\subsection{Direct and resolved photon at HERA}

Measurements of $F_2^\gamma$ constrain the quark distribution, while
the gluons are badly determined. The first indications that
gluons have to be present in the photon came from an analysis of jets
in photon-photon interactions performed by the AMY
collaboration~\cite{amy-gluon}. They showed that the only way they can
explain the data is by introducing some gluons in the photon. However
it was at HERA that the picture of a direct and resolved photon was
most clearly seen. At HERA? How does one study the structure of the
photon at HERA? Did we not study the structure of the proton with the
help of the photon in the previous section? As we said in the
introduction to the present section, we consider now the structure of
a quasi-real photon. In this case the probe is one of the large
transverse momentum partons from the proton, and the photon plays the
role of the probed target.

\begin{figure}[hbt]
\begin{minipage}{7cm}
\begin{center}
  \includegraphics [bb= 227 260 576 699,width=6cm,totalheight=8cm]
  {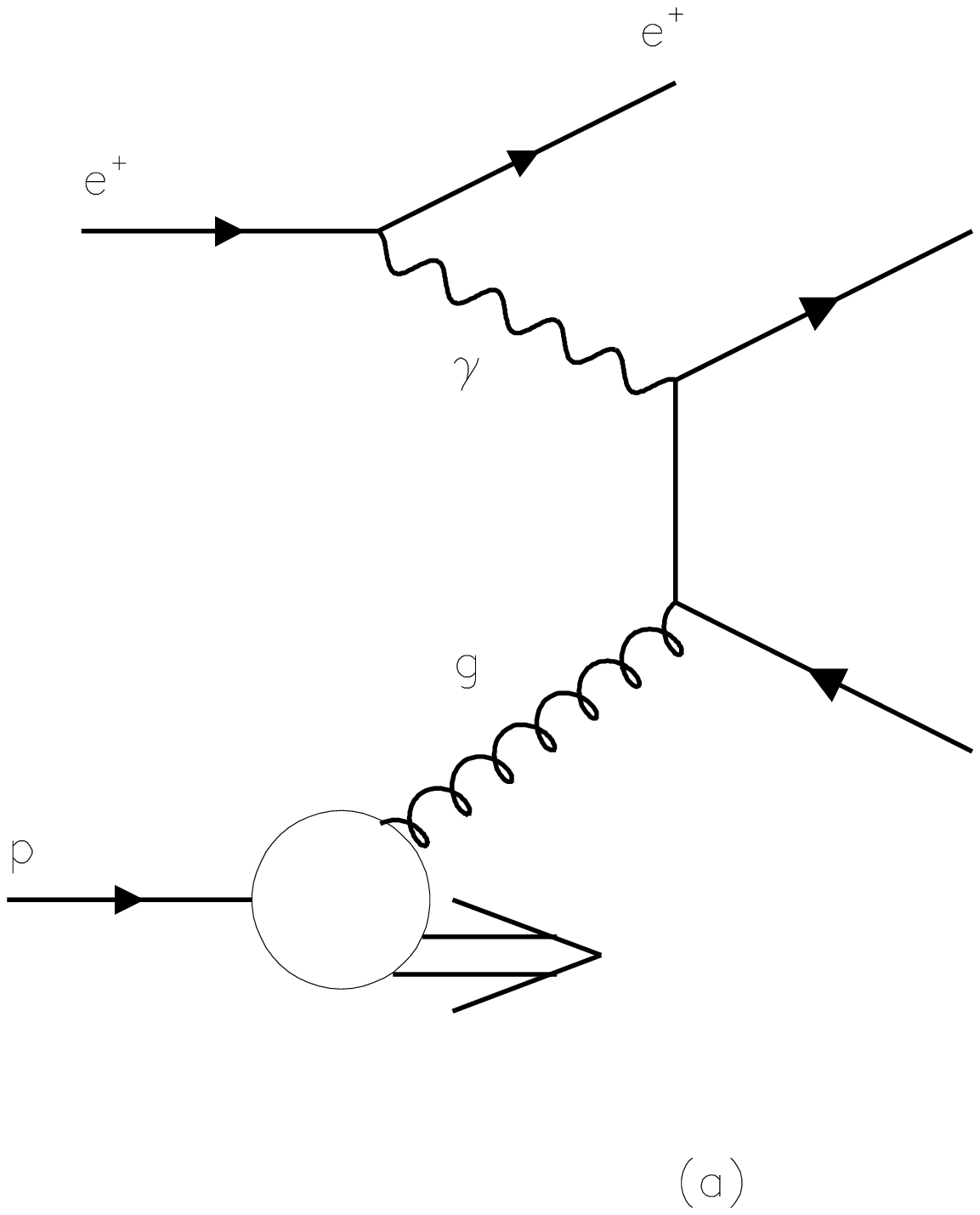}
\end{center}
\end{minipage}
\begin{minipage}{7cm}
\begin{center}
  \includegraphics [bb= 117 260 576 699,width=8cm,totalheight=10cm]
  {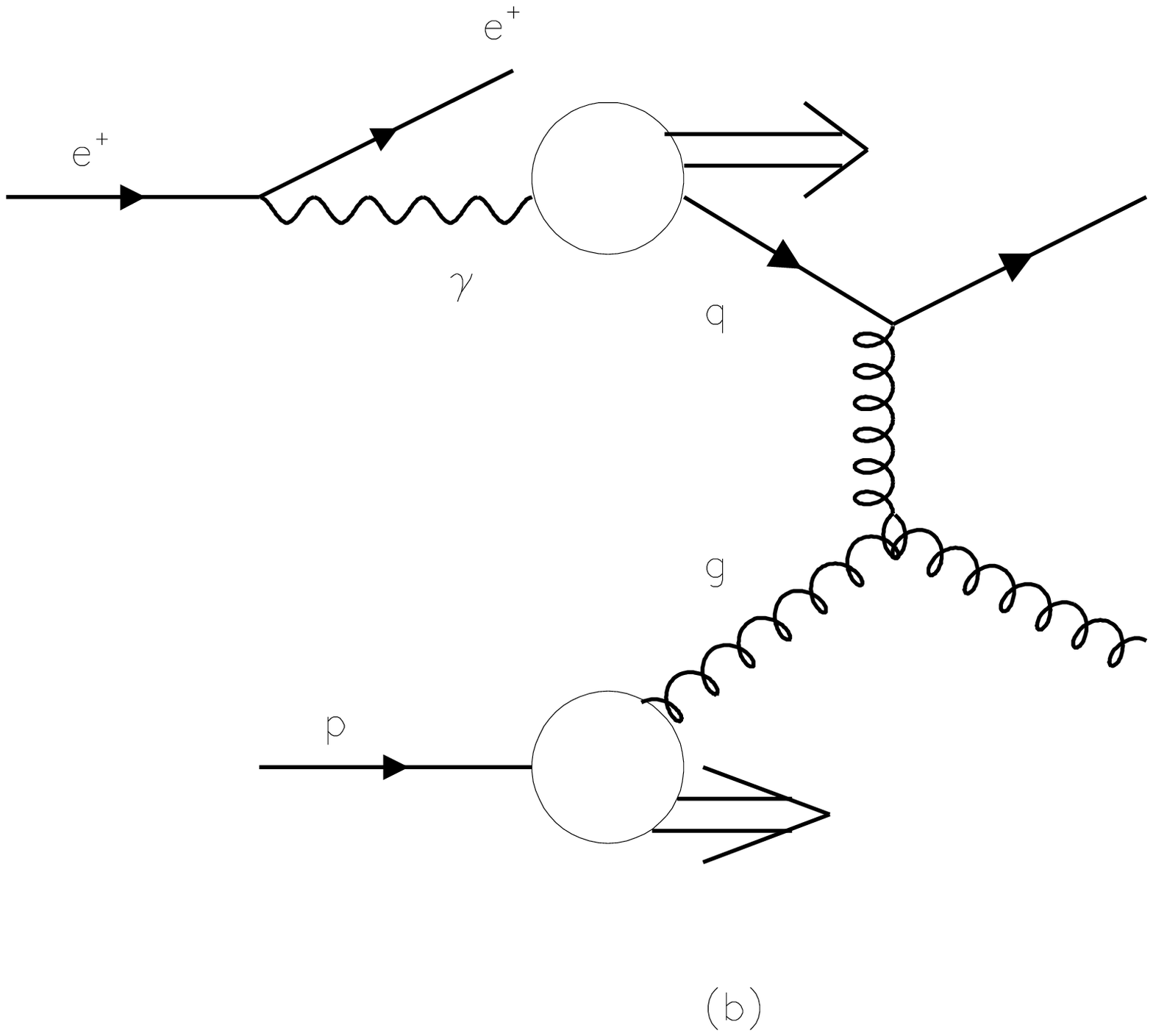}
\end{center}
\end{minipage}
\vspace{-.5cm}
\caption {\it
{Examples of leading order QCD (a) `direct' and (b) `resolved' dijet
production diagrams.  }}
\label{fig:direct-resolved}
\end{figure}

The diagrams in figure~\ref{fig:direct-resolved} describe what is
meant in leading-order (LO) by direct and resolved photon in an
example of photoproduction of dijets. 
Diagram (a) describes the case in which all of the photon energy is
involved in the dijet production. In diagram (b) the photon first
resolves into partons; one of them interacts with a parton from the
photon to produce a dijet while the rest remain as a photon
remnant. In this case only part of the initial photon energy
participates in the dijet production. If we denote by $x_\gamma$ the
fraction of the photon energy participating in the dijet production,
we expect $x_\gamma$ = 1 for the direct reaction and $x_\gamma <$ 1
for the resolved case.  One can calculate an observable
$x_\gamma^{obs}$ which will gives us a good approximation of the
fraction $x_\gamma$,
\begin{equation}
x_\gamma^{obs} = \frac{E_T^{j1} e^{-\eta^{j1}} +
E_T^{j2} e^{-\eta^{j2}}} {2E_\gamma}.
\end{equation}
The transverse energies of the outgoing jets are denoted
by $E_T^{ji}$ and their pseudorapidities~\footnote{The pseudorapidity
of a particle is defined as $\eta = -\ln \tan{\frac{\theta}{2}}$,
where $\theta$ is the particle production angle.} by $\eta^{ji}$.

\begin{figure}[hbt]
\begin{center}
  \includegraphics [width=\hsize,totalheight=8cm]
  {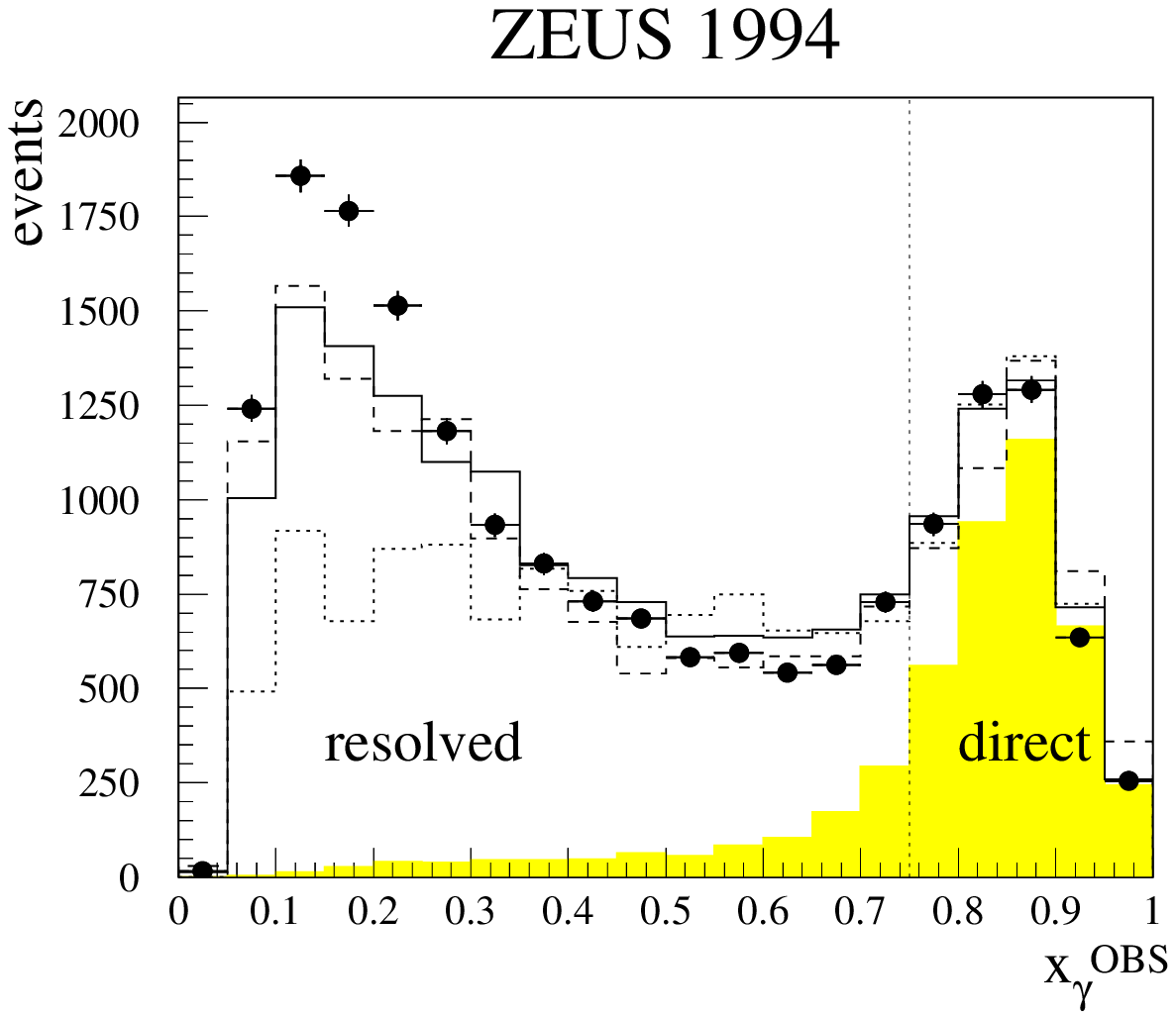}
\end{center}
\vspace{-.5cm}
\caption {\it
{The $x_\gamma^{obs}$ distribution as obtained from photoproduction of
dijet events. The shaded area are the expectations of the distribution
of this variable from the generation of direct photon events. The
dotted vertical line is the border of an operational definition of
direct and resolved photon events.  }}
\label{fig:xg}
\end{figure}
Figure~\ref{fig:xg} displays the variable $x_\gamma^{obs}$ as measured
by the ZEUS collaboration~\cite{zeus-xg}.  The data show a large
enhancement at low $x_\gamma$ values coming from events in which only
a small part of the photon energy participated in the production of
the two jets. We called these resolved photon events in the above
discussion. However, there is in addition a second peak around
$x_\gamma^{obs} \approx$ 0.9, indicating that the whole of the photon
participated in the dijet production, meaning it was a direct photon
event. As we said above, $x_\gamma^{obs}$ is an estimator of
$x_\gamma$ and therefore is somewhat smaller than 1. In order to see
that the large $x$ enhancement indeed comes from direct photon
interactions, such events were generated in a Monte Carlo program and
the results are displayed as a shaded band in figure~\ref{fig:xg}. It
is clear from this that most of the large $x$ enhancement comes from
direct photon events. How would one choose these events? We have a
distribution in the figure and have to make an operational definition
as to what we call direct and resolved events. The dotted line at
$x_\gamma^{obs}$ = 0.75 is defined as the division line between direct
($>$ 0.75) and resolved ($<$ 0.75) events.

\begin{figure}[hbt]
\begin{center}
  \includegraphics [bb = 19 400 535 700,width=\hsize,totalheight=8cm]
  {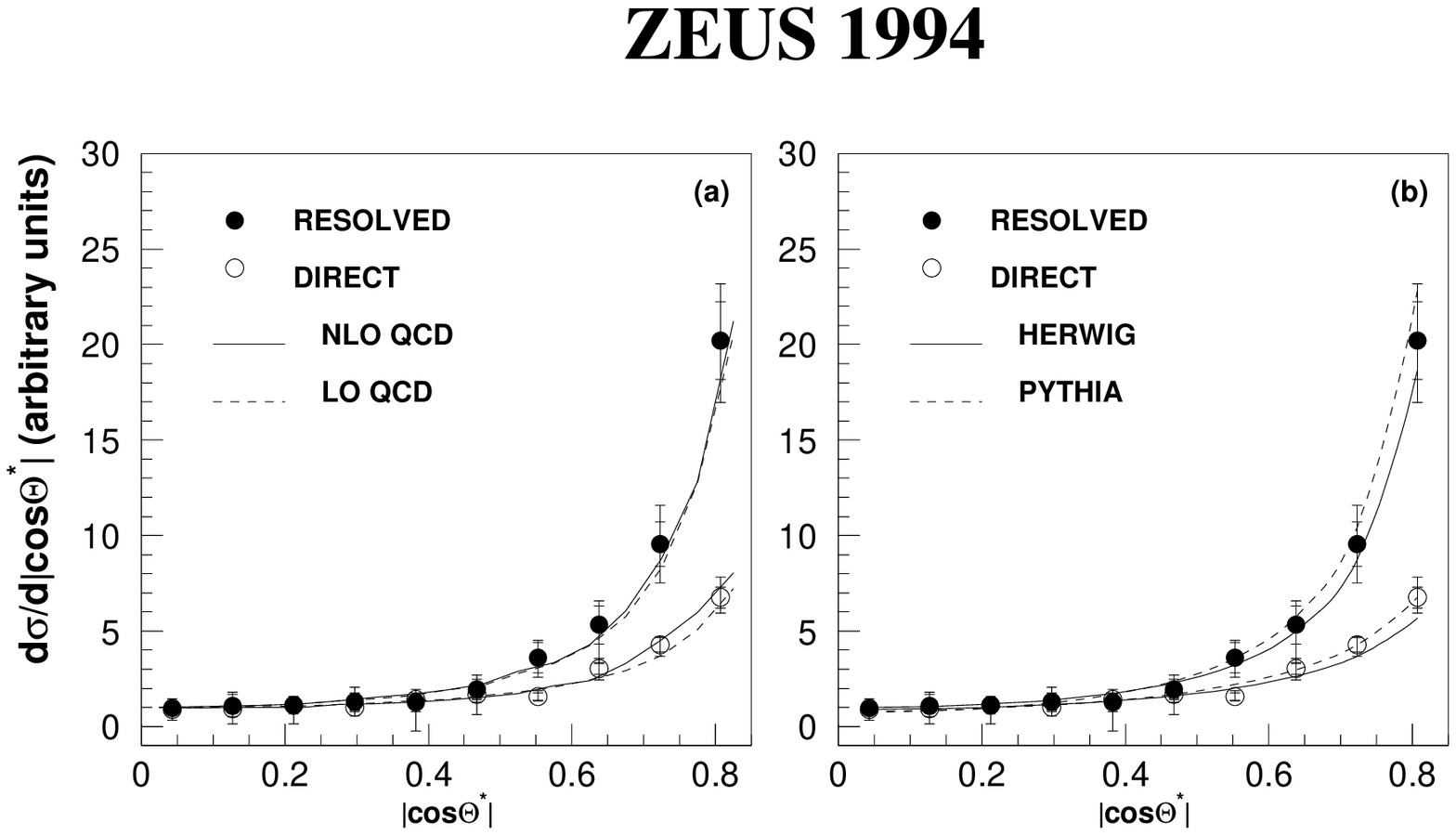}
\end{center}
\vspace{-.5cm}
\caption {\it 
{ The angular distribution $\frac{d\sigma}{d\cos\theta}$ of dijet
events produced by resolved and by direct photon events.  In (a) the
data are compared to leading-order (LO) and next-to-leading-order
(NLO) QCD calculations at the parton level. In (b) the comparison of
the data is done at the hadron level with two Monte Carlo generators.
}}
\label{fig:ang}
\end{figure}

How can one check independently that this definition is sensible? Let
us look again at the two diagrams in figure~\ref{fig:direct-resolved}. In
case of the direct photon diagram (also called `boson-gluon fusion')
the exchanged particle in the reaction $\gamma g \to q q$ is a quark,
while in the resolved case, the exchange particle in the reaction $q g
\to q g$ is a gluon. In case of the quark exchange, we have a 
spin $\frac{1}{2}$ propagator while in the gluon exchange case there is
a spin 1 propagator. These different propagators lead to different
angular distributions of the two outgoing quarks which are the
`parents' of the dijet. In case of the quark exchange one expects,
\begin{equation}
\frac{d\sigma}{d\cos\theta} \sim \frac{1}{1 - |\cos\theta|},
\end{equation}
while for the spin 1 exchange the angular distribution should be,
\begin{equation}
\frac{d\sigma}{d\cos\theta} \sim \frac{1}{(1 - |\cos\theta|)^2}.
\end{equation}
The quark exchange should dominate the direct photon events while the
gluon exchange will dominate the resolved photon events. By using the
above definition of the $x_\gamma^{obs}$ cut, one can choose samples of
direct and resolved photon events and study their angular
distribution. The results of this study~\cite{zeus-ang} are displayed
in figure~\ref{fig:ang}.
The data fulfill the expectations of our definition above. The events
which were chosen as resolved photon events have a steeper angular
distribution than the direct dijet events. The data are compared at
two levels. The diagrams and the angular distribution considerations
were done at the parton level, while the measured data are at the
hadron level. In case of the jet the expectations are that the jet
`remembers' the direction of its `parent' parton. Thus the comparison
of the data is once done to a parton level calculation and then to a
hadron level simulation. In figure~\ref{fig:ang}(a) the comparison of
the data is to a leading-order (LO) and to a next-to-leading-order
(NLO) QCD calculation. In figure~\ref{fig:ang}(b) the comparison is
done with two Monte Carlo generators who give the angular
distributions at the hadron level. The agreement with the data is very
good in all cases. We can thus conclude that our direct and resolved
definition using the $x_{\gamma}^{obs}$ cut is a reasonable one. 

\subsection{The gluon density in the photon}

We have already said this but let us repeat: the structure function
constrains the quarks. The information about the gluon distribution is
indirectly obtained through the evolution equations. However the
photon structure function data are not precise enough for a good
determination of the gluon density in the photon.

\begin{figure}[hbt]
\begin{center}
  \includegraphics [width=\hsize,totalheight=8cm]
  {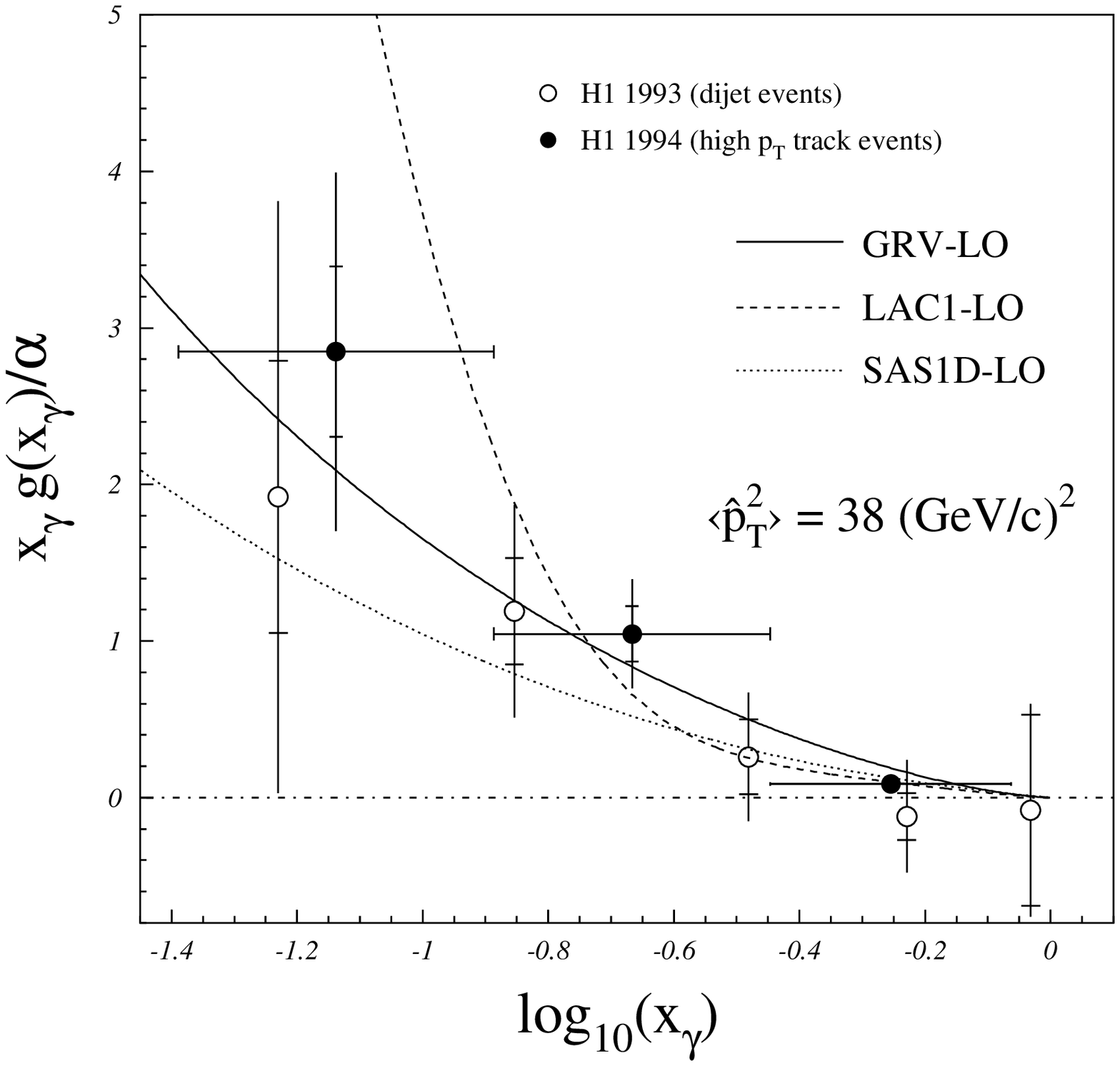}
\end{center}
\vspace{-.5cm}
\caption {\it
{ The gluon density in the photon as function of $x_\gamma$ at a scale
of 38 GeV$^2$. The different curves are the expectations of different
parameterizations of parton distributions in the photon.  }}
\label{fig:h1-gluong}
\end{figure}
An interesting method to overcome this difficulty was carried out by
the H1 collaboration~\cite{h1-gluong}. They use the high transverse
momentum charged tracks in photoproduction events to reconstruct the
$x_\gamma$ distribution. Since the quark densities are constrained by
the $F_2^\gamma$ measurements and thus relatively well known, their
contribution to the $x_\gamma$ distribution are taken from the
expectations of the photon parton parameterizations and are subtracted
to give the gluon density as function of $x_\gamma$. The result of
this analysis is shown in figure~\ref{fig:h1-gluong} for the case
where the hard scale, taken from the average squared transverse
momentum of all charged tracks, was 38 GeV$^2$. The data shows a rise
of the gluon density with decreasing $x_\gamma$, a similar trend as in
the proton case. The data of the analysis described above is compared
to a similar analysis in which the $x_\gamma$ was reconstructed using
dijet photoproduction events.The two results are consistent with each
other. The data are also in general agreement with three
parameterizations of the parton distributions in the photon
(GRV-LO~\cite{grv-lo-g}, LAC1-LO~\cite{lac}, SaS1D-LO~\cite{sas1d}).

\subsection{The structure of virtual photons}

So far we discussed the structure of a real photon. Do virtual photons
also have structure? The only measurement of the structure function of
a photon with virtuality of about 0.4 GeV$^2$ was carried out by the
PLUTO Collaboration~\cite{pluto-virtg} in $e^+ e^-$ collisions at a
probing scale of 5 GeV$^2$. The cross section of $e^+ e^-$
interactions in which both leptons are detected is falling fast and
thus such a measurement is difficult. At HERA however, one can study
the structure of virtual photons by measuring the $x_\gamma$
distributions for events in different ranges of the photon virtuality
$Q^2$. The presence of a structure of virtual photons will show itself
in events having a low $x_\gamma$ value. This will signal resolved
virtual photons (see discussion of figure~\ref{fig:direct-resolved}).

The ZEUS collaboration~\cite{zeus-virtg} used data in three different
ranges of $Q^2$. In the first, the scattered lepton was measured in
the luminosity electron calorimeter, ensuring that the photon is
quasi-real with a median $Q^2$ = $10^{-3}$ GeV$^2$. The second sample
included events in which the scattered positron was measured in the
BPC, yielding photons with virtuality in the region $0.1 < Q^2 < 0.7$
GeV$^2$. In the third sample the scattered lepton was detected in the
main calorimeter and was in the range $1.5 < Q^2 < 4.5$ GeV$^2$.
\begin{figure}[hbt]
\begin{center}
  \includegraphics [width=5cm,totalheight=8cm]
  {figs/xgq2-1.ps}
  \includegraphics [width=5cm,totalheight=8cm]
  {figs/xgq2-2.ps}
  \includegraphics [width=5cm,totalheight=8cm]
  {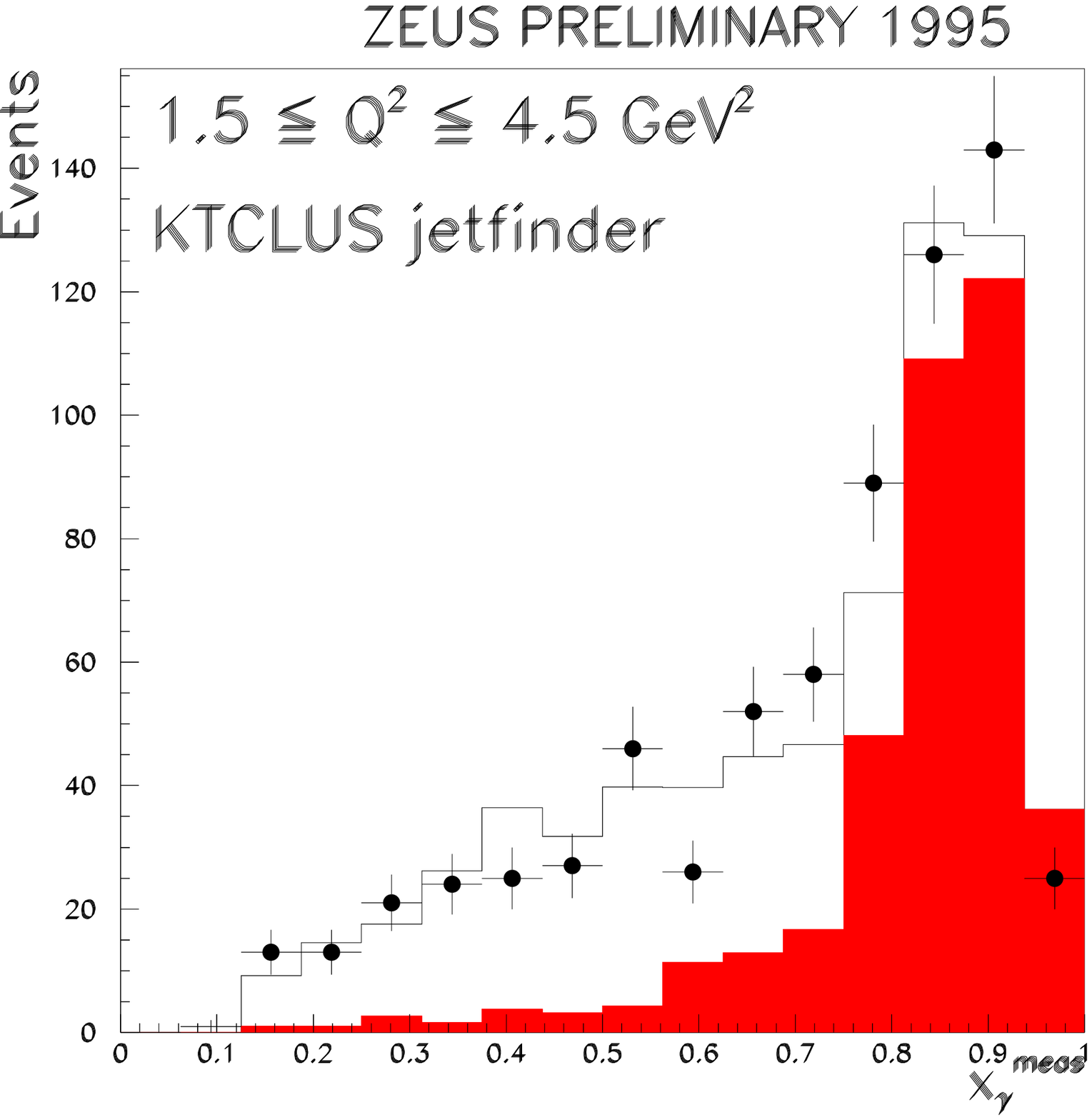}
\end{center}
\vspace{-.5cm}
\caption {\it
  { The $x_gamma$ distributions for three regions of $Q^2$ as
    displayed in the figure. The unshaded histograms are the sum of
    resolved and direct photon contributions as calculated from a LO
    Monte Carlo generator. The shaded histograms are only the LO
    direct photon contributions.  }}
\label{fig:xg-q2}
\end{figure}
The dijets were used to calculate the value of $x_\gamma$ in each
$Q^2$ region, results of which are displayed in
figure~\ref{fig:xg-q2}. The data show the presence of resolved photon
events in the low $x_\gamma$ region while the enhancement at high
$x_\gamma$ is due to direct photon events. The histograms in the
figures are the sum of resolved and direct photon contributions as
calculated from a LO Monte Carlo generator. The shaded histograms are
only the LO direct photon contributions.

\begin{figure}[hbt]
\begin{minipage}{6cm}
\begin{center}
  \includegraphics [width=\hsize,totalheight=6cm]
  {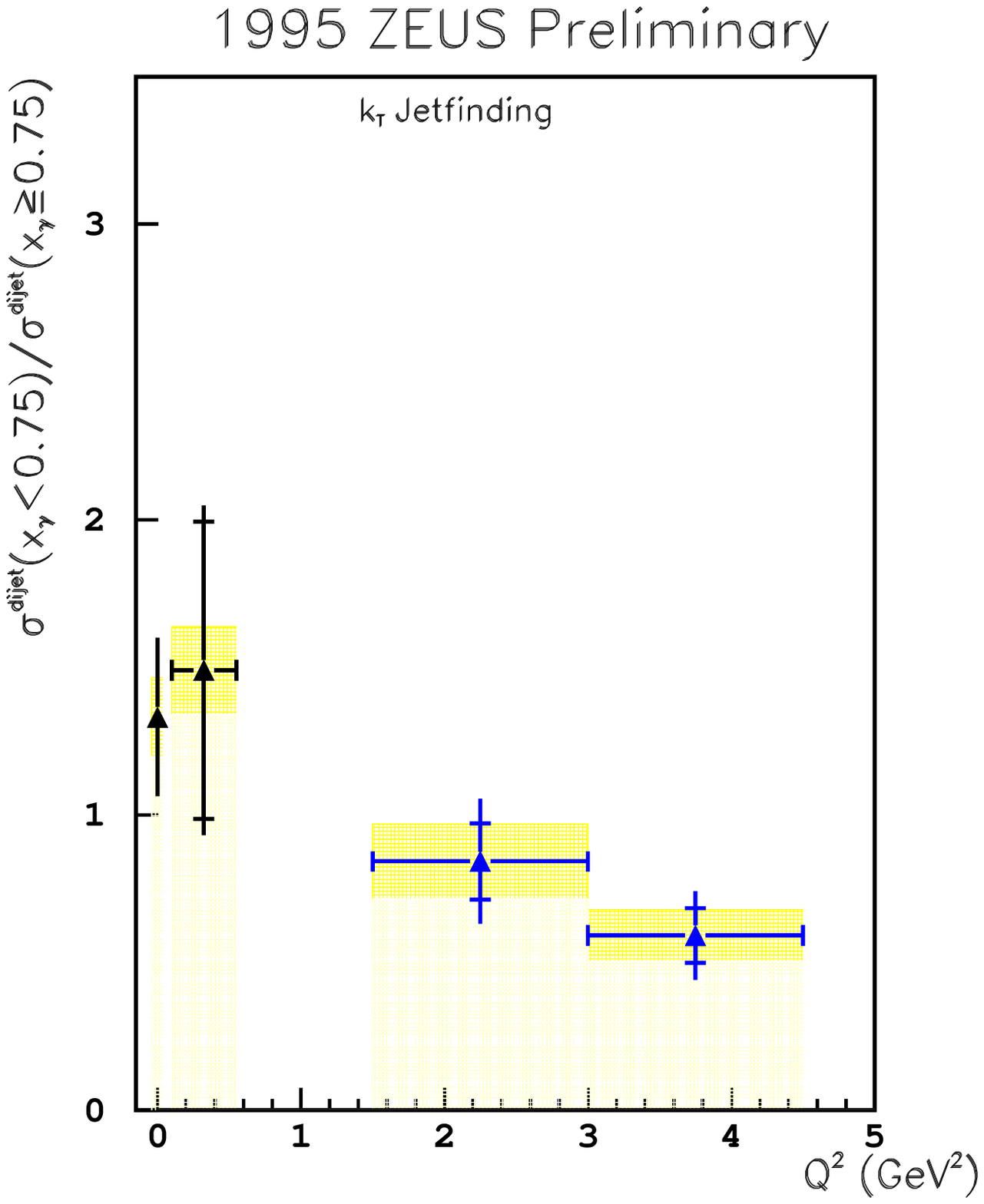}
\end{center}
\vspace{-.5cm}
\caption {\it
{ The ratio of resolved to direct photon cross sections as a function
of the photon virtuality $Q^2$.  }}
\label{fig:rg-q2}
\end{minipage}
\hspace{3mm}
\begin{minipage}{9.5cm}
\begin{center}
  \includegraphics [width=\hsize,totalheight=10cm]
  {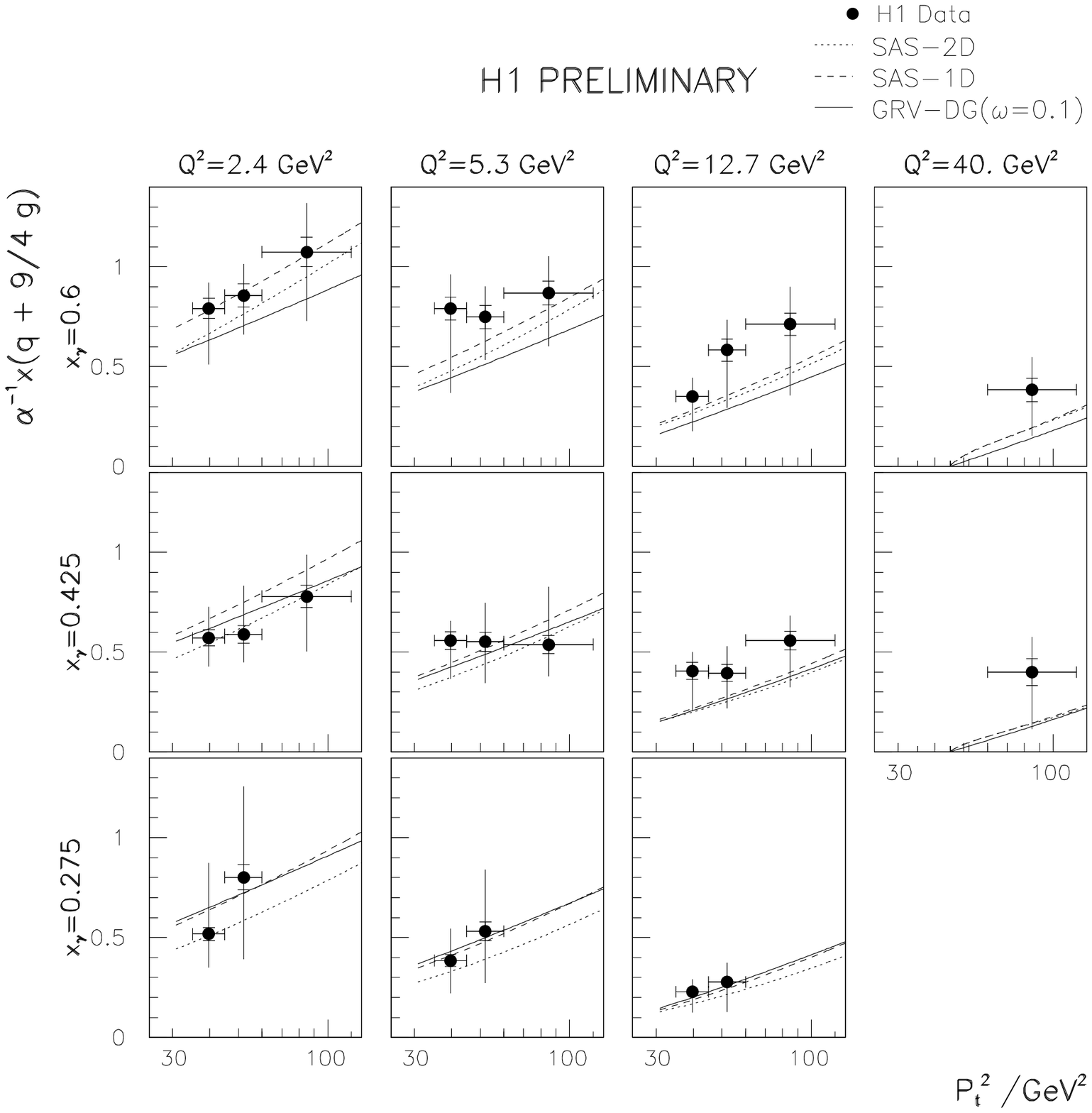}
\end{center}
\vspace{-.5cm}
\caption {\it
{ The effective parton density of the photon as function of the
squared of the parton transverse momentum $P_t^2$ for different values
of $Q^2$ and $x_\gamma$. The lines are expectations of different
parameterizations of parton distributions in the photon.
}}
\label{fig:h1-eff-partons}
\end{minipage}
\end{figure}

Using the data and the operational definition of $x_\gamma <$ 0.75 as
resolved photon events, one can calculate the ratio of resolved to
direct photon events as function of the virtuality of the photon.
This ratio is displayed in figure~\ref{fig:rg-q2} and is decreasing
with increasing $Q^2$. Note however that there is still an appreciable
cross section of resolved photon events even at a virtuality of $Q^2$
= 4.5 GeV$^2$. It should be noted that in this study the minimum
transverse energy $E_T$ of each of the two jets was 6.5 GeV$^2$, thus
making sure that the probing scale (estimated by $E_T^2$) is much
larger than the virtuality $Q^2$ of the probed photon.

The H1 collaboration extracted in a similar study~\cite{h1-virtg} an
effective parton density of the virtual photon, shown in
figure~\ref{fig:h1-eff-partons}. Here too dijet events have been used
in a wide range of the photon virtuality, $1.6 < Q^2 < 80$ GeV$^2$,
with the requirement that the probing parton transverse momentum
$P_t^2$ is always much larger than $Q^2$. The effective parton density
seems to rise with $P_t^2$ in all regions of $Q^2$, a behaviour which
is in general agreement with expectations from parameterizations of
parton distributions in the photon.

\subsection{Who is probing whom?}

The results we just obtained in the last subsection are somewhat
alarming. We understand that a quasi-real photon has structure which
is built during the interaction. Now we see that the same can be true
for virtual photons, even with virtualities of some tens of GeV$^2$.
How can a virtual photon develop a structure through fluctuation? Is
the fluctuation time long enough? What about
equation~(\ref{eq:tfluc-virtg})? It turns out that the calculation of
the fluctuation time for low $x$ gives~\cite{ioffe,afs},
\begin{equation}
t_f \approx \frac{1}{2 m_p x}.
\end{equation}
So at low $x$ even a photon of virtuality $Q^2 \sim 10^3$ GeV$^2$ can
fluctuate! And here comes the alarming part: what are we measuring in
a DIS interaction at low $x$?  The simple picture of DIS is
complicated at low $x$ by the long chain of gluon and quark ladders
which describes the process in QCD.  In this long chain of partons
along the ladder, where does one draw the line? Does one study the
structure of the proton? of the photon? of both? How should one
interpret the DIS measurement? Who is probing whom?

It is clear that physics can not be frame dependent~\cite{bj94}.  Thus
it must be that both descriptions are correct and reflect the fact
that cross sections are Lorentz invariant but time development is
not~\cite{lonya}. This means that it shouldn't matter whether one
interprets the cross section measurements as yielding the proton or
the photon structure function. By extracting one of them from the
cross section measurement, there should be a relation allowing to  
obtain the other. 

Here we have a problem. We have seen, at least as far as a real photon
is concerned, that its structure function behaves very differently
from that of the proton one. For instance, the $Q^2$ scaling violation
is positive in the photon case for all values of $x$, while for the
proton they change from positive to negative scaling violations as one
moves to higher $x$ values. So how can the proton structure function
$F_2^p$ and that of the photon, $F_2^\gamma$, be related?

Actually at low $x$ the two structure functions can be related. By
assuming Gribov factorization~\cite{gribov-fact} to hold also for a
virtual photon one can show~\cite{al-fact} that,
\begin{equation}
F_2^\gamma(x,Q^2) = F_2^p(x,Q^2) \frac{\sigma_{\gamma p}(W^2)} 
{\sigma_{p p}(W^2)}.
\label{eq:fac-f2}
\end{equation}
This last equation connects the proton and the real photon structure
function at low $x$. By measuring one of them, the other can be determined
through relation~(\ref{eq:fac-f2}).

\subsection{$F_2^\gamma$ at low $x$}

\begin{figure}[hbt]
\begin{center}
  \includegraphics [bb= 8 25 520 516,width=\hsize,totalheight=10cm]
  {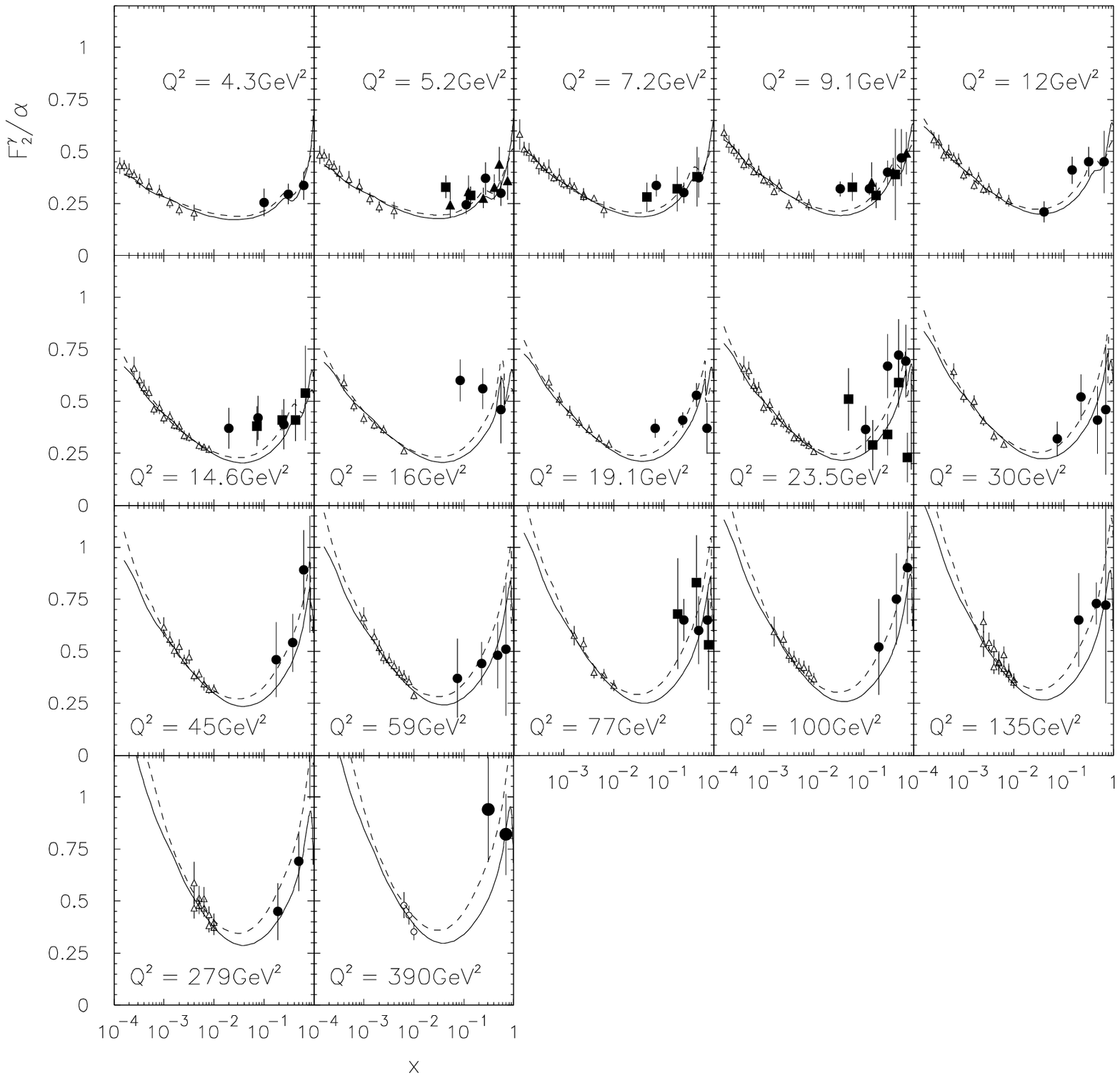}
\end{center}
\vspace{-.5cm}
\caption {\it
{ The photon structure function as function of $x$ for fixed $Q^2$
values as indicated in the figure. The full points are the direct
measurements and the open triangles are those obtained from $F_2^p$
through the Gribov factorization relation.  The full line is the
result of the present HO fit and the dashed line is that of the LO
parameterization.
 }}
\label{fig:newf2}
\end{figure}
Relation~(\ref{eq:fac-f2}) allows the use of well measured quantities
like total cross sections and the proton structure function $F_2^p$ to
predict the values of the photon structure function $F_2^\gamma$ in
the region of low $x$ where equation~(\ref{eq:fac-f2}) is expected to
be valid.  Since this is also the region where direct measurements of
the photon structure function are difficult and not available, the use
of~(\ref{eq:fac-f2}) provides a way to `obtain' $F_2^\gamma$ `data'
and use them as an additional source, on top of the direct
measurements of $F_2^\gamma$, to constrain the parton distributions in
the photon. This was done in LO in~\cite{gal-lo}, where for the total
cross sections of $\gamma p$ and $p p$ the DL~\cite{dl}
parameterization has been used. Recently~\cite{gal-ho} the same method
was applied in a higher-order (HO) treatment and the results are
displayed in figure~\ref{fig:newf2}.  The directly measured photon
structure function data appear as full points, while the data obtained
through the use of the Gribov factorization relation are displayed as
open triangles.  All the low $x$ data coming from the proton
structure function have been scaled to the value of $Q^2$ which is
indicated in the figure.  The solid curves are the results of the HO
parameterization. For comparison we show as dashed lines the result of
the LO parameterization which is very similar to the HO one, with the
difference between the two growing as $Q^2$ increases.

\subsection{Configurations of photon fluctuation}

\begin{figure}[hbt]
\begin{center}
  \includegraphics [bb= 60 215 479 398,width=\hsize,totalheight=4cm]
  {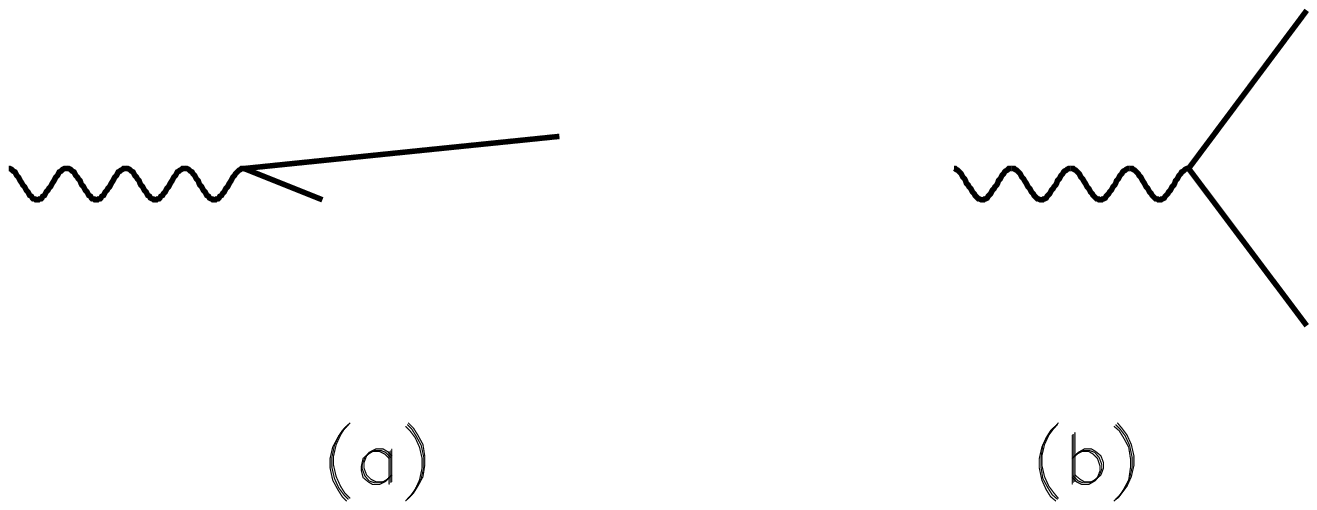}
\end{center}
\vspace{-.5cm}
\caption {\it
{Fluctuation of the photon into a $q \bar{q}$ pair in (a)
asymmetric small $k_T$ configuration, (b) into a symmetric large $k_T$
configuration.}}
\label{fig:ajm} 
\end{figure}

The photon can fluctuate into typically two configurations. A large size
configuration will consist of an asymmetric $q\bar{q}$ pair with each
quark carrying a small transverse momentum $k_T$ (fig. \ref{fig:ajm}(a)).
For a small size configuration the pair is symmetric, each quark having a
large $k_T$ (fig. \ref{fig:ajm}(b)). One expects the asymmetric large
configuration to produce 'soft' physics, while the symmetric one would   
yield the 'hard' interactions.

In the aligned jet model (AJM)~\cite{ajm} the first
configuration dominates while the second one is the 'sterile combination'
because of color screening. In the photoproduction case ($Q^2$ = 0), the
small $k_T$ configuration dominates. Thus one has large color forces
which produce the hadronic component, the vector mesons, which finally   
lead to hadronic non--perturbative final states of 'soft' nature. The
symmetric configuration contributes very little. In those cases
where the photon does fluctuate into a high $k_T$ pair, color
transparency suppresses their contribution.

In the DIS regime ($Q^2 \neq$ 0), the symmetric contribution becomes bigger.
Each such pair still contributes very little because of color
transparency, but the phase space for the symmetric configuration
increases. However the asymmetric pair still contribute also to the DIS
processes. In fact, in the quark parton model (QPM) the fast quark
becomes the current jet and the slow quark interacts with the proton
remnant resulting in processes which look in the $\gamma^* p$ frame just
like the 'soft' processes discussed in the $Q^2$ = 0 case. So there  
clearly is an interplay between soft and hard interactions also in the   
DIS region.

\subsection{What have we learned about the photon?}

We can summarize our present knowledge about the structure of the
photon in the following way:

\begin{itemize}
\item Real photons have structure which is developed in the
  interactions through the fluctuation of the photon into a $q
  \bar{q}$ pair. There are clear signs of the direct and resolved
  photon processes at HERA.
\item The photon and proton structure functions can be related at low
  $x$.
\item Virtual photons also develop a structure.
\item The nature of the interaction, soft or hard, is determined by
  the configuration of the photon fluctuation. This creates regions in
  physics in which there is an interplay between soft and hard
  interactions.
\end{itemize}

\section{The structure of the Pomeron}
\label{sec:pomeron}

The soft and hard interplay mentioned in the last section brings us in
a natural way to the subject of diffraction. We know from
hadron-hadron interactions that diffraction~\cite{predazzi} is a
`soft' phenomena. It is described by the exchange of the Pomeron
trajectory. We already mentioned earlier that Donnachie and
Landshoff~\cite{dl} determined this trajectory as,
\begin{equation}
\alpha_{\pom}(t) = 1.08 + 0.25 t.
\end{equation}
The intercept was determined from a fit to all hadron-hadron total
cross section data and the slope $\alpha^\prime_{\pom}$ was
determined~\cite{alpha-prime} from $p p$ elastic scattering data.
What does HERA tell us about diffraction? Is it a soft phenomenon
described by Regge phenomenology? Is it a hard process calculable in
pQCD? Or both? Let us first look at inclusive studies in DIS.

\subsection{Inclusive studies in diffractive DIS processes.}

We have seen in the introduction that the discovery of large rapidity
gap events in DIS came as a surprise.  
The reason was that our
intuition about DIS is based on the quark-parton model and on the QCD
evolution.  It is difficult to see how in such a picture the struck
parton will not radiate gluons so as to create large rapidity gap
events which are not exponentially suppressed.  That is why
none of the Monte-Carlo
generators which were written for DIS physics at HERA included
diffractive type of events. 
On the other hand, from the
proton rest frame point of view, one could naturally expect large
rapidity gap events for example in the aligned-jet model (shown to
hold also in QCD~\cite{fs-physrep}).

However, even before HERA started producing data, Ingelman and
Schlein~\cite{ingelman-schlein} hypothesized that the Pomeron has a
partonic structure like any other hadron. They suggested that the
structure of the Pomeron can be studied in a similar DIS process as
for the proton, in the reaction $e p \to e p X$ which was described in
figure~\ref{fig:dis-pomeron}. That diagram is very similar to the one
of a $\gamma^* \gamma$ interaction in which the process at the lepton
vertex is looked upon as a source of photons and is factorized in the
calculation of the $\gamma^*
\gamma$ cross section of the diagram. Also in the Pomeron case one
assumes that the proton acts as a source of a flux of Pomerons which
are probed by the $\gamma^*$. In the interpretation of the process as
probing the structure of the Pomeron one thus assumes that the cross
section for the process $ e p \to e p X$ can be factorized into the
contribution coming from $e {\pom} \to e X$ and the one from the
vertex $ p \to {\pom} p$ which produces the flux of Pomerons.

The assumption of factorization is used to develop the whole formalism
of the Pomeron structure function which can be subjected to the same
DGLAP equations as the proton, given the fact that the QCD
factorization theorem has been proven~\cite{collins-qcd-fact} to hold
also in inclusive DIS diffractive processes. These inclusive processes
have been measured by both the H1~\cite{h1-incl-diff} and the
ZEUS~\cite{zeus-incl-diff} collaborations and the cross section has
been analyzed as a product of the Pomeron flux and the Pomeron
structure function.  Contrary to the case of the proton, the
scaling violation of the Pomeron structure function did not change
sign when moving from low to high $\beta$ values. The variable $\beta$
for the Pomeron case has a similar meaning as the Bjorken $x$ for the
proton.

\begin{figure}[hbt]
\begin{minipage}{8cm}
\begin{center}
  \includegraphics [width=7.5cm,totalheight=10cm]
  {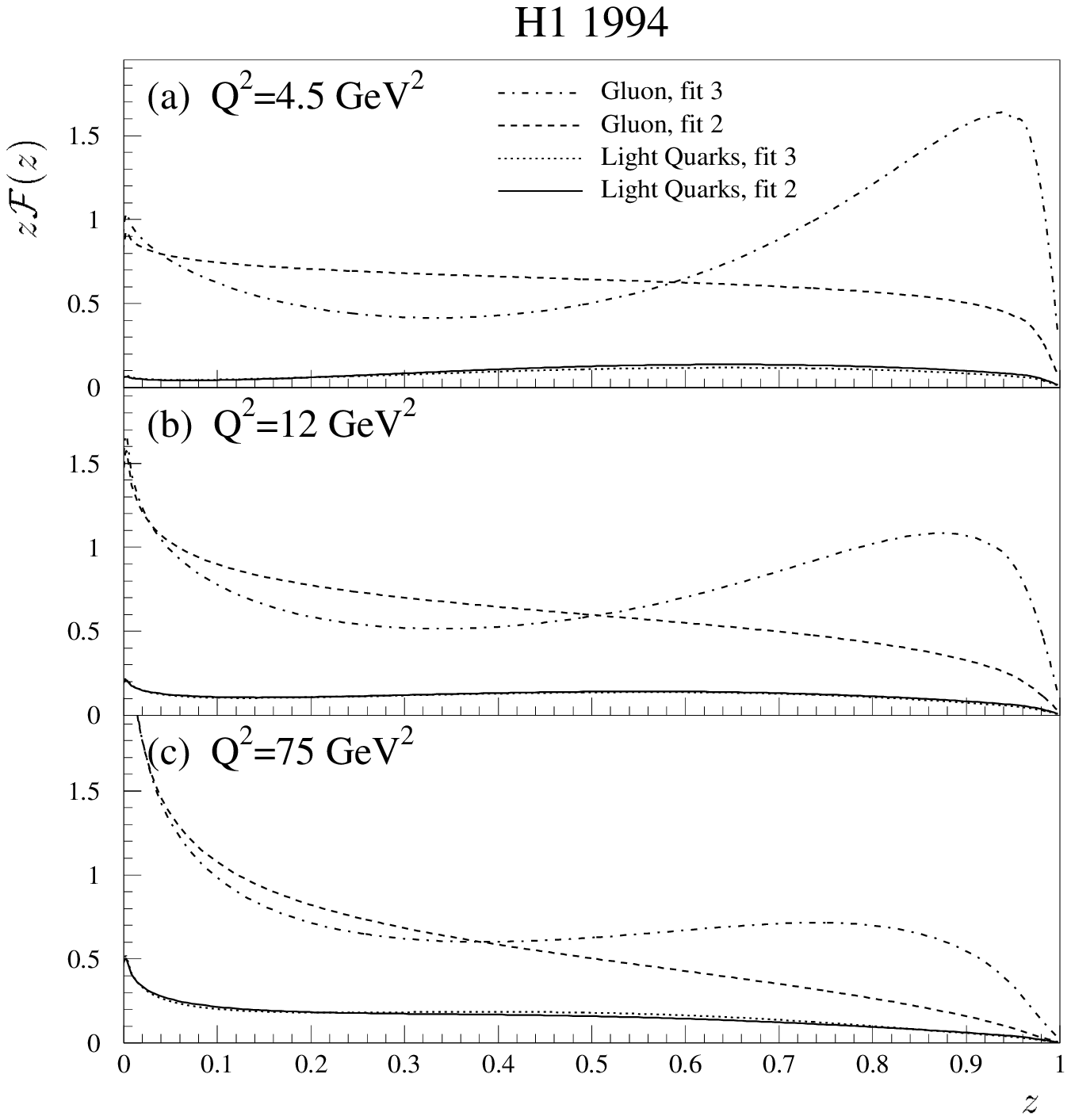}
\end{center}
\caption {\it 
{ Quark and gluon momentum distributions in the Pomeron as a function
of the fraction $z$ of the Pomeron momentum, for different fits to the
measurements of the diffractive structure function in DIS. The results
are evolved to different $Q^2$ values as denoted in the figure.
 }}
\label{fig:pompartons}
\end{minipage}
\hspace{3mm}
\begin{minipage}{8cm}
\begin{center}
  \includegraphics [width=7.5cm,totalheight=8cm]
  {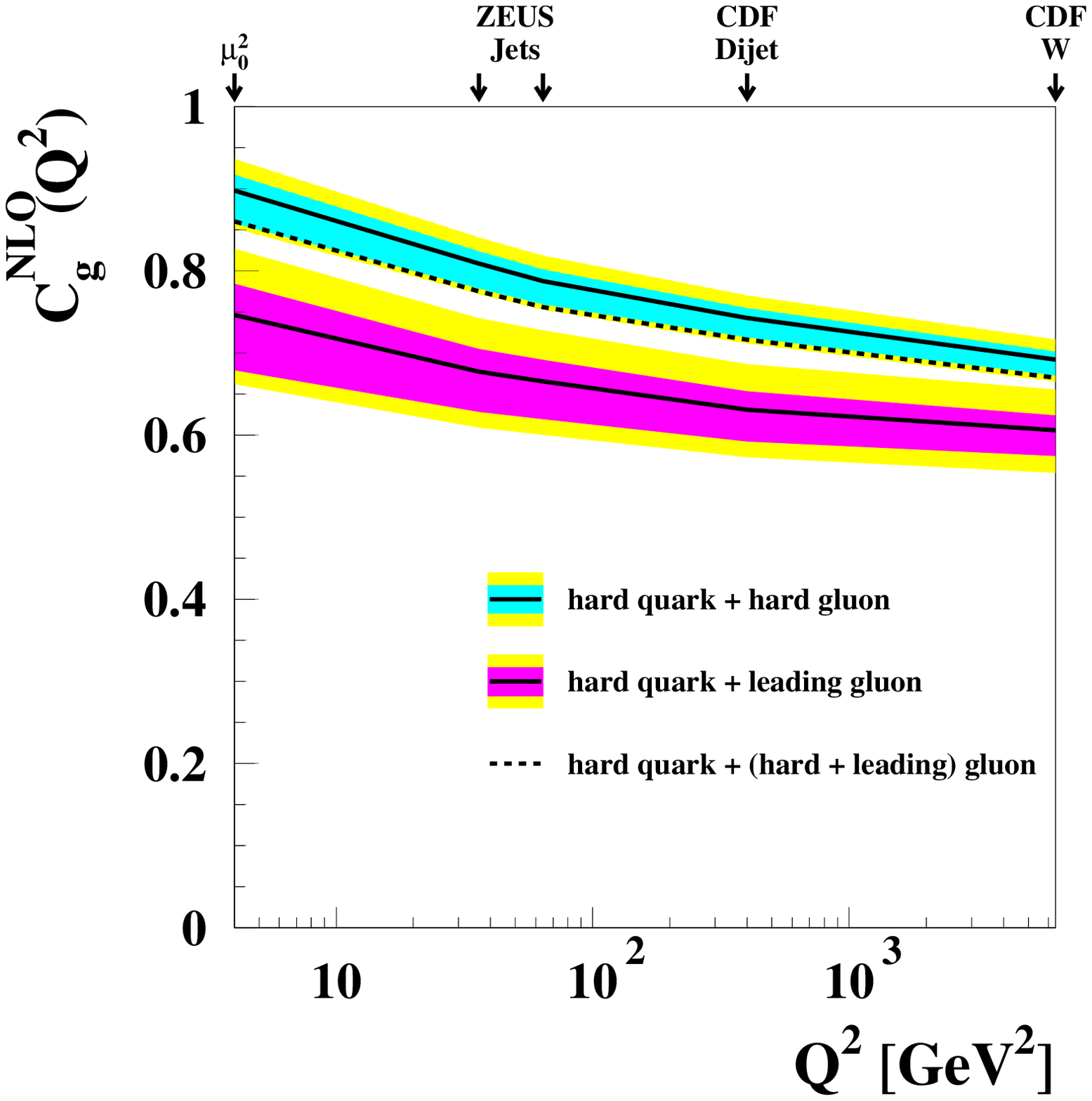}
\end{center}
\vspace{-.8cm}
\caption {\it
{ Fraction $c_g$ of the Pomeron momentum carried by gluons as a
function of $Q^2$, determined from a NLO QCD fit to the diffractive
structure function in DIS and the diffractive dijet cross section in
photoproduction.
 }}
\label{fig:cg}
\end{minipage}
\end{figure}

From the results of such a factorization analysis one gets two main
results. From the flux factor one gets information about the Pomeron
trajectory. From the Pomeron structure function QCD analysis one gets
a determination of the parton distributions in the Pomeron. Let us
start with the latter; in figure~\ref{fig:pompartons} the resulting
parton distributions at different $Q^2$ values are plotted as function
of the fraction $z$ of the Pomeron momentum carried by the struck
parton.
The plot shows results of two different fits. However without going
into details, one can clearly see that the gluons carry a dominant
part of the Pomeron momentum.
This can also be seen from figure~\ref{fig:cg} where the fraction
$c_g$ of the Pomeron momentum carried by the gluons, as obtained from
a NLO QCD fit, is plotted as a function of $Q^2$. In spite of the
slight decrease with $Q^2$, the gluons seem to carry about 60-80\% of
the Pomeron momentum.

\begin{figure}[hbt]
\begin{center}
  \includegraphics [width=\hsize,totalheight=4.4cm]
  {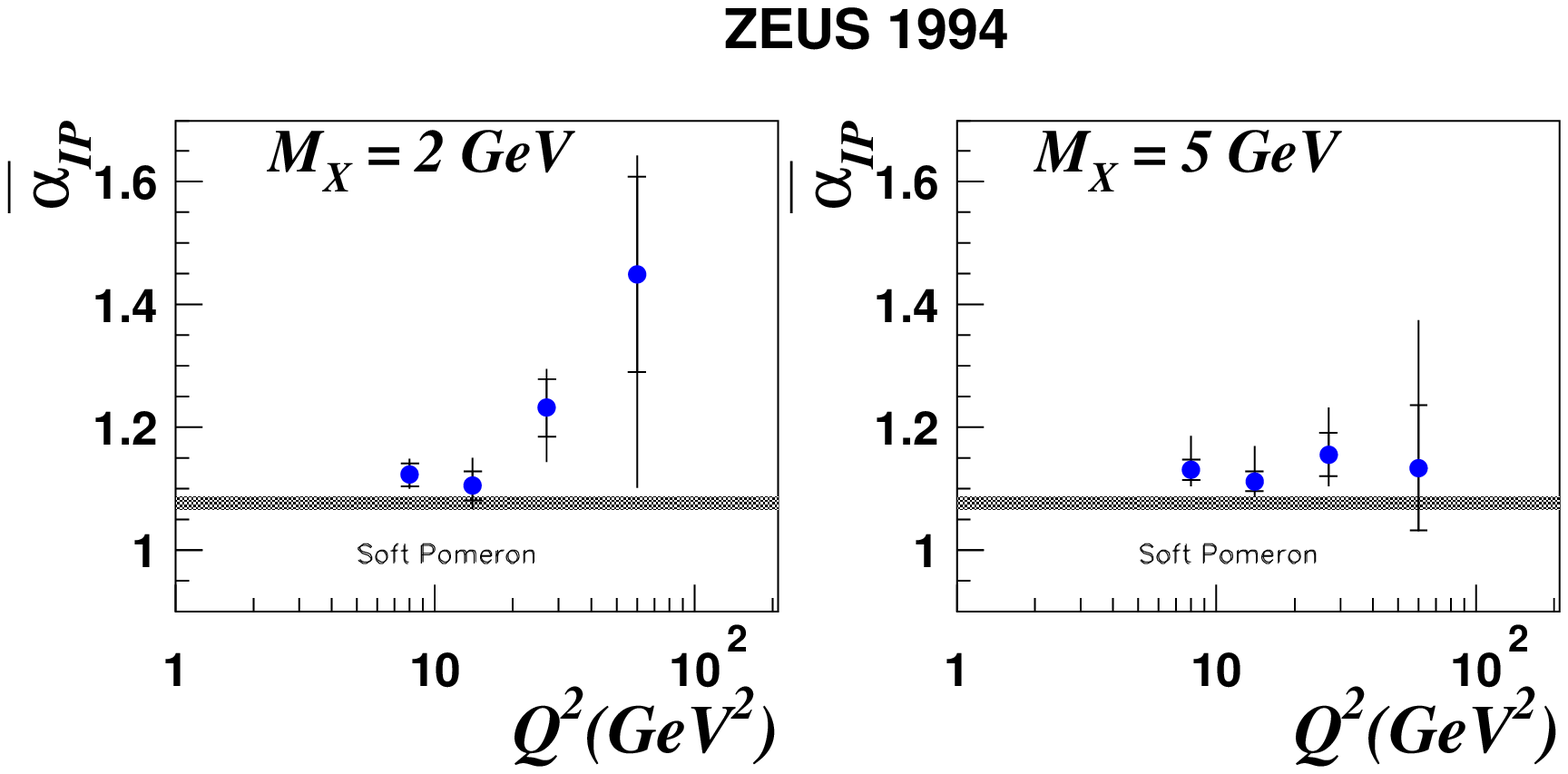}
\end{center}
\vspace{-.9cm}
\caption {\it
{ The value of the $t$ averaged Pomeron trajectory,
$\overline{\alpha_{\pom}}$, as a function of $Q^2$ and for two mass
ranges $M_X$ as derived from the fit to the $W$ dependence of the
cross section. The shaded area is the result expected for the soft
Pomeron exchange assuming $\alpha^\prime_{\pom}$ = 0.25 GeV$^{-2}$. }}
\label{fig:alpha-pom-dis}
\end{figure}
As for the Pomeron trajectory, the H1~\cite{h1-incl-diff} analysis
yields the following result,
$\alpha_{\pom}(0) = 1.203 \pm 0.020({\rm stat}) \pm 0.013({\rm syst}) 
^{+0.030}_{-0.035}({\rm model})$,
a value which seems to be $Q^2$ independent in the range $0.4 < Q^2 <$
75 GeV$^2$.  The ZEUS~\cite{zeus-incl-diff} analysis of inclusive
diffraction processes is limited to diffractive masses $M_X <$ 15
GeV. From the study of the energy dependence of the diffractive cross
section, a $t$-averaged Pomeron trajectory $\overline{\alpha_{\pom}}$
is obtained.  The results are displayed in
figure~\ref{fig:alpha-pom-dis} as a function of $Q^2$ for two regions
of $M_X$. Though there seems to be some signs for a $Q^2$ dependence
in the low mass region, a higher statistics measurement at higher
$Q^2$ is needed for firm conclusions. One can however conclude that
the trajectory is higher than that expected from soft processes.

What can one conclude from these results? The inclusive DIS
diffractive cross section seems to be factorizable and one obtains a
partonic picture of the Pomeron in which gluons carry the dominant
fraction of the Pomeron momentum. The resulting Pomeron trajectory
seems to be different than that of the DL soft Pomeron. Actually if
the trajectory is $Q^2$ dependent, it is not a Regge pole. The $Q^2$
dependence is a feature of pQCD, indicating the important role of hard
physics.  What does it mean as far as the nature of the diffractive
process is concerned?  Are we seeing a hard diffractive process? Is it
all hard? Is there here too an interplay of hard and soft processes?

\subsection{Exclusive diffractive vector meson production}

We will try to look into the questions raised in the earlier
subsection by studying simpler systems. Actually we shall start with
the most inclusive process - the total cross section. We have already
seen that Regge phenomenology expects $\sigma_{tot} \sim
s^{\alpha(0)-1}$ and therefore for a soft process dominated by the DL
Pomeron one expects,
\begin{equation}
\sigma_{tot} \sim s^{0.08} \sim W^{0.16}.
\end{equation}
This behaviour is indeed found to hold for $\sigma_{tot}(\gamma p)$ as
we have shown in figure~\ref{fig:gptot} and therefore making it a
predominantly soft process.

\begin{figure}[hbt]
\begin{minipage}{8cm}
\begin{center}
  \includegraphics [width=7.5cm,totalheight=5cm]
  {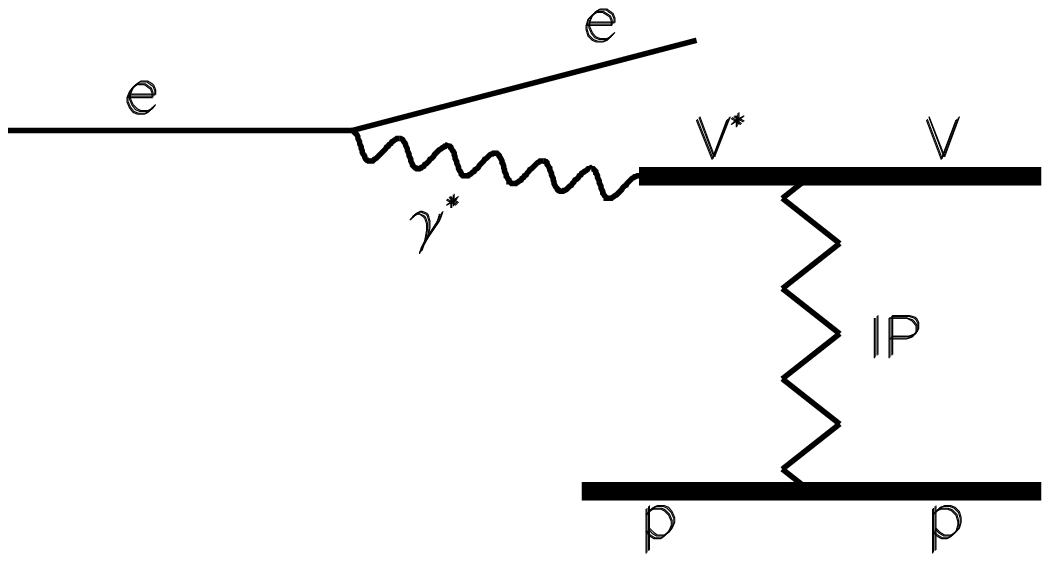}
\end{center}
\vspace{-1.5cm}
\caption {\it
{ A diagram describing diffractive production of vector mesons by an
exchange of a Pomeron. }}
\label{fig:vm-soft}
\end{minipage}
\hspace{3mm}
\begin{minipage}{8cm}
\begin{center}
  \includegraphics [width=7.5cm,totalheight=5cm]
  {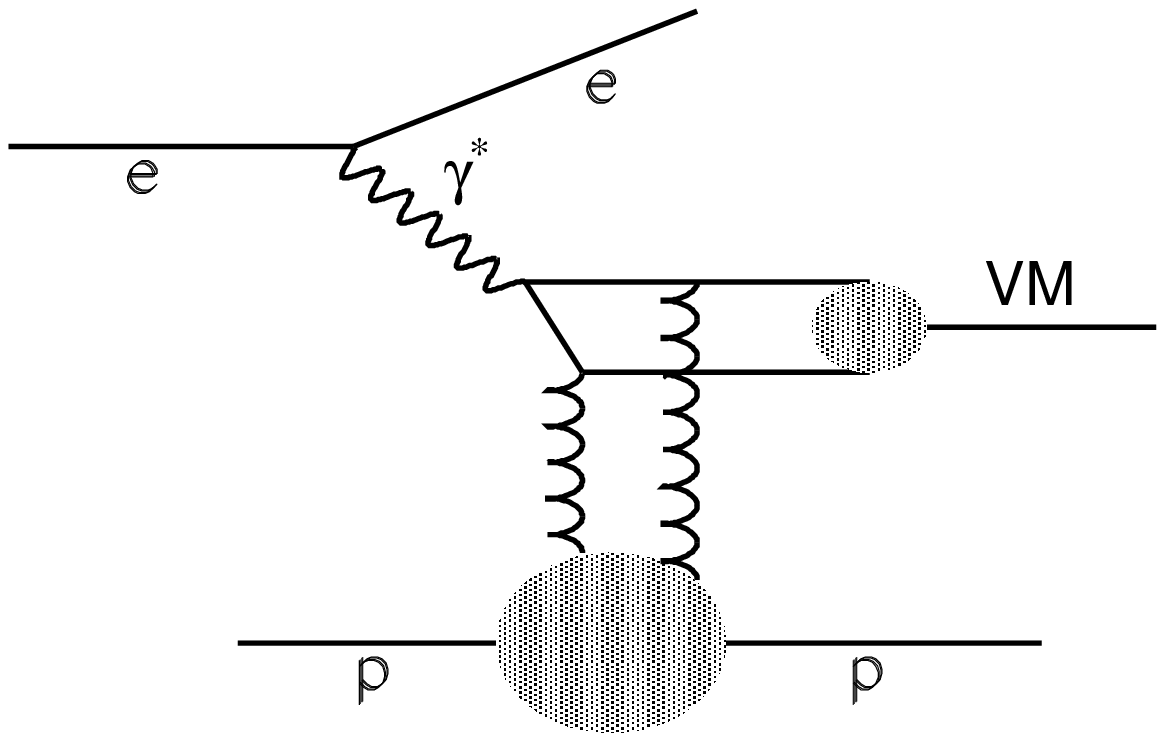}
\end{center}
\vspace{-1.5cm}
\caption {\it
{ A diagram describing diffractive production of vector mesons as two
gluon exchange. }}
\label{fig:vm-pom-diag}
\end{minipage}
\end{figure}

What does one expect in this picture for the behaviour of the
elastic~\footnote{In photoproduction the reaction is called `elastic'
  while in DIS it is named `exclusive' vector meson production. We
  shall denote both cross sections by $\sigma_{el}$.}  cross section
with energy? In photoproduction the elastic cross section refers to
vector meson production, as described by the diagram in
figure~\ref{fig:vm-soft}. The photon first fluctuates into a virtual
vector meson which scatters elastically off the proton. This is a
diffractive process dominated by a Pomeron exchange.  
In this picture, the use of the optical theorem would predict that the
elastic cross section $\sigma_{el}$ should behave as,
\begin{equation}
\sigma_{el} \sim W^{4\alpha(t)-4} \sim W^{0.22}.
\end{equation} 
The power of $W$ is less than 0.32 because of the slope of the Pomeron
trajectory, taken here as $\alpha_{\pom}^\prime$ = 0.25 GeV$^{-2}$.

How would one describe the production of vector mesons in a hard
process?  The virtual photon fluctuates into a symmetric $q\bar{q}$
pair which exchange two gluons with the proton and turn into a vector
meson~\cite{ryskin,brodsky}.  This is described diagrammatically in
figure~\ref{fig:vm-pom-diag}. In this case the $W$ behaviour of the
cross section is dictated by the $x$ behaviour of the gluon density
and since the latter shows a steep increase as $x$ decreases, one thus
expects a steep increase of the cross section as $W$ increases. Can we
quantify this? We saw that at low $x$ the proton structure function
behaves like $F_2^p \sim x^{-\lambda}$ and since at low $x$ the rise
is driven by the gluons, this is also the behaviour of the gluon
density. Translated into $W$ dependence, we expect for a hard process,
\begin{equation}
\sigma_{el} \sim W^{4\lambda}.
\end{equation}
Since we have seen in the proton section that $\lambda$ is $Q^2$
dependent, this means that we expect the $W$ dependence of a hard
process also to be $Q^2$ dependent.

\begin{figure}[hbt]
\begin{center}
  \includegraphics [bb= 18 106 538 713,width=\hsize,totalheight=8cm]
  {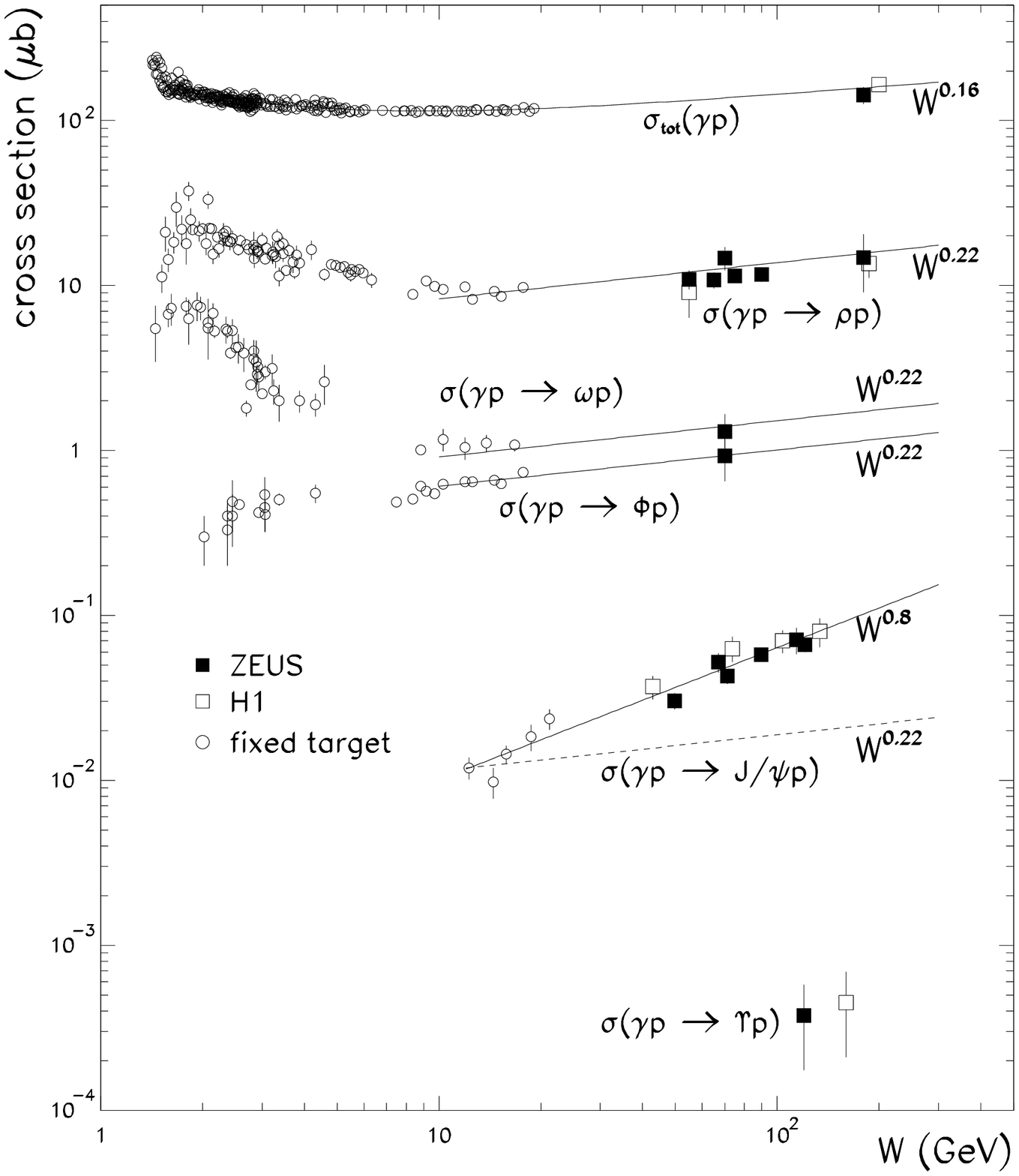}
\end{center}
\vspace{-.5cm}
\caption {\it
{ The elastic photoproduction cross section of vector mesons as
function of $W$. The total photoproduction cross section
$\sigma_{tot}$ is also shown for comparison. 
 }}
\label{fig:sigvm}
\end{figure}

What is the experimental situation? The cross section data for the
elastic vector meson photoproduction are displayed in
figure~\ref{fig:sigvm}. For comparison also the data of
$\sigma_{tot}(\gamma p)$ are shown together with the line describing
the $W^{0.16}$ behaviour. 
The light vector mesons $\rho^0$, $\omega$ and $\Phi$ have an energy
dependence which is well described by $W^{0.22}$ as expected from a
diffractive process mediated by the soft Pomeron. This behaviour
changes drastically for the $J/\psi$ vector meson, where the $W$
behaviour is much steeper, $\sim W^{0.8}$, indicative of a hard
process. What has happened? Why this change from a soft behaviour to a
hard one? In the total cross section case, the processes are dominated
by low scales. The same is true for the light vector mesons.  However
in case of the $J/\psi$, the heavy quark mass produces a large enough
scale for the reaction to become hard.  Does it mean that the process
is completely calculable in pQCD?  We will return to this question
later.

Following the above logic, we can now check if we observe a steep $W$
dependence in other exclusive processes where a hard scale is
present. In case of the light vector meson, the hard scale has to be
provided by the virtuality of the photon.
\begin{figure}[hbt]
\begin{center}
  \includegraphics [width=\hsize,totalheight=8cm]
  {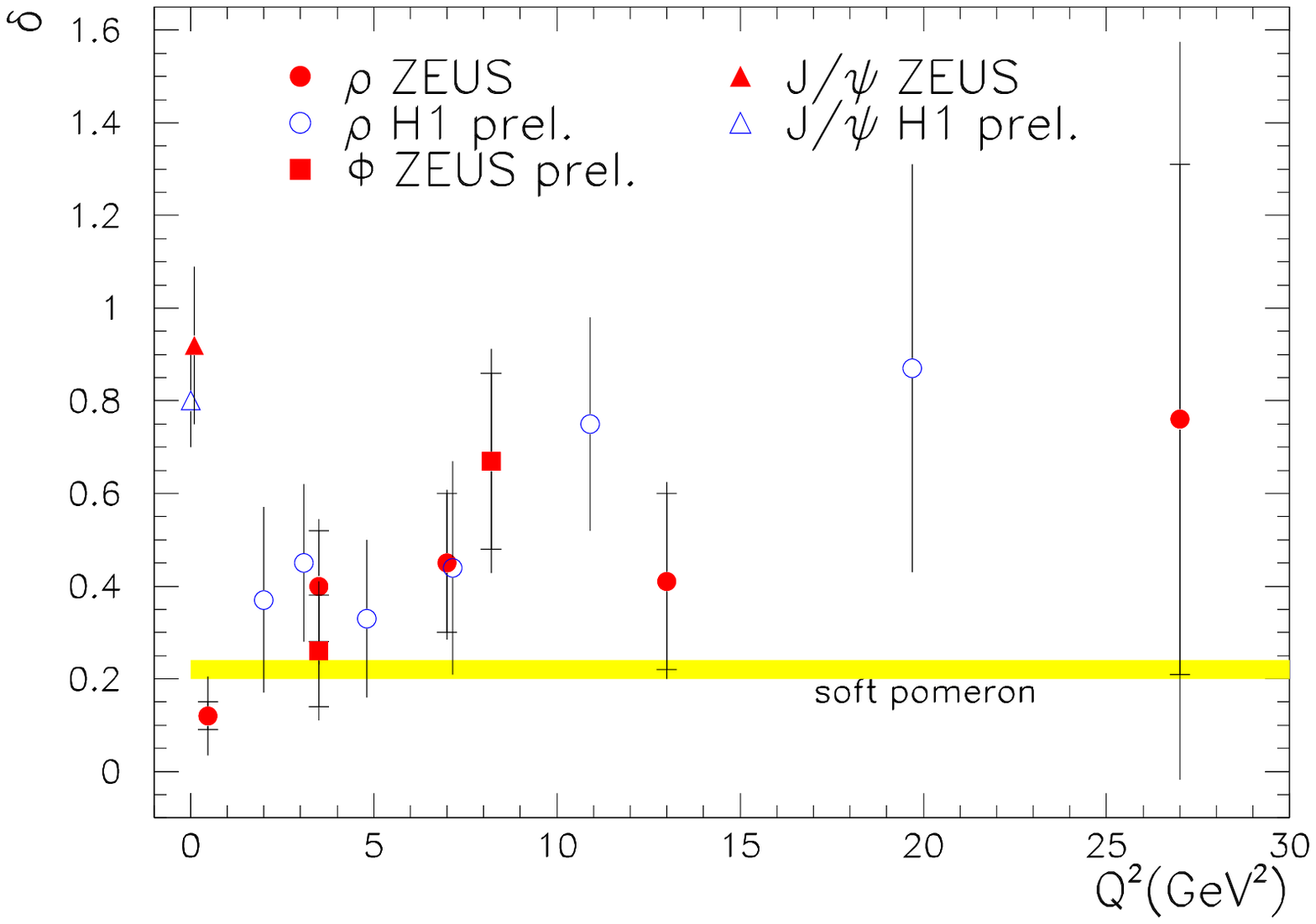}
\end{center}
\vspace{-.5cm}
\caption {\it 
{ The power $\delta$ of the energy dependence of the cross section of
diffractive vector meson production, $\sigma \sim W^\delta$, as
function of $Q^2$ for different vector mesons, as indicated in the
figure.
  }}
\label{fig:vm-wdep}
\end{figure}
The elastic (exclusive) cross section data have been fitted to the
expression $\sigma \sim W^\delta$. In order to avoid normalization
problems in comparing data from different experiments, the fits of the
photoproduction $J/\psi$ data as well as that of the DIS exclusive
vector meson data was done by using only the HERA data. The results of
the fit are displayed in figure~\ref{fig:vm-wdep} which shows the
dependence of $\delta$ on $Q^2$.  For the $\rho^0$ and the $\Phi$
vector mesons, the value of $\delta$ shows the tendency of an increase
with $Q^2$, though the errors on $\delta$ are still quite large.  One
sees at $Q^2$ = 0 the high value of $\delta$ for the case of the
$J/\psi$. At higher $Q^2$ there is not enough HERA data on $J/\psi$ to
perform a $W$-dependence fit for obtaining the value of
$\delta$. However the higher $Q^2$ $J/\psi$ data can be described with
the same value of $\delta$ as the one obtained at $Q^2$ = 0.

\subsection{Gribov diffusion}

The conclusion from the last figure seems to be that the presence of a
large scale ($\sim$ 10-20 GeV$^2$) causes a process to become hard. We
have however to define more accurately what we mean by a hard
process~\cite{yura}.  {\it A process is said to be hard if it is
driven by parton interactions and therefore calculable in pQCD.}  One
of the requirements is indeed a steeper $W$ dependence. Is this
enough? We saw that $\sigma_{tot}(\gamma^* p)$ has a steep behaviour
as $Q^2$ increases. Are all processes at higher $Q^2$ calculable in
pQCD? Can we calculate the complete process of $J/\psi$
photoproduction in pQCD?

\begin{figure}[hbt]
\begin{center}
  \includegraphics [bb= 105 247 487 600,clip,width=\hsize,totalheight=5cm]
  {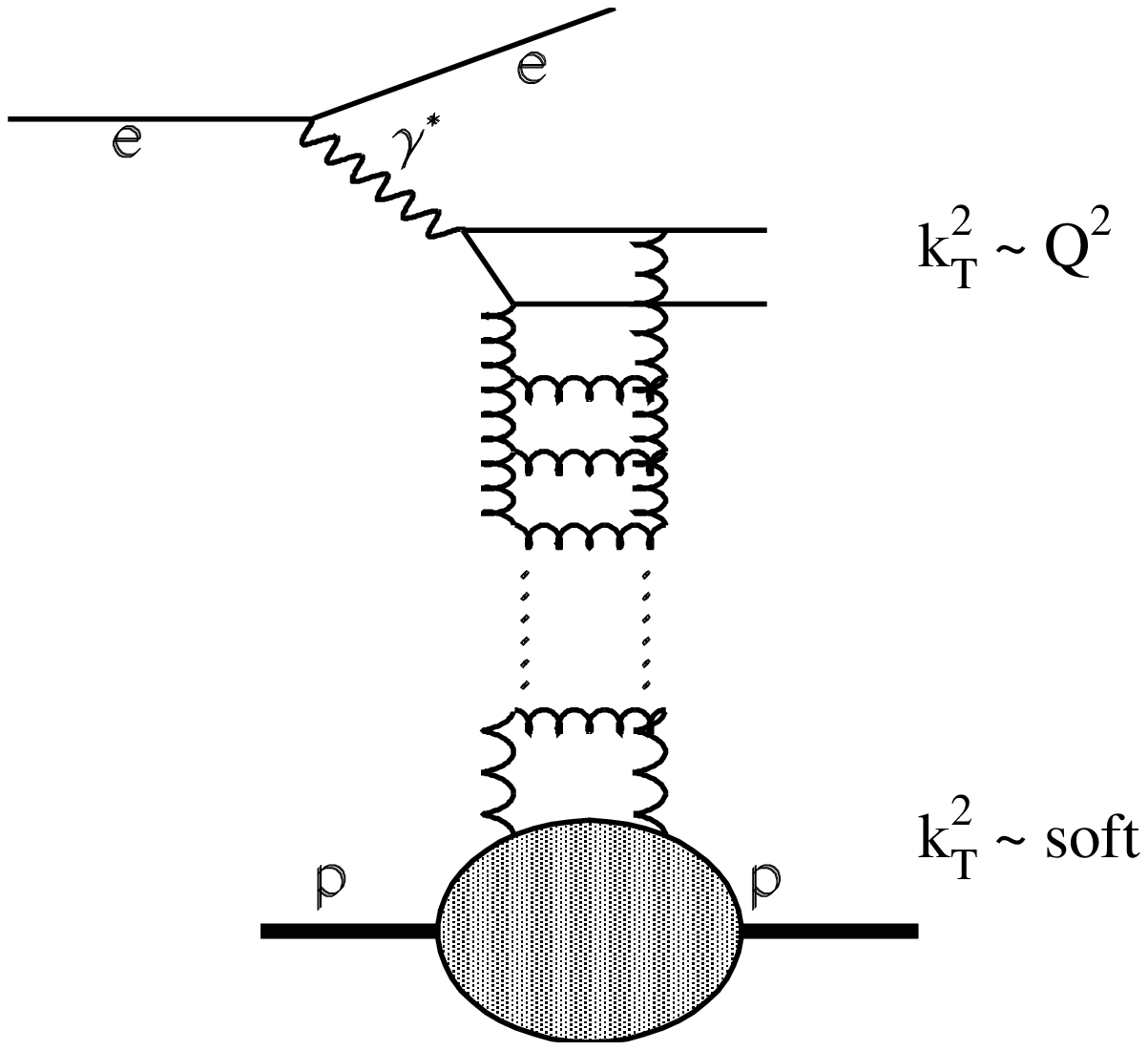}
\end{center}
\vspace{-.5cm}
\caption {\it
{ A diagram describing a gluon ladder in a diffractive process.
 }}
\label{fig:ladder}
\end{figure}
Let us look again at the diagram describing two gluon exchange in
figure~\ref{fig:vm-pom-diag}. The virtual photon fluctuates into two
high $k_T$ quarks. Although in the diagram there are only two gluons
getting down to the proton, we actually have a whole ladder due to the
large rapidity range available at these high $W$ energies (see
figure~\ref{fig:ladder}). During the trip from the virtual photon
vertex down to the proton, the average $k_T$ of the gluons gets
smaller, the configuration larger and we get into the region of low
$k_T$ physics governed by non-perturbative QCD. This process is called
Gribov diffusion~\cite{gribov-pom}(see also~\cite{forshaw-ross}). Thus
a process can start off as a hard process at the photon vertex but
once it arrives at the proton it gets a `soft' element which makes the
process non calculable in pQCD.  The average $k_T$ of the partons in
the process can be estimated by the slope of the trajectory since
$\alpha^\prime \sim 1/<k_T>$. Is Gribov diffusion always present? If
the answer is positive, we will always have a `soft' element in the
process which will prevent a full pQCD calculation. One way to look
for an answer is to determine the Pomeron trajectory in a given
process and look at the value of the slope of the trajectory. For a
hard process as defined above we would expect $\alpha^\prime \ll$ 0.25
GeV$^{-2}$.

\subsection{Determination of the Pomeron trajectory.}

Regge phenomenology~\cite{collins-regge} connects the differential
cross section of a two-body process with the leading exchanged
trajectory as follows,
\begin{equation}
\frac{d\sigma}{dt} = f(t)(W^2)^{[2\alpha(t) -2]},
\label{eq:traj}
\end{equation}
where $f(t)$ is a function of $t$ only. Thus by studying the $W$
dependence of $d\sigma/dt$ at fixed $t$ values, one can determine
$\alpha(t)$. If in addition the trajectory is assumed to be linear,
\begin{equation}
\alpha(t) = \alpha(0) + \alpha^\prime t,
\end{equation}
the intercept and slope of a trajectory can be obtained by fitting the
measured $\alpha(t)$ values to a linear form.

In order to use this method for the determination of the Pomeron
trajectory, one needs to find processes where the Pomeron is the
dominating exchanged trajectory. Furthermore, for a good determination
of the values of $\alpha(t)$ one needs data in a large range in
$W$ at a given $t$. The new HERA data allows such an analysis for the
reactions $\gamma p \to \rho^0 p$, $\gamma p \to \Phi p$, and $\gamma p
\to J/\psi p$. The elastic photoproduction of $\rho^0$ is dominated by
Pomeron exchange for $W >$ 8 GeV. For the $\Phi$ and $J/\psi$ elastic
photoproduction reaction, the Pomeron is the only possible trajectory
which one can exchange and thus one can use also the low $W$ data (of
course after moving far enough above threshold effects).

\begin{figure}[hbt]
\vspace{-0.3cm}
\begin{center}
  \includegraphics [bb= 24 103 532 744,width=5.5cm,totalheight=9cm]
  {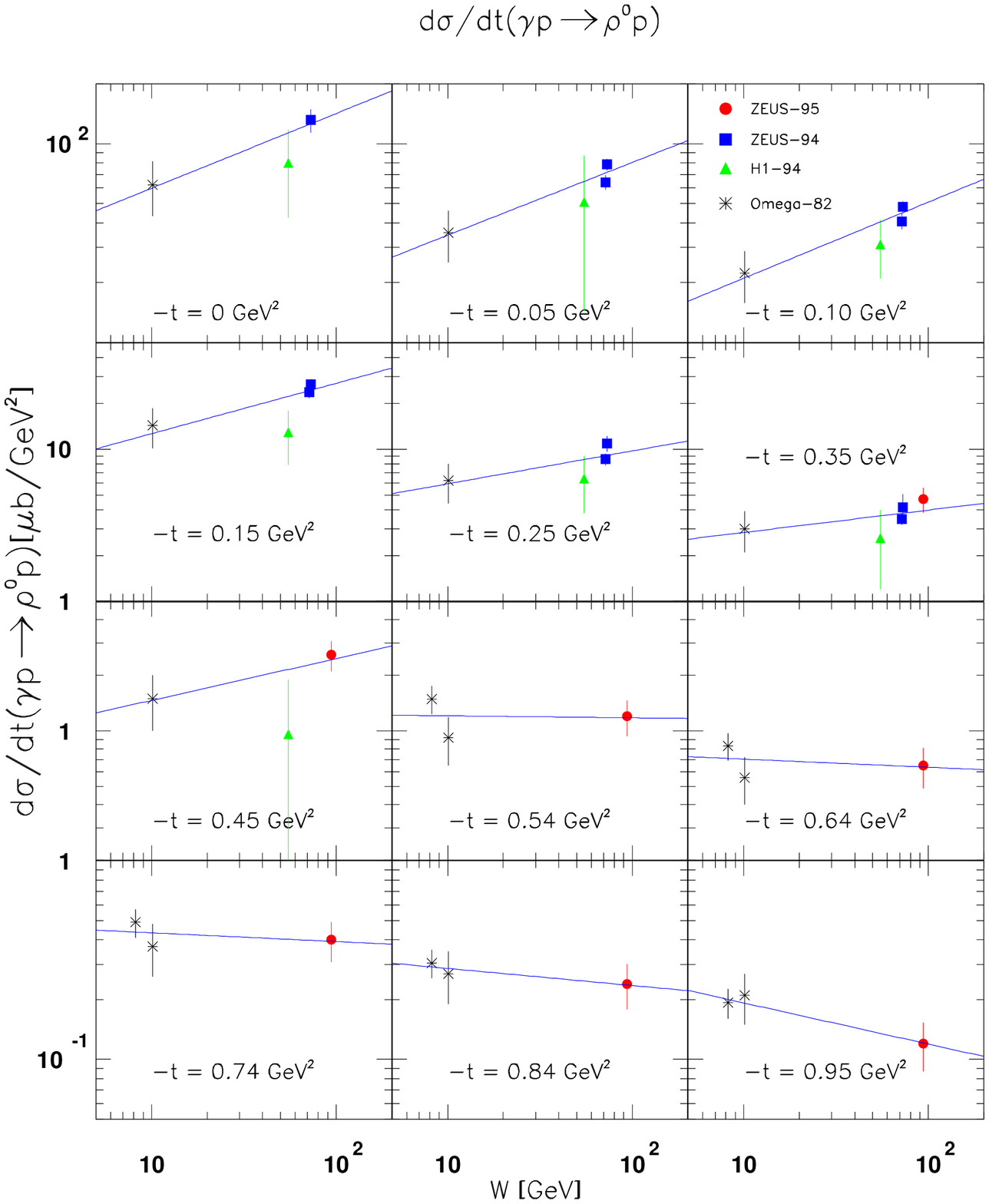}
  \includegraphics [bb= 24 103 532 744,width=5.5cm,totalheight=9cm]
  {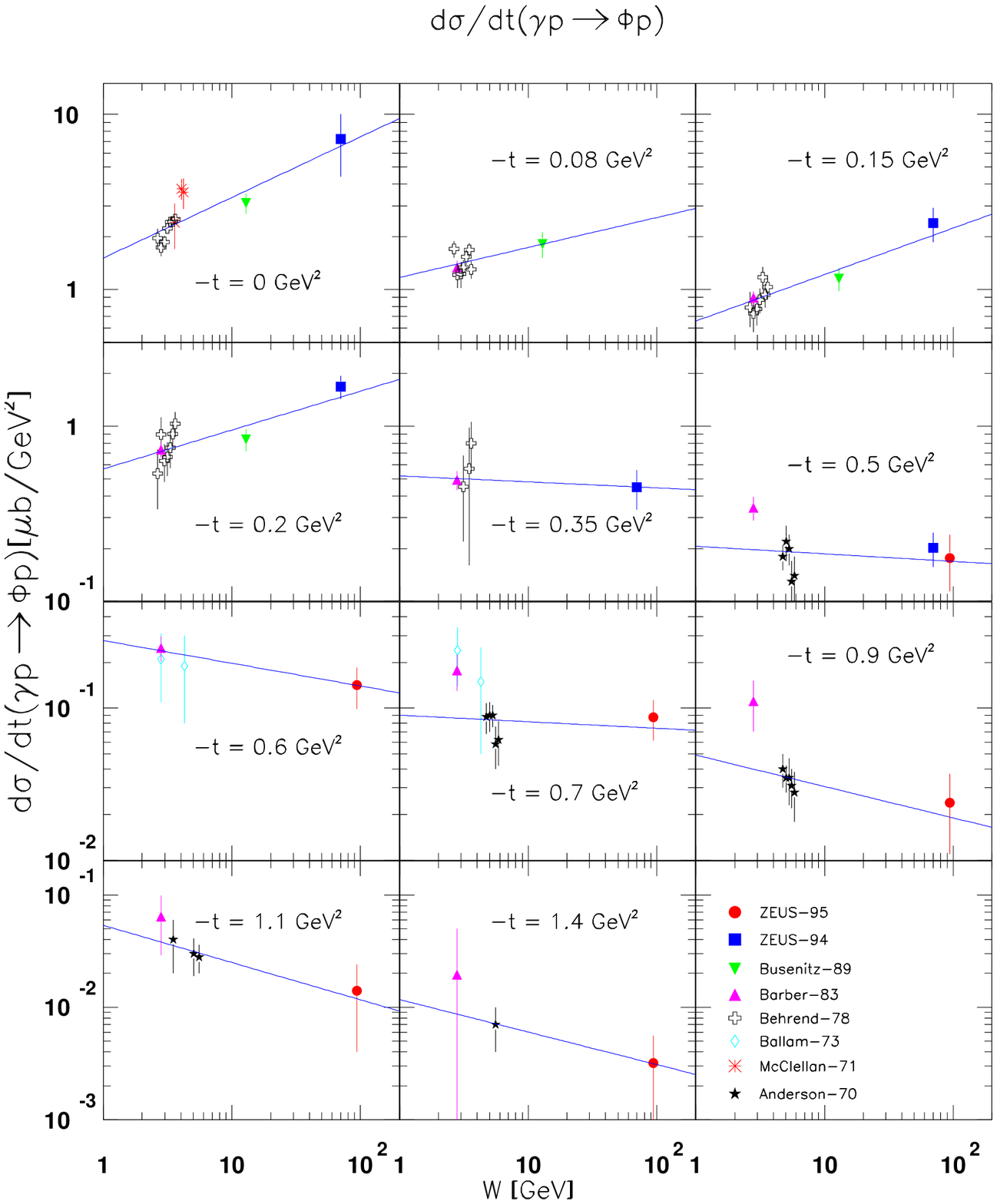}
  \includegraphics [bb= 24 103 532 744,width=5.5cm,totalheight=9cm]
  {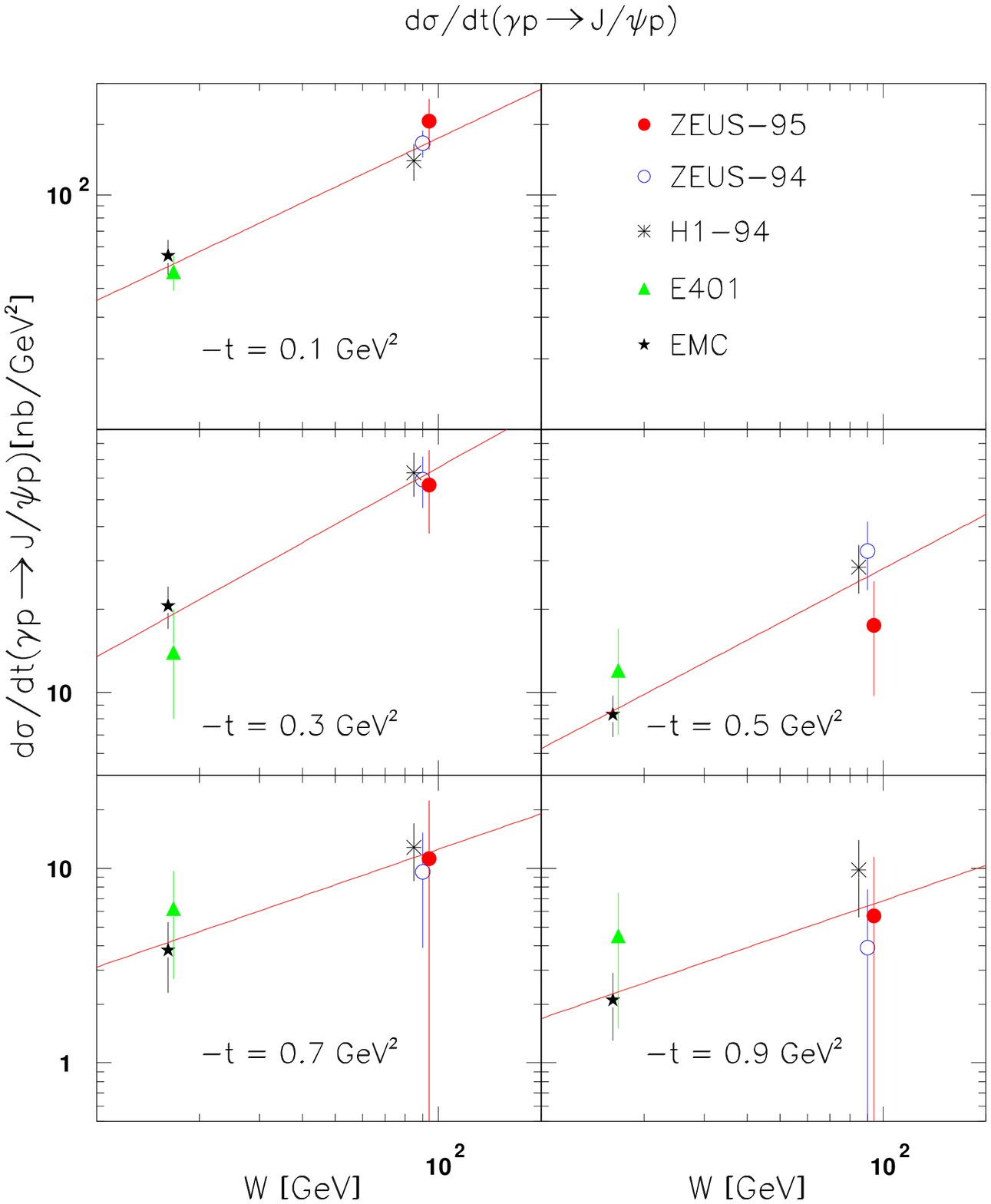}
\end{center}
\vspace{-.5cm}
\caption {\it
{
Fits to $d\sigma/dt \sim (W^2)^{2\alpha(t)-2}$ for  the reactions $\gamma p
\to \rho^0 p$ (left figure), $\gamma p \to \phi p$ (center figure) and
$\gamma p \to J/\psi p$ (right figure). 
 }}
\label{fig:dsdt-vm}
\end{figure}

\begin{figure}[h]
\begin{center}
  \includegraphics [bb= 24 103 532 744,width=5.5cm,totalheight=9cm]
  {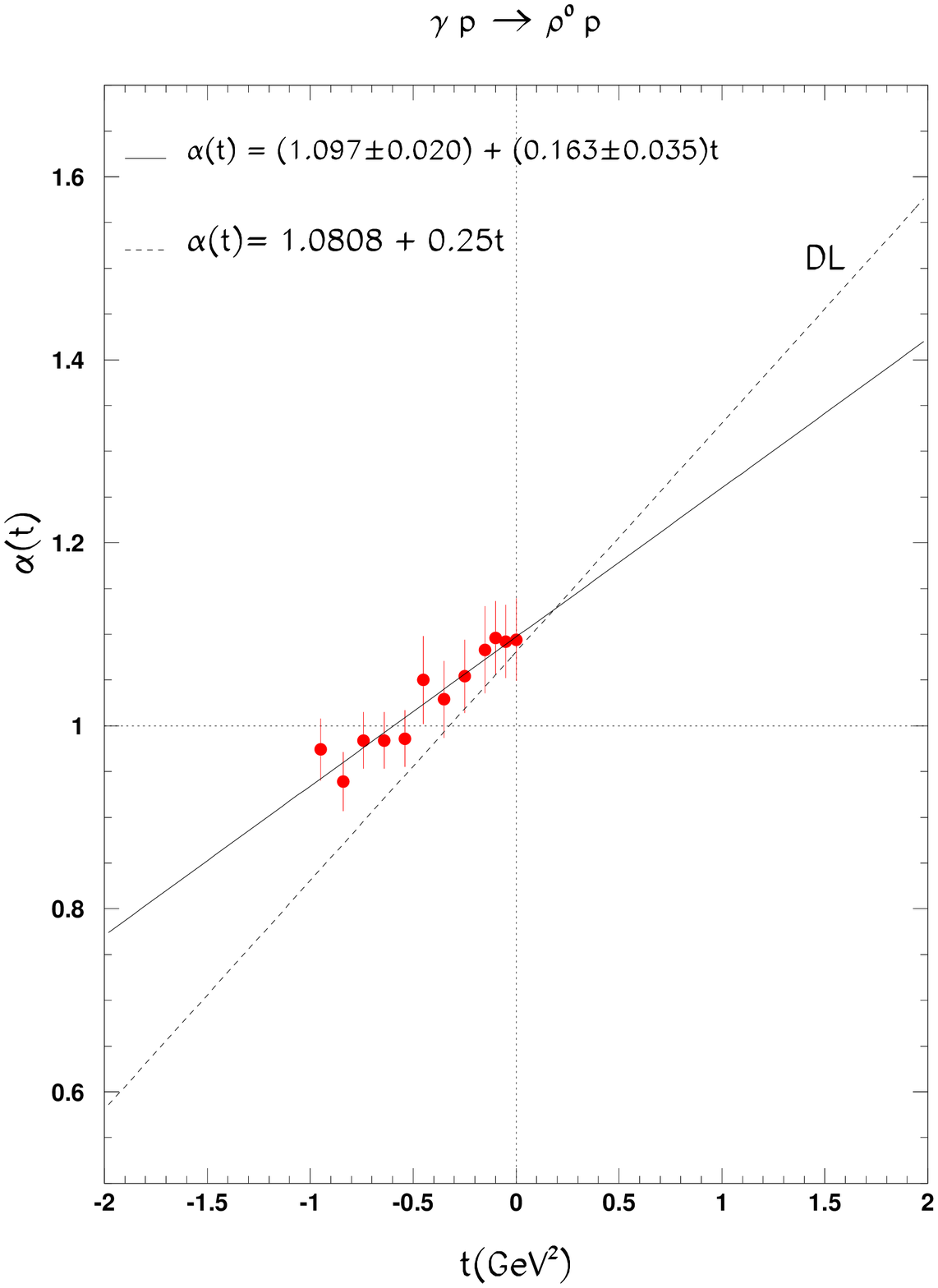}
  \includegraphics [bb= 24 103 532 744,width=5.5cm,totalheight=9cm]
  {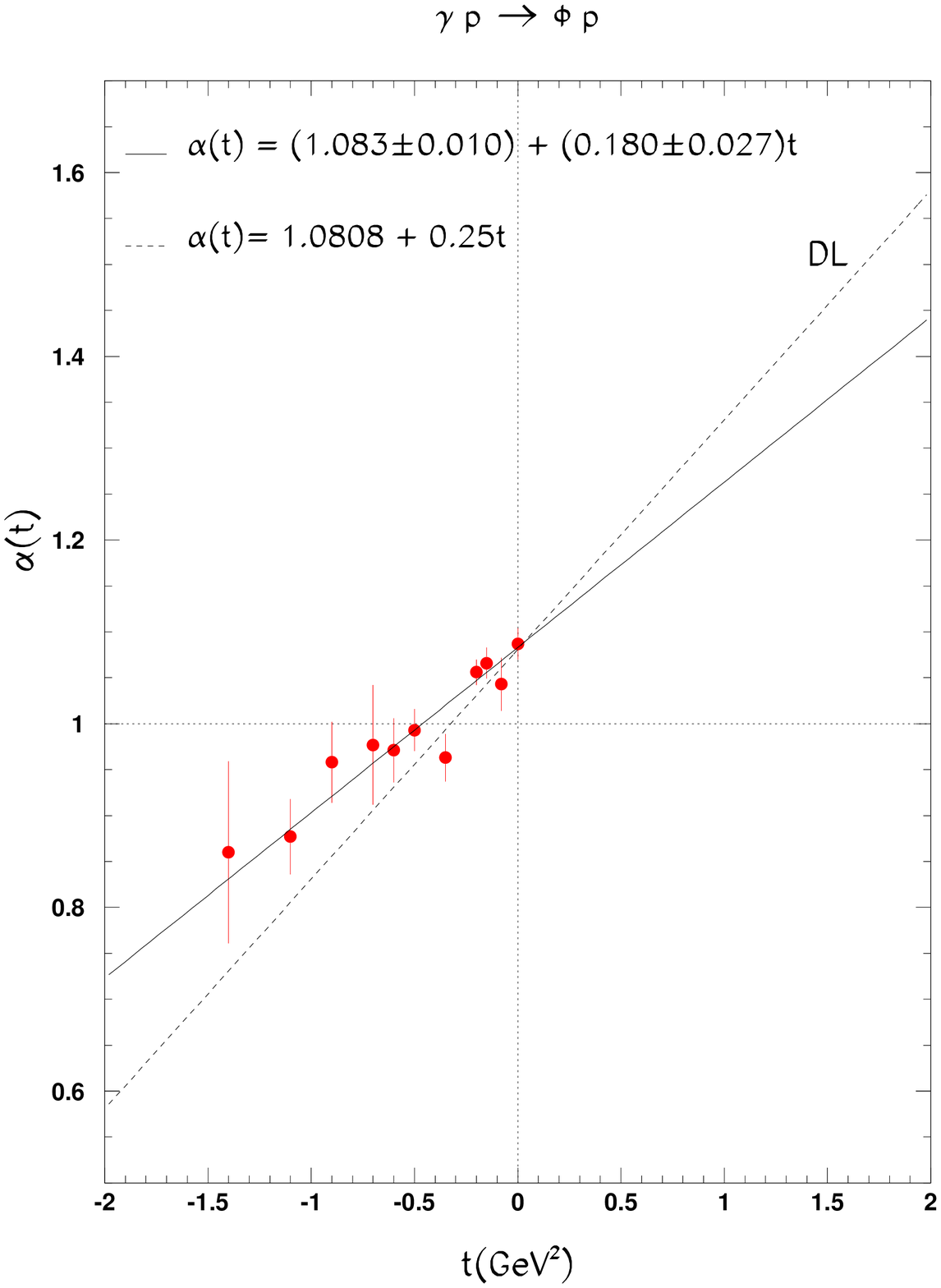}
  \includegraphics [bb= 24 103 532 744,width=5.5cm,totalheight=9cm]
  {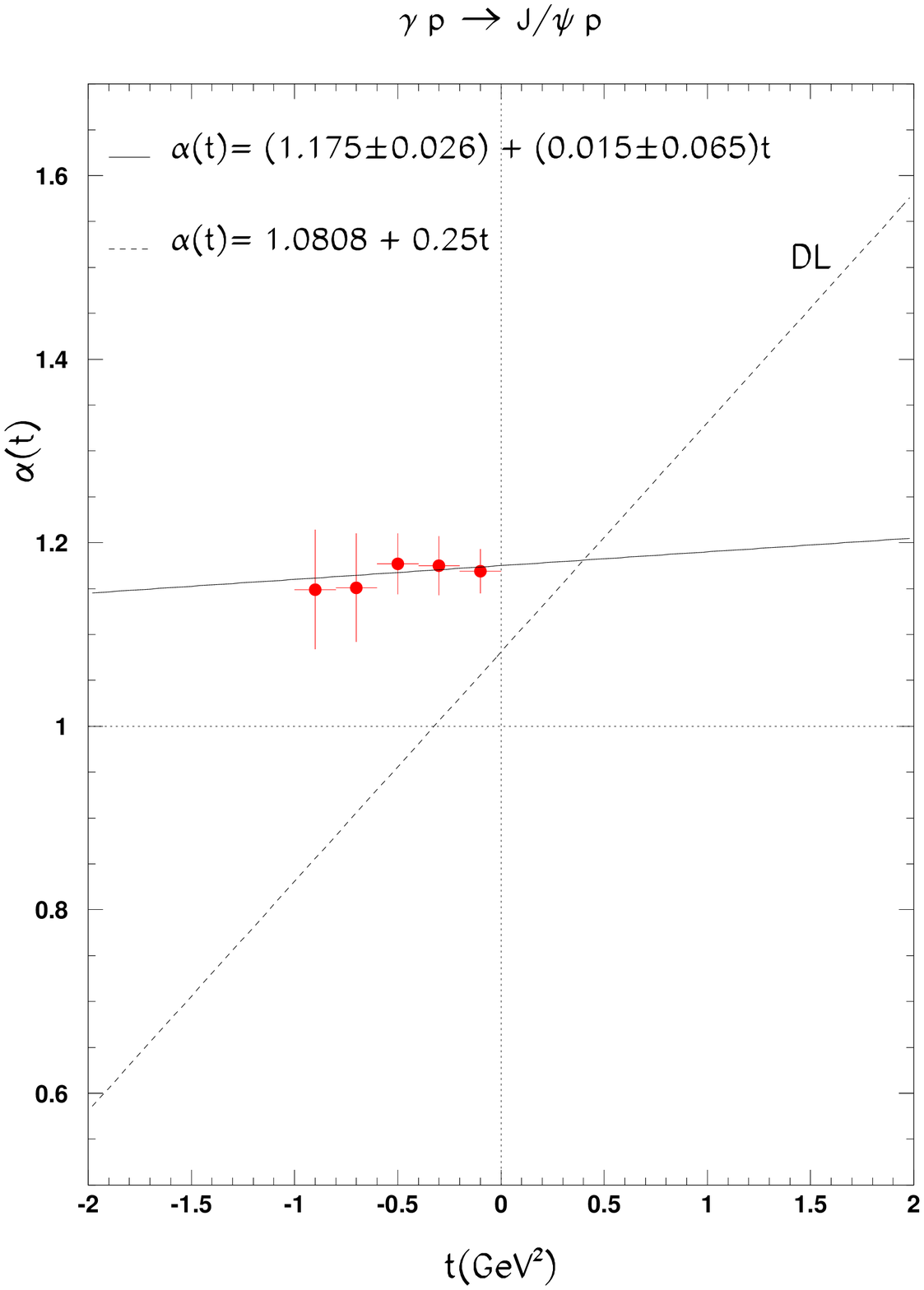}
\end{center}
\vspace{-.5cm}
\caption {\it 
{ Determination of the Pomeron trajectory from the reactions $\gamma p
\to \rho^0 p$ (left figure), $\gamma p \to \phi p$ (center figure) and
$\gamma p \to J/\psi p$ (right figure). In each of the three figures,
the dots are the values of the trajectory as determined in the earlier
figure and the full line is the result of a linear fit to these
values. The Pomeron trajectory as determined by DL is shown for
comparison as a dashed line.  
}}
\label{fig:a-vm}
\end{figure}

The determination of the Pomeron trajectory in the elastic
photoproduction of $\rho^0$, $\Phi$ and $J/\psi$ was carried out by
the ZEUS collaboration~\cite{zeus-pom-traj}.
Figure~\ref{fig:dsdt-vm} shows the differential cross section data
used in this analysis for the three vector mesons. At each $t$ value a
fit to expression~(\ref{eq:traj}) was performed and a value for
$\alpha(t)$ was obtained. 
These values are shown in figure~\ref{fig:a-vm}.
 The resulting
trajectories are,
\begin{itemize}
\item
$\gamma p \to \rho^0 p$: \ \ \ \ \ $\alpha(t) = (1.097 \pm 0.020) +
(0.163 \pm 0.035) t$,
\item
$\gamma p \to \Phi p$: \ \ \ \ \ $\alpha(t) = (1.083 \pm 0.010) +
(0.180 \pm 0.027) t$,
\item
$\gamma p \to J/\psi p$: \ \  $\alpha(t) = (1.175 \pm 0.026) +
(0.015 \pm 0.065) t$.
\end{itemize}
The following observations can be made: in case of the light vector
mesons $\rho^0$ and $\Phi$ the intercepts are in good agreement with
the DL value of 1.08. The slopes, however, are significantly different
from the value of 0.25 GeV$^{-2}$ but still far enough from 0 in order
to count as `soft' processes in which Gribov diffusion is present.

In case of the $J/\psi$ the value of the slope is close to 0, as was
already shown previously~\cite{al-shrinkage}. This indicates that for
this reaction Gribov diffusion is unimportant in the present $W$
range. In other words, for the process $\gamma p \to J/\psi p$ the
average $k_T$ of the partons involved in the exchange remains large
and therefore the process is a `hard' one, governed by the so-called
`perturbative Pomeron' and fully calculable in perturbative QCD.

\subsection{Universal Pomeron?}

The results of the last subsection lead us to question the
universality of the Pomeron trajectory. If all data are interpreted in
terms of the exchange of a Pomeron trajectory, we need as many
trajectories as reactions.  In the DIS region the
$\sigma_{tot}(\gamma^* p)$ data can not be described by a $Q^2$
independent intercept. Whenever there is a large scale present, the
$W$ dependence is steeper than expected from the DL Pomeron.  The
direct determination of the Pomeron trajectory indicates that even in
processes which are shown to be of a predominantly `soft' nature, the
concept of a universal Pomeron is not borne out by the data.

We can try again to look at the large rapidity gap processes from the
point of view of the different configurations into which the photon
fluctuates. The small configuration enables to resolve partons in the
proton, while the large configuration does not. The exchanged chain of
partons, responsible for the strong interactions between the partons
at the photon and those at the proton vertex, turn out to be
predominantly gluons in both cases. However because of the different
initial configuration of the photon fluctuation, there is no reason
why their behaviour should be identical in both cases. There are many
theoretical models, inspired by this picture, which attempt to
describe the diffractive processes and the reader is referred to a
summary in~\cite{mcdermot-gena}.

\subsection{What have we learned about the Pomeron?}

It is a bit difficult to summarize our knowledge about the Pomeron as
presented in this section. If indeed it is not a universal concept
then it is not clear what is the meaning of the results described
above. Let us nevertheless list the main points:
\begin{itemize}
\item There are processes in DIS reactions leading to large rapidity
  gap events. These events are interpreted as diffractive reactions in
  which a Pomeron is being exchanged. When viewed as a DIS on an
  Ingelman-Schlein type of Pomeron, most of its momentum is carried by
  gluons. The intercept of such a Pomeron is different from the DL
  Pomeron. 
\item The properties of the Pomeron exchanged in elastic
  photoproduction of $J/\psi$ is also different from the DL
  Pomeron. Its intercept is larger, while it slope is much smaller
  than that of the DL Pomeron.
\item The concept of a universal Pomeron is not borne out even by data
  coming from `soft' processes.
\end{itemize}


\section{Some simple-minded questions}

We discussed the structure of the proton, the photon and the
Pomeron. Let us start from the end. We know very little about the
Pomeron. We know some of its quantum numbers: it has charge
conjugation $C$ = +, it has positive parity $P$ = +, it has zero
isospin $I$ = 0. What is its mass?  We do not know. Is it a particle?
We do not know. Are there particles lying on the Pomeron trajectory?
There are some suggestions that there might exist a glueball candidate
which could be on the Pomeron trajectory. Can we however talk about a
universal trajectory given the different slopes and intercepts one
seems to measure?

We know much more about the photon. It is an elementary gauge vector
boson with definite spin, parity, charge conjugation $J^{PC} =
1^{--}$.  We are talking about the structure of the photon, though we
understand that this structure is built up during the interaction. We
thus talk about the photon structure function, parton distributions in
the photon and understand that when it interacts with another hadron
it has a probability to either interact with it directly (direct
photon) or first turn into a state where it is composed of partons
which then interact with the hadron (resolved photon). Although this
picture is borne out by experimental data we still have some
difficulty with the concept of the structure of the photon. The photon
is not a particle that we can stop and look at its inside.

This brings us to the proton. Here the concept of the structure of the
proton comes very naturally. The DIS experiments show clearly that
the proton is composed of partons. How many partons are there in the
proton? We don't know, but we know that the higher we go in energy the
density of these partons increases. Where in the proton are these
partons located? Are they spread over the proton? Are they
concentrated in a small part of the proton? What kind of experiment
should we do to get some answer to this? We always say that DIS is
just the continuation of Rutherford's experiment. Well, he found that
the nucleus is concentrated in a very small part of the Atom. What can
we say about the partons inside the proton? Lonya Frankfurt's
answer~\cite{lonya} was that we have first to go to the proton rest
frame. The reason is that the infinite momentum frame picture does not
give direct information on the space location of the partons. This is
because the light-cone description of the Feynman parton model does
not explore the space-time location of partons at all. Within the
infinite momentum frame description the variable $x$ has no direct
relation to the space coordinate of a parton but is related to a
combination of the energy and momentum of a parton.

On the contrary, the proton rest frame picture contains rather direct
information about the location of partons in space-time. The key
formula which relates both descriptions is that derived by
Ioffe~\cite{ioffe},
\begin{equation} 
l = \frac{1}{2 m_p x} \approx \frac{0.1 \ {\rm fm}}{x},
\label{eq:length}
\end{equation}
giving the relation between Bjorken-$x$ and the distance $l$ in the
direction of the exchanged photon. It follows from equation
(\ref{eq:length}) that partons with $x >$ 0.1 are in the interior of
the proton. All partons with $x <$ 0.1 have no direct relation to the
structure of the proton. They do not belong to the proton ! In this
picture all the sea quarks found at small $x$ are to a large extent
the property of the photon wave function. Thus the popular comparison
which I have used here with the Rutherford experiment is wrong for the
HERA kinematics, though it was right to do so for the early SLAC DIS
experiment which obtained data for $x >$ 0.1. Does this mean that in
order to learn from HERA about the structure of the proton we should
concentrate on the high $x$ physics? 

The above argumentation actually suggests that the study of the low
$x$ region will improve our knowledge about the details of the
interaction. In order to find new features about the structure of the
proton we should look into the high $x$ region with better
resolution. The HERA high luminosity upgrade program will provide this
opportunity.

Following the confusing questions above I find that the best way to
finish this talk is with a quotation which I found in the book of
Yndurain~\cite{yndurain}. Alphonse X (The Wise, 1221--1284), who was
King of Castillo and Leon, had the Ptolemaic system of epicycles
explained to him. His reaction was the following:

{\it `If the Lord Almighty had consulted me before embarking upon
creation, I should have recommended something simpler.'}

\section{Acknowledgment}

It is a pleasure to acknowledge fruitful discussions with
Halina Abramowicz, Yuri Dokshitzer, Lonya Frankfurt and Mark Strikman. 
Many thanks to Vladimir Petrov and the local organizing committee for
inviting me to this pleasant and special workshop.

This work was partially supported by the German--Israel Foundation
(GIF), by the U.S.--Israel Binational Foundation (BSF) and by the
Israel Science Foundation (ISF).


\begin{thebibliography}{99}
\bibitem{leader-predazzi}For an excellent textbook on this subject see
E. Leader, E. Predazzi, {\it An introduction to gauge theories and
modern particle physics}, Cambridge University Press, 1996.
\bibitem{halzen-martin} F. Halzen, A.D. Martin, {\it Quarks and Leptons: 
An introductory course in modern particle physics}, Wiley \& Sons, 1984. 
\bibitem{first-f2} SLAC-MIT Collab, E.D. Bloom et al., \prl{23}{1969}{930}; \\
M. Breidenbach et al., \prl{23}{1969}{935}.
\bibitem{scaling-violation} CDHS Collab., J.G.H. De Groot et al., 
\zp{C1}{1979}{143}.
\bibitem{hera} B.H. Wiik, {\it Electron-proton colliding beams, the
    physics program and the machine}, Proc. 10$^{th}$ SLAC Summer
  Institute, p. 233, 1982;\\
G.A. Voss, Proc. First Euro Acc. Conf., Rome, p 7, 1988.
\bibitem{slac} L.W. Whitlow et al., \pl{B282}{1992}{475}.
\bibitem{bcdms} BCDMS Collab., A.C. Benvenuti et al., \pl{B223}{1989}{485}.
\bibitem{e665} E665 Collab., M.R. Adams et al., \prev{D54}{1996}{3006}.
\bibitem{nmc} NMC Collab., M. Arneodo et al., \np{B483}{1997}{3}. 
\bibitem{h1f2} H1 Collab., I. Abt et al., \np{B407}{1993}{515};\\
  T. Ahmed et al., \np{B439}{1995}{471};\\ T. Ahmed et al.,
  \np{B470}{1996}{3};\\ C. Adloff et al., \np{B497}{1997}{3}.
\bibitem{zeusf2} ZEUS Collab., M. Derrick et al.,
  \pl{B316}{1993}{412};\\ M. Derrick et al., \zp{C65}{1995}{379};\\
  M. Derrick et al., \zp{C69}{1995}{607};\\ M. Derrick et al.,
  \zp{C72}{1996}{399};\\ J. Breitweg et al., \pl{B407}{1997}{432};\\
  J. Breitweg et al., {\it ZEUS results on measurement and
    phenomenology of $F_2$ at low $x$ and low $Q^2$}, DESY 98-121, 1998.
\bibitem{hand} L.N. Hand, \prev{129}{1963}{1834}.
\bibitem{vdm} J.J. Sakurai, \ap{11}{1960}{1}.
\bibitem{ioffe} B.L. Ioffe, \pl{B30}{1969}{123}.\\ See also V.N. Gribov, 
B.L. Ioffe, I.Ya. Pomeranchuk, \sjnp{2}{1966}{549}; \\
B.L. Ioffe, \jetp{9}{1969}{97}; \ \ \jetp{10}{1969}{90}; \\
V.N. Gribov, \spj{30}{1970}{709}.
\bibitem{dl} A. Donnachie, P.V. Landshoff, \pl{B296}{1992}{227}.
\bibitem{Pom} I.Ya. Pomeranchuk, \spj{7}{1958}{499}.
\bibitem{h1-gptot} H1 Collab., T. Ahmed et al., \pl{B299}{374}.
\bibitem{zeus-gptot} ZEUS Collab., M. Derrick et al., \pl{B293}{465}.
\bibitem{allm} H. Abramowicz et al., \pl{B269}{1991}{465}.
\bibitem{zeus-lrg} ZEUS Collab., M. Derrick et al., \pl{B315}{1993}{481}.
\bibitem{h1-lrg} H1 Collab., T. Ahmed et al., \np{B429}{1994}{477}.
\bibitem{afs} H. Abramowicz, L. Frankfurt, M. Strikman, \shep{11}{1997}{51}. 
\bibitem{gribov-pom} V.N. Gribov, \jetp{41}{1961}{667}. (Reprinted in 
{\it Regge theory of low-$p_T$ hadronic interactions}, L. Caneschi
(ed.) pp. 22-23).
\bibitem{levin} For a recent review see E. Levin, {\it An introduction to 
Pomerons}, DESY 98-120, 1998.
\bibitem{gribov-morrison} D.R.O. Morrison, \prev{165}{1968}{1699}.
\bibitem{rmp} For a recent review on HERA physics, see H. Abramowicz,
  A. Caldwell, {\it HERA Physics}, to be published in {\em Review of
    Modern Physics}.
\bibitem{zeus-Wmass} ZEUS Collaboration, J. Breitweg et al., {\it
    Measurement of high-$Q^2$ charged-current DIS cross sections at HERA},
paper 751 submitted to the XXIX International Conference on
High Energy Physics, Vancouver, 23-29 July, 1998.
\bibitem{pdg} Particle Data Group, R.M. Barnett et al., \epj{C3}{1998}{1}.
\bibitem{h1-highx-q2} H1 Collab., C. Adloff et al., \zp{C74}{1997}{191}.
\bibitem{zeus-highx-q2} ZEUS Collab., J. Breitweg et al., \zp{C74}{1997}{207}.
\bibitem{bruce-lp97} B. Straub, {\it New results on neutral and
    charged current scattering at high $Q^2$ from H1 and ZEUS}, XVIII
  Int. Symp. Lepton Photon Interactions, Hamburg, 1997. 
\bibitem{wolf} For a recent review see G. Wolf, \shep{12}{1998}{1}.
\bibitem{h1-vc}H1 Collaboration, {\it
    Measurement of neutral and charged current cross sections at high $Q^2$},
paper 533 submitted to the XXIX International Conference on
High Energy Physics, Vancouver, 23-29 July, 1998.
\bibitem{zeus-vc}  ZEUS Collaboration, J. Breitweg et al., {\it
    Measurement of high-$Q^2$ neutral-current DIS cross sections at HERA},
paper 752 submitted to the XXIX International Conference on
High Energy Physics, Vancouver, 23-29 July, 1998.
\bibitem{cteq4d} CTEQ Collab., H.L. Lai et al., \prev{D55}{1997}{1280}.
\bibitem{qcd-factorization} J.C. Collins, D.E. Soper, G. Sterman, \np{B261}
{1985}{104}.
\bibitem{dglap} V.N. Gribov, L.N. Lipatov, \sjnp{15}{1972}{438,675};\\
G. Altarelli, G. Parisi, \np{B126}{1977}{298};\\ Yu.L. Dokshitzer,
\spj{46}{1977}{641}.
\bibitem{bfkl} E.A. Kuraev, L.N. Lipatov, V.S. Fadin, \spj{44}{1976}{443};\\
E.A. Kuraev, L.N. Lipatov, V.S. Fadin, \spj{45}{1977}{199};\\
Y.Y. Balitski, L.N. Lipatov, \sjnp{28}{1978}{822}.
\bibitem{f2review} For a recent review on the nucleon structure
  function, see A.M. Cooper-Sarkar, R. Devenish, A. De Roeck, 
\ijmp{A13}{1998}{3385}.
\bibitem{mrst} A.D. Martin et al., 
{\it Parton distributions: a new global analysis}, hep-ph/9803445, 1998.
\bibitem{h1-lambda} H1 Collab., C. Adloff et al., \np{B497}{1997}{3}.
\bibitem{grv94} M. Gluck, E. Reya, A. Vogt, \zp{C67}{1995}{433}.
\bibitem{ckmt} A. Capella et al., \pl{B337}{1994}{358}.
\bibitem{bk} J. Kwiecinski, B. Badelek, \zp{C43}{1989}{251}; \ \ \pl{B295}
{1992}{263}.
\bibitem{aby} K. Adel, F. Bareiro, F.J. Yndurain, \np{B495}{1997}{221}.
\bibitem{al-lowq2} A. Levy, {\it Low-$x$ Physics at HERA}, in {\it
    Lectures on QCD}, F. Lenz, H. Griesshammer, D. Stoll (eds.),
pp. 347-477, Springer-Verlag 1997;\ \ DESY 97--013 (1997).
\bibitem{allm97} H. Abramowicz, A. Levy, {\it The ALLM parameterization of 
$\sigma_{tot}(\gamma^* p)$: an update}, DESY 97--251, hep-ph/9712415 (1997).
\bibitem{dlq2} A. Donnachie, P.V. Landshoff, \np{B244}{1984}{322}.
\bibitem{stefan-lp97} S. Soldner-Rembold, {\it The structure of the photon},
  XVIII Int. Symp. Lepton Photon Interactions, Hamburg, 1997.
\bibitem{f2g-para} M. Gluck, E. Reya, A. Vogt, \prev{46}{1992}{1973};\\
G.A. Schuler, T. Sj\"ostrand, \zp{C68}{1995}{607};\\
L.E. Gordon, J.K. Storrow, \zp{C56}{1992}{307}.
\bibitem{lep-gg} 
DELPHI Collab., P. Abreu et al.,
\zp{C69}{1995}{209};\\ OPAL \ Collab., K. Ackerstaff et al.,
\zp{C74}{1997}{33};\\ \ \  \ K. Ackerstaff et al., \pl{B411}{1997}{387};\\
ALEPH \ Collab., Submitted to Photon '97 and LP97;\\ L3 Collab.,
M. Acciarri et al., \pl{B436}{1998}{403}.
\bibitem{maria-photon} M. Krawczyk, A. Zemburski, M. Staszel, {\it Survey 
of recent data on photon structure functions and resolved photon processes}, 
DESY 98-013, 1998.
\bibitem{amy-gluon} AMY Collab., R. Tanaka et al., \pl{B277}{1992}{215}.
\bibitem{zeus-xg} ZEUS Collab., M. Derrick et al., \epj{C1}{1998}{109}. 
\bibitem{zeus-ang} ZEUS Collab., M. Derrick et al., \pl{B384}{1996}{401}.
\bibitem{h1-gluong} H1 Collab., C. Adloff et al., {\it Charged Particle Cross 
Sections in Photoproduction and Extraction of the Gluon Density in the
Photon}, DESY-98-148 (1998).
\bibitem{grv-lo-g} M. Gluck, E. Reya, A. Vogt, \prev{46}{1992}{1973}.
\bibitem{lac} H. Abramowicz, K. Charchula, A. Levy, \pl{B269}{1991}{458}.
\bibitem{sas1d} G.A. Schuler, T. Sj\"ostrand, \pl{B376}{1996}{193}.
\bibitem{pluto-virtg} PLUTO Collab., Ch. Berger et al., \pl{B142}{1984}{119}.
\bibitem{zeus-virtg} ZEUS Collab., J. Breitweg et al., {\it Dijet cross 
sections in $\gamma p$ interactions using real and virtual photons at
HERA}, paper 816 submitted to the XXIX International Conference on
High Energy Physics, Vancouver, 23-29 July, 1998.
\bibitem{h1-virtg} H1 Collab., {\it Measurement of di-jet cross-sections 
in low $Q^2$ deep-inelastic scattering processes at HERA and
extraction of an effective parton density of the virtual photon},
paper 544 submitted to the XXIX International Conference on High
Energy Physics, Vancouver, 23-29 July, 1998.
\bibitem{bj94} J.D. Bjorken {\it Proceedings of the International Workshop 
on DIS and related subjects}, Eilat, 1994, p. 151.
\bibitem{lonya} L. Frankfurt,  private communication.
\bibitem{gribov-fact} V.N. Gribov, L.Ya. Pomeranchuk, \prl {8} {1962} {343}.
\bibitem{al-fact} A. Levy, \pl{B404}{1997}{369}.
\bibitem{ajm} J.D. Bjorken, {\it Proceedings of the International Symposium
on Electron and Photon interactions at High Energies}, Cornell, 1971,
p. 281.
\bibitem{gal-lo} H. Abramowicz, E. Gurvich, A. Levy, \pl{B420}{1998}{104}.
\bibitem{gal-ho} H. Abramowicz, E. Gurvich, A. Levy, {\it Next to leading 
order parton distributions in the photon from $\gamma^*\gamma$ and
$\gamma^* p$ scattering}, paper 684 submitted to the XXIX International
Conference on High Energy Physics, Vancouver, 23-29 July, 1998.
\bibitem{alpha-prime} G. Jaroszkiewicz, P.V. Landshoff, \prev{D10}{1974}{170}; 
\\ P.D.B. Collins, F.D. Gault, A. Martin, \np{B80}{1974}{135}.
\bibitem{predazzi} For a recent review see E. Predazzi, {\it Diffraction: 
past, present and future}, hep-ph/9809454, 1998.
\bibitem{fs-physrep} L.L. Frankfurt, M.I. Strikman, \prep{160}{1988}{235}.
\bibitem{ingelman-schlein} G. Ingelman, P.E. Schlein, \pl{B152}{1985}{256}.
\bibitem{collins-qcd-fact} J.C. Collins, \prev{D57}{1998}{3051}.
\bibitem{h1-incl-diff} H1 Collab., C. Adloff et al., \zp{C76}{1997}{613}.
\bibitem{zeus-incl-diff} ZEUS Collab., J. Breitweg et al., {\it Measurement 
of the diffractive cross section in deep inelastic scattering using
ZEUS 1994 data}, DESY 98-084, to be published in {\em Europ. Phys. J.}, 1998.
\bibitem{ryskin} M.G. Ryskin, \sjnp{52}{1990}{529};\\
M.G. Ryskin, M. Besancon, {\it Heavy photon dissociation in deep inelastic 
scattering}, in {\it Proceedings of Physics at HERA}, W. Buchmueller, 
G. Ingelman (eds.), p. 215, 1991.
\bibitem{brodsky} S.J. Brodsky et al., \prev{D50}{1994}{3134}.
\bibitem{yura} See discussion in Yu.L. Dokshitzer, {\it QCD, theoretical 
issues}, to appear in {\it Proceedings of the HEP EPS Conference},
Jerusalem, August 1997, hep-ph/9801372, 1998.  
\bibitem{forshaw-ross} J.R. Forshaw, D.A. Ross, {\it Quantum
    Chromodynamics and the Pomeron}, Cambridge University Press, 1997.
\bibitem{collins-regge} P.D.B. Collins, {\it An introduction to Regge
    theory and High Energy Physics}, Cambridge University Press, 1977.
\bibitem{zeus-pom-traj} ZEUS Collab., {\it Study of vector meson
    production at large $|t|$ at HERA and determination of the Pomeron 
trajectory}, paper 788 submitted to the XXIX International
Conference on High Energy Physics, Vancouver, 23-29 July, 1998.
\bibitem{al-shrinkage} A. Levy, \pl{B424}{1998}{191}.
\bibitem{mcdermot-gena} M.F. McDermott, G. Briskin, {\it Diffractive 
structure functions in DIS}, in {\it Proceedings of Future Physics at HERA},
G. Ingelman, A. De Roeck, R. Klanner (eds.), p. 691, 1996. 
\bibitem{yndurain} F.J. Yndurain, {\it Quantum Chromodynamics. 
An introduction to the theory of quarks and gluons.}, Springer-Verlag,
1983.
\end{thebibliography}
\end{document}